\documentstyle[aps,prb,psfig]{revtex}
\begin{document}
\title{Interactions and Weak Localization: Perturbation Theory and Beyond}
\author{Dmitrii S. Golubev$^{1,3}$ and Andrei D. Zaikin$^{2,3}$}
\address{$^{1}$ Physics Department, Chalmers University of Technology,
S-41296 G\"oteborg, Sweden\\
$^2$ Forschungszentrum Karlsruhe, Institut f\"ur Nanotechnologie,
76021 Karlsruhe, Germany\\
$^3$ I.E.Tamm Department of Theoretical Physics, P.N.Lebedev
Physics Institute, Leninskii pr. 53, 117924 Moscow, Russia}

\maketitle

\begin{abstract}
We establish an explicit correspondence between perturbative and nonperturbative
results in the problem of quantum decoherence in disordered conductors. We demonstrate
that the dephasing time $\tau_{\varphi}$ cannot be unambiguously extracted
from a perturbative calculation. We show that the effect of the
electron-electron interaction on the magnetoconductance is described by the
function $A_d(t)\exp (-f_d(t))$. The dephasing time is determined by 
$f_d(t)$, i.e. in order to evaluate $\tau_{\varphi}$ it is sufficient to 
perform a nonperturbative analysis with an exponential accuracy.
The effect of interaction on the pre-exponent $A_d(t)$ is important
if one calculates the interaction-dependent part of the weak localization 
correction for strong magnetic fields.
The zero temperature dephasing time drops out of this
correction in the first order due to the exact cancellation of the linear in 
time $T$-independent contributions from the exponent and the
pre-exponent.
Nonlinear in time $T$-independent contributions do not cancel out already in
the first order of the perturbation theory and yield an additional contribution 
to dephasing at 
all temperatures including $T=0$.

\end{abstract}



\section{Introduction}

Recent experiments by Mohanty, Jariwala and Webb \cite{Webb} strongly indicate 
an intrinsic nature of
a low temperature saturation of the electron decoherence time $\tau_{\varphi}$
in disordered conductors \cite{AAK,AAK1}. It was argued in Refs. \onlinecite{Webb,MW} 
that zero
point fluctuations of electrons could be responsible for a finite dephasing
at low temperatures. These as well as various other recent experimental
results attract a lot of attention to the fundamental role of interactions in
disordered mesoscopic systems.

A theory of the above phenomenon \cite{Webb} was proposed in our papers
\cite{GZ1,GZ2}.
We demonstrated that
electron-electron interactions in disordered systems can indeed be responsible
for a nonzero electron decoherence rate down to $T=0$. Our results \cite{GZ1,GZ2}
are in a good agreement with experimental findings \cite{Webb}.
We also argued \cite{GZ98} that this interaction-induced decoherence has the
same physical nature as in the case of a quantum particle interacting with a bath
of harmonic oscillators \cite{CL,Weiss}.

The low temperature saturation of the decoherence rate $1/\tau_{\varphi}$ on a level
predicted in Refs. \onlinecite{GZ1,GZ2} has serious theoretical consequences. Therefore
it is not surprising that these predictions initiated
intensive theoretical debates \cite{AGA,AAG1,GZr1,AAG2,GZr2,CI,VA,IFS,RSC}. In contrast
to Refs. \onlinecite{Webb,MW,GZ1,GZ2,GZ98}, various authors \cite{AGA,AAG1,AAG2,CI,IFS,RSC}
argued that interaction-induced electron dephasing at $T \to 0$ is not possible. 
Vavilov and Ambegaokar (VA) \cite{VA} argued the quantum correction to the classical result
\cite{AAK,AAK1} should be small at least in the limit $T\tau_{\varphi} \gg 1$.

It should be emphasized that the above discussion goes far beyond the problem
of electron dephasing only. This discussion is important for a general understanding
of the role of the electron-electron interactions in mesoscopic systems
at low temperatures. According e.g. to Aleiner, Altshuler and Gershenzon (AAG) 
\cite{AAG2} this role is
merely to provide a (temperature dependent) renormalization of a disordered
potential of impurities. Within this picture, at sufficiently low $T$ (when the effect of
thermal fluctuations is small and can be neglected) electrons propagate in an
effective inhomogeneous {\it static} potential which should be determined
self-consistently in the presence of Coulomb interaction. If so, electron
scattering on such a static potential is not any different from that on
static impurities and, hence, cannot lead to dephasing.
Our results \cite{GZ1,GZ2} suggest a different picture, according to which {\it dynamical}
effects are important at all temperatures down to $T=0$ and the high frequency
``quantum'' modes with $\omega >T$ {\it do} contribute to dephasing.

Two main arguments supporting the
first (``static'') picture are usually discussed
\cite{AGA,AAG1,AAG2,CI,VA,IFS,RSC}. The first argument is quite general and
is not necessarily related to electrons in a disordered metal. One
can argue \cite{AGA,AAG2,IFS} that a particle with energy $\sim T$ cannot excite harmonic
oscillators with frequencies $\omega >T$ and, hence the latter will at most lead to 
renormalization effects. It is easy to observe, however, that this argument
explicitly contradicts to the exact results
obtained e.g. within the Caldeira-Leggett model \cite{CL}. There
even at $T=0$ the off-diagonal elements of the particle density matrix decay at
a finite length set by interaction. This effect is due to {\it all} high 
frequency
modes of the effective environment, i.e. the picture is by no means ``static''
(see also Refs. \onlinecite{GZ98,GZr1,GZr2} for further discussion).

One can also modify the above argument and conjecture \cite{VA} that the system of electrons 
can behave differently from a bosonic
one \cite{CL} because of the Pauli principle which restricts scattering space for
electrons at low $T$ and, hence, their ability to exchange energy. Again, this argument
contradicts to the well known results obtained for {\it fermionic} systems.
E.g. it is well established \cite{Naz,SZ} that tunneling electrons exchange energy with
the effective environment (formed by other electrons in the leads) even at $T=0$.
This exchange results in the temperature independent broadening of
the effective energy distribution $P(E)$ for tunneling electrons \cite{Naz}.
This so-called ``P(E)-theory'' was verified in many experiments \cite{many}.
A close formal and physical similarity between the theory \cite{Naz} and our
analysis \cite{GZ2} is discussed in Ref. \onlinecite{GZPE}.

The second argument against the possibility of the interaction-induced saturation
of $\tau_{\varphi}$ is purely formal. It is based on a perturbative 
calculation by AAG \cite{AAG2}. These authors claimed \cite{AAG1,AAG2}
that the results of this calculation explicitly contradict to our results \cite{GZ1,GZ2}
and, hence, the latter are incorrect. However, a convincing comparison between the
two calculations was not presented. Furthermore, it our Reply
\cite{GZr1,GZr2} we pointed out that the origin of the above controversy lies deeper, 
and the AAG's suggestion that our calculation is ``profoundly incorrect'' 
can not be taken seriously. We argued that both approaches {\it do agree}
on a perturbative level and the key difference
between them is that our calculation \cite{GZ2} is nonperturbative
while the analysis by AAG does not go beyond the first order in the interaction and,
on top of that, involves additional approximations not contained in our paper \cite{GZ2}.
For instance, for the exactly solvable Caldeira-Leggett model we demonstrated 
\cite{GZr1,GZr2} that within the perturbative approach involving analogous
approximations one arrives at incorrect results and misses the effect of 
quantum decoherence at low temperatures.

Motivated by this discussion as well as by the fundamental importance of the problem 
we have undertaken an additional
analysis of the effect of interaction-induced decoherence in disordered metals. This
analysis will help us to demonstrate the actual relation
between our approach \cite{GZ2} and that of AAG \cite{AAG2}.
Since it is hardly possible to settle a calculational dispute without
presenting sufficiently many details, in this paper we made an effort to
provide the reader with such details of our calculation.  

The structure of the paper is as follows. In section 2 we will demonstrate a principal
insufficiency of the perturbation theory in the interaction for the problem
of quantum dephasing. We will argue that $\tau_{\varphi}$ cannot be unambiguously
extracted even from a correct perturbative calculation. In section 3 we extend our
nonperturbative calculation \cite{GZ2}. We will carry out a complete analysis of the
problem with the exponential accuracy. We will also present semi-quantitative arguments
which, however, will be sufficient in order to understand the effect of interaction on
the pre-exponent.  In section 4 we perform a detailed perturbative
calculation and demonstrate that at low $T$ some previous perturbative results
are based on several insufficient approximations, the main of which is the
golden rule approximation. We also establish an explicit relation between nonperturbative
\cite{GZ2} and perturbative \cite{AAG2} calculations.
A close formal similarity between the problem in question and the exactly solvable
Caldeira-Leggett model is analyzed in section 5. In section 6 we briefly summarize
our main observations. For the sake of convenience we will briefly announce 
the main steps of our calculation in the beginning of each section. Some further 
technical details are presented in Appendices A, B, C and D. In Appendix E we 
discuss the results \cite{CI,VA,IFS}.

\section{Insufficiency of the perturbation theory}

In this section we will demonstrate a principal insufficiency of a perturbative
(in the interaction) approach to the problem of quantum dephasing. In the subsection
A we will present some general remarks concerning the role of the perturbation theory
for the problem of a quantum mechanical particle interacting with
other quantum degrees of freedom. In the subsection B we discuss
the relation between perturbative and nonperturbative calculations 
of the magnetoconductance and the decoherence time $\tau_{\varphi}$
in disordered conductors.

\subsection{General remarks}

The time evolution of
the density matrix of such a particle is defined by the following equation:
\begin{equation}
\rho(t,x_{1f},x_{2f})=\int dx_{1i}dx_{2i}\;\; J(t,x_{1f},x_{2f},x_{1i},x_{2i})
\rho(0,x_{1i},x_{2i}),
\label{rho}
\end{equation}
where $x$ is the particle coordinate. The kernel $J$ depends on the 
Feynman-Vernon influence functional \cite{FV,FH} and contains the full information
about the effect of interaction. This kernel can formally be expanded in
powers of the interaction strength
\begin{equation}
J(t)=\sum_{n=0}^{\infty}J^{(n)}(t).
\label{jnt}
\end{equation}
The ``noninteracting'' kernel $J^{(0)}$ does not change the state of the system
(provided its initial state is an eigenstate of the noninteracting Hamiltonian) and
in this sense it is equivalent to the unity operator. All other terms
of this expansion grow with time the faster the larger the number $n$ is. As a result
in general all these terms (\ref{jnt}) become important for sufficiently long times. Hence,
the perturbation theory in the interaction (which amounts to keeping only several first
terms of the expansion (\ref{jnt})) is equivalent to the short time expansion of the
exact density matrix. Thus in general this perturbation theory cannot correctly
describe the long time behavior of the interacting system no matter how weak the
interaction is.

Keeping only the term $J^{(1)}$ in (\ref{jnt}) under some additional
assumptions one can express the probability $W_{ii}$ for the particle $x$
to remain in its initial state as
\begin{equation}
W_{ii}(t)= 1-\int_0^tdt_1\int_0^{t_1}dt_2K(t_1,t_2),
\label{wii}
\end{equation}
where the kernel $K(t_1,t_2)$ can be derived from the influence functional \cite{FH}
and will not be specified here. In equilibrium one usually has $K(t_1,t_2)\equiv K(t_1-t_2)$.
Eq. (\ref{wii}) applies at short times, when
the second term is still much smaller than unity. But even in
this limit the correct information can be missed by insufficient approximations.
For instance, the frequently used approximation amounts to retaining only the $\omega =0$
term in the kernel $K_{\omega}= \int (d \tau /2\pi)K(\tau )\exp ( i \omega \tau )$.
Within this so-called golden rule approximation one finds
\begin{equation}
W_{ii}(t)= 1-\Gamma t, \;\;\;\;\;\; \Gamma = \pi K_{\omega=0}.
\label{wii1}
\end{equation}
Furthermore, assuming that the effect of higher order terms in the expansion (\ref{jnt})
can be accounted for by exponentiating the last term in (\ref{wii1}) one immediately
arrives at $W_{ii}(t)=\exp (-\Gamma t)$.

Obviously the above set of approximations is justified only in special cases. For
instance, the golden rule approximation can work only provided the kernel $K(\tau )$ decays
rapidly as compared to other relevant time scales in the problem. This could be the
case e.g. at sufficiently high temperatures. In general, and especially in the low
temperature limit, the golden rule approximation (\ref{wii1}) fails. And it is
particularly dangerous to combine the {\it short} time perturbative expansion with the
{\it long} time golden rule approximation. E.g. if $K_{\omega =0}$ happens to be zero,
it would follow from (\ref{wii1}) that the particle will stay in its initial state
forever even in the presence of interaction. Obviously this cannot be the case.
The exponential decay of the probability $W_{ii}$ in time is also an artifact of
the golden rule approximation. In general the time dynamics of an interacting system
is much more complicated, and it should be determined from eq. (\ref{rho}).

In eq. (\ref{rho}) is usually implied that the initial density matrix $\rho(0)$
does not coincide with the exact reduced equilibrium density matrix
for the {\it interacting} system. The standard approach is simply 
to factorize the initial density matrix \cite{FV,FH}, i.e. to represent it as
a product of the particle density matrix
$\rho(0,x_{1i},x_{2i})$ and the equilibrium density matrix of all other
degrees of freedom. In this case, even if initially both the particle and the
environment were in their noninteracting ground states at $T=0$,
the relaxation process occurs because the factorized density matrix
does not describe the ground state of the interacting system. One could 
question the relevance of such initial conditions e.g. to the problem
of electron transport in disordered conductors in the presence of interaction.
Indeed, in this case the density matrix is never factorized and no time
evolution can be expected for the equilibrium density matrix of the whole 
interacting system. 
Hence, at $T=0$ in equilibrium no relaxation should occur. 

In order to clarify the situation let recall the formal expression for 
the conductivity (see the eq. (\ref{sigma00}))
\begin{equation}
\sigma\propto \int\limits_{-\infty}^t dt'\int
d\bbox{r}_{1i}d\bbox{r}_{2i}J(t-t',d\bbox{r}_{1f},d\bbox{r}_{2f},
d\bbox{r}_{1i},d\bbox{r}_{2i})\rho_{\rm eff}(\bbox{r}_{1i},\bbox{r}_{2i}),
\end{equation}
where $\rho_{\rm eff}(\bbox{r}_{1i},\bbox{r}_{2i})=
(\bbox{r}_{1i}-\bbox{r}_{2i})\rho_{0}(\bbox{r}_{1i},\bbox{r}_{2i})$ and
$\rho_{0}$ is the equilibrium electron density matrix. We observe
that the effective initial density matrix $\rho_{\rm eff}$ in this expression 
is strongly perturbed at all $T$ as compared to $\rho_{0}$ due to the 
factor $(\bbox{r}_{1i}-\bbox{r}_{2i})$. Therefore relaxation always takes
place in our problem. Since for a dissipative system 
relaxation times do not depend on the initial conditions, one can safely assume
the initial density matrix to be, for instance, factorized. Actually the same
assumption is used within the diagrammatic approach: a complete equivalence between
eq. (\ref{sigma00}) (factorized density matrix) and the diagrammatic
expression for the conductance \cite{AAG2} was demonstrated in Appendix A 
in the first order in the interaction.

\subsection{Magnetoconductance}

The weak localization correction to the conductivity
of a disordered metal can be expressed in the following form
\begin{equation}
\delta\sigma_d(H)=-\frac{2e^2D}{\pi}\int\limits_{\tau_e}^{\infty}dt
A_d(t)\exp\left(-t/\tau_H-f_d(t)\right),
\label{dsigWL}
\end{equation}
$\tau_e=l/v_F$ is the elastic electron mean free time.
The function $f_d(t)$ increases with time and describes the Cooperon decay
due to interaction ($f_d(t)$ equals to zero without interaction).
The presence of the magnetic field $H$ causes an additional
decay on a time scale $\sim \tau_H$. By varying the magnetic field
and thus $\tau_H$ (which decreases with increasing $H$) one can 
extract information about the interaction-induced decoherence directly from the
magnetoconductance measurements.

The pre-exponential function $A_d(t)$ without interaction is
$A_d^{(0)}(t)=1/(4\pi Dt)^{d/2}$. In the presence of interaction the
function $A_d(t)$ will, of course, depend on the interaction as well.
As it is demonstrated below, this dependence is can be ignored while
calculating the decoherence time $\tau_{\varphi}$ which should only
be extracted from the function $f_d(t)$ in the exponent of (\ref{dsigWL}).
This is the procedure of Ref. \onlinecite{GZ2}. However,
the dependence of the pre-exponent $A_d(t)$ on the interaction {\it is}
important if one wants to recover the subleading in $\tau_H/\tau_{\varphi}$
term in the expression for $\delta \sigma_d (H)$ in the limit of a
strong magnetic field $\tau_H \ll \tau_{\varphi}$. In this case
only short times $t \lesssim \tau_H$ contribute to the integral (\ref{dsigWL})
and it is sufficient to perform a short time expansion of both $\exp (-f_d(t))$
and $A_d(t)$. This expansion mixes terms important and unimportant for
dephasing and in general makes it {\it impossible} to extract
correct information about the dephasing time $\tau_{\varphi}$ from
the perturbation theory even in the limit of strong magnetic
fields $\tau_H \ll \tau_{\varphi}$.

In order to illustrate this conclusion let us restrict ourselves
to a quasi-1d case. The expression (\ref{dsigWL}) may then be
rewritten as follows
\begin{equation}
\delta\sigma_1(H) = -\frac{e^2\sqrt{D}}{\pi^{3/2}}
\int\limits_0^{+\infty} \frac{dt}{\sqrt{t}} e^{-t/\tau_H}F(t/\tau_{\varphi}),
\label{1}
\end{equation}
where the function $F(t/\tau_{\varphi})$ accounts for the interaction.
Note, that the function $F$ can (and in general does) depend not only on one but on
several parameters $F=F(t/\tau_1,t/\tau_2,...,t/\tau_n)$. In this section
we will assume that $F$ depends on only one parameter $\tau_{\varphi}$. This is
sufficient for our purposes.

In the absence of interaction $F \equiv 1$ and the divergence in the
integral (\ref{1}) is cut at times $t \sim \tau_H$. In this case from (\ref{1})
we reproduce the well known result
\begin{equation}
\delta\sigma^{(0)}_1 = -\frac{e^2}{\pi}\sqrt{D\tau_H}.
\label{H}
\end{equation}
For large $\tau_H$ (i.e. for $H \to 0$) the result (\ref{H}) diverges
and the effect of interaction should be taken into account. Provided
in the long time limit the function $F$ decays faster than $1/\sqrt{t}$
the integral (\ref{1}) converges even for $1/\tau_H=0$ and we get
\begin{equation}
\delta\sigma_1 = -a\frac{e^2}{\pi}\sqrt{D\tau_\varphi},
\label{2}
\end{equation}
where the prefactor $a \sim 1$ which
depends on  the function $F$. The precise definition of $a$ 
is of little practical interest since this prefactor can always be removed
by rescaling of $\tau_{\varphi}$. Of importance, however, is to
describe the behavior of the function $F(t/\tau_{\varphi})$ 
at $t \sim \tau_{\varphi}$. This
allows to determine the magnitude of the dephasing time $\tau_{\varphi}$. Clearly,
a nonperturbative analysis in the interaction is needed in order to
determine the function $F$ at times $t \sim \tau_{\varphi}$ simply
because there exists no small parameter in the problem. E.g. if one would
formally decrease the interaction strength, the magnitude of the
dephasing time $\tau_{\varphi}$ would increase, but one would never
avoid the necessity to determine the function $F$ at $t \sim \tau_{\varphi}$.
Thus the problem of finding the decoherence time in disordered conductors
is {\it nonperturbative for any interaction strength}. In this respect
the statement of Ref. \onlinecite{IFS} that ``since $\tau_{\varphi}$ is
much longer than the elastic scattering time, the dephasing is weak and
there is no need to invoke nonperturbative ideas'' remains puzzling to us.
Indeed, at times of order $\tau_{\varphi}$ the dephasing is strong by definition,
and it is not clear how the condition $\tau_{\varphi} \gg \tau_e$ might
help to avoid ``nonperturbative ideas''. 

Observing this problem AAG \cite{AAG2} suggested to consider the
limit of strong magnetic fields $\tau_H \ll \tau_{\varphi}$, for which
the integral (\ref{1}) converges already at times $t \sim \tau_H$
much shorter than $\tau_{\varphi}$. In this case the weak localization correction
can be calculated perturbatively in the interaction or, equivalently,
by means a short time expansion of the function $F$. The recipe to evaluate the
dephasing time from the perturbation theory suggested by AAG
can be summarized as follows.

In the zero order in the interaction we have $F=1$ and the magnitude of the weak
localization correction (\ref{H}) increases as $\sqrt{\tau_H}$ with increasing $\tau_H$.
If, expanding in the interaction, one would recover the term $\propto \sqrt{\tau_H}$,
this term could just be added to (\ref{H}) and interpreted as an interaction-induced
renormalization effect of the bare parameters. The presence of such a term would
imply that $F$ is not anymore equal to one but acquires some
interaction correction. Nevertheless no time dependence of $F$ and, hence, no dephasing
occurs in this case and therefore the terms $\propto \sqrt{\tau_H}$ are not
``dangerous''. If, however, the first order conductance correction is found to increase
with $\tau_H$ {\it faster} than $\sqrt{\tau_H}$ and to have an opposite with respect
to $\delta\sigma^{(0)}_1$ (\ref{H}) (i.e. positive) sign, this would already
mean that the function $F$ depends on time (decays with increasing $t$) due to
interaction and, hence, nonzero dephasing occurs.
Then, if such ``dephasing'' terms are recovered within this
perturbative procedure, one should look at a temperature dependence of such terms.
If these terms are present at a finite $T$, but decrease and vanish as temperature
approaches zero this would imply that interaction does not cause any dephasing at $T=0$.
If $T$-independent positive terms growing faster than $\sqrt{\tau_H}$ are recovered
one would be able to conclude that nonzero dephasing occurs at $T=0$ already
within the first order perturbation theory in the interaction.

We are going to demonstrate that the above perturbative strategy {\it in principle}
cannot be used to correctly obtain the dephasing time $\tau_{\varphi}$ for
any magnetic field even though the correction $\delta \sigma_H$ {\it can} be
evaluated perturbatively in the limit $\tau_H \ll \tau_{\varphi}$.

To begin with, we note that already the terms $\propto \sqrt{\tau_H}$ can
easily cause troubles provided they give a (positive) contribution to
$\delta \sigma_1(H)$ large as compared to the magnitude of the zero
order term (\ref{H}). In fact, the presence of  terms
$\propto \sqrt{\tau_H}$ just implies that their time dependence saturates
already at short times $t \lesssim \tau_H$. If this saturated value turns out
to exceed the zero order term, this would only indicate the breakdown of the
perturbation expansion in the interaction and, hence, no definite conclusion
from this expansion can be drawn.

An even much more important problem is that the form of the function $F(t)$
in (\ref{1}) cannot be recovered from the perturbation theory at all. It is quite
obvious that the first order perturbative terms will depend only on the derivative
$F'(0)$. Although in the limit $\tau_H \ll \tau_{\varphi}$ the value
$\delta \sigma_1(H)$ can be calculated perturbatively in the interaction,
this would yield no information about the dephasing time $\tau_{\varphi}$.
Such information can be extracted only if one {\it assumes}
some particular form of the function $F(t/\tau_{\varphi})$. But this form should
be {\it found} rather than assumed. This task can be accomplished only if one goes beyond
the perturbation theory.

Let us consider several different functions
$F(t/\tau_{\varphi})$. Perhaps the most frequent choice of this function is based
on the assumption about purely exponential decay of the phase correlations,
in which case one has
\begin{equation}
F(t/\tau_{\varphi}) = \exp (-t/\tau_{\varphi}).
\label{1a}
\end{equation}
As it was already discussed above, this form of the function $F$ 
follows directly from the
golden rule approximation. Substituting (\ref{1a}) into (\ref{1}) in the limit
of weak magnetic fields  $\tau_H \gg \tau_{\varphi}$ one immediately arrives at the
result for the weak localization correction of the form (\ref{H}) with
$\tau_H$ substituted by $(1/\tau_H+1/\tau_{\varphi})^{-1}$. In the limit
of weak magnetic fields  $\tau_H \gg \tau_{\varphi}$ the result (\ref{2}) with $a=1$
is recovered. In the opposite limit $\tau_H \ll \tau_{\varphi}$ eqs. (\ref{1})
and (\ref{1a}) yield
\begin{equation}
\delta\sigma_1 - \delta\sigma^{(0)}_1 \simeq
\frac{e^2}{2\pi}\frac{\sqrt{D}\tau_H^{3/2}}{\tau_{\varphi}},
\label{2a}
\end{equation}
where $\delta\sigma^{(0)}_1$ is defined in (\ref{H}). Another possible choice
of the function $F$ can be
\begin{equation}
F(t/\tau_{\varphi}) = \exp (-(t/\tau_{\varphi})^{3/2}).
\label{1b}
\end{equation}
The reason for such a choice will become clear later. The
substitution of (\ref{1b}) into (\ref{1}) again yields the result (\ref{2})
(with $a=2\Gamma(1/3)/3\sqrt{\pi}\simeq 1.0076$, $\Gamma (x)$ is the Euler gamma-function)
in the limit  $\tau_H \gg \tau_{\varphi}$,
while in the opposite limit  $\tau_H \ll \tau_{\varphi}$ from (\ref{1}) and
(\ref{1b}) one obtains
\begin{equation}
\delta\sigma_1- \delta\sigma^{(0)}_1
\simeq
\frac{e^2\sqrt{D\tau_H}}{\pi^{3/2}}\left(\frac{\tau_H}{\tau_{\varphi}}\right)^{3/2}.
\label{2b}
\end{equation}
Comparing (\ref{2a}) and (\ref{2b}) we observe that for strong magnetic fields
the interaction corrections to the leading order term (\ref{H}) are different
depending on the choice of the function $F$, even though for weak magnetic fields
both choices (\ref{1a}) and (\ref{1b}) yield the same result (\ref{2}) with only
slightly different values of a numerical prefactor $a$.

The magnetoresistance data are frequently fitted to the formula \cite{AAK1}
\begin{equation}
\delta\sigma_1(H)=\frac{e^2}{\pi}\sqrt{D\tau_\varphi}
\frac{{\rm Ai}(\tau_\varphi/\tau_H)}{{\rm Ai}'(\tau_\varphi/\tau_H)},
\label{Airy}
\end{equation}
where Ai$(x)$ is the Airy function. In the limit $\tau_H\gg\tau_\varphi$ this equation again reduces to (\ref{2}) with
the factor $a=-{\rm Ai}(0)/{\rm Ai}'(0)\simeq 1.372$. In the opposite limit
$\tau_H \ll \tau_{\varphi}$ one finds
\begin{equation}
\delta\sigma_1 - \delta\sigma_1^{(0)} \simeq\frac{e^2\sqrt{D\tau_H}}{4\pi}
\left(\frac{\tau_H}{\tau_\varphi}\right)^{3/2}.
\label{Airy1}
\end{equation}
We observe the equivalence between (\ref{2b}) and
(\ref{Airy1}) up to a numerical prefactor of order one.

Finally, let choose the trial function $F$ in the following form:
\begin{equation}
F(t/\tau_{\varphi}) = \frac{e^{-t/\tau_\varphi}\sqrt{bt}}
{\sqrt{\tau_\varphi(1-e^{-bt/\tau_\varphi})}},
\label{1c}
\end{equation}
where $b$ is a numerical coefficient of order one. Combining (\ref{1}) and (\ref{1c})
we find
\begin{equation}
\delta\sigma_1(H) = -\frac{e^2}{\pi}\sqrt{D\tau_\varphi}\frac1{\sqrt{b}}
\frac{\Gamma \left(\frac1{b}\left(1+\frac{\tau_{\varphi}}{\tau_H}\right)\right)}
{\Gamma \left(\frac12+\frac1{b}\left(1+\frac{\tau_{\varphi}}{\tau_H}\right)\right)}.
\label{20}
\end{equation}
In the limit
$\tau_H \ll \tau_{\varphi}$ one can expand this equation in powers of
$\tau_H/\tau_{\varphi}$ and get
\begin{equation}
\delta\sigma_1 - \delta\sigma^{(0)}_1 \simeq
\frac{4-b}{4}\frac{e^2\sqrt{D\tau_H}}{2\pi}\frac{\tau_H}{\tau_{\varphi}},
\label{2c}
\end{equation}
while in the limit  $\tau_H \gg \tau_{\varphi}$ from (\ref{1}), (\ref{1c}) one again
recovers eq. (\ref{2}) with a slightly modified numerical prefactor $a$ (which
now also depends on the value $b$). Absorbing $a$ by a proper redefinition of
$\tau_{\varphi}$ one can plot the result (\ref{1}) for the trial functions
(\ref{1a}), (\ref{1c}) (for different values of $b$) and eq. (\ref{Airy}) depending
on the magnetic field (or $\tau_H$).  These plots are presented in Fig. 1. We
observe that all four plotted functions are very close to each other (e.g. the
maximum deviation between (\ref{Airy}) and $\delta \sigma_1(H)$ obtained from (\ref{1c})
with $b=4$ does not exceed 0.01). If one would fit the experimental data for
the magnetoconductance with any of these four functions one (i) would not be
able to distinguish between them within typical error bars and (ii) would
obtain {\it the same} value $\tau_{\varphi}$ for all these functions (up to a
prefactor $a \sim 1$ absorbed in $\tau_{\varphi}$ anyway). In other words, the
results for $\tau_{\varphi}$ extracted from fitting the experimental data to
several different functions $F(t/\tau_{\varphi})$ will be practically
insensitive to the particular form of $F$ as long as its decay at long times is
sufficiently fast to provide an effective cutoff for the integral (\ref{1}) at
$t \sim \tau_{\varphi}$.

At the same time if one tries to extract $\tau_{\varphi}$ from the perturbation theory
in the interaction one immediately arrives at ambiguous and contradictory results.
Let us, for example, consider the perturbative result of AAG
\begin{equation}
\delta\sigma_1 - \delta\sigma^{(0)}_1 \propto T\tau_H^2
\label{3a}
\end{equation}
(see e.g. eq. (4.8b) of Ref. \onlinecite{AAG2}) and, following the above paper,
assume an exponential decay of correlations (\ref{1a}). In this case the dephasing time
$\tau_{\varphi}$ is obtained from a direct comparison of eq. (\ref{2a}) (equivalent to
eq. (4.3b) of Ref. \onlinecite{AAG2} or eq. (3) of Ref. \onlinecite{AAG1}) with
eq. (\ref{3a}) (or eq. (4.8b) in Ref. \onlinecite{AAG2}). One obtains
\begin{equation}
1/\tau_{\varphi}^{AAG} \propto T\sqrt{\tau_H}
\label{3b}
\end{equation}
(cf. eq. (4.9b) of Ref. \onlinecite{AAG2}). The result (\ref{3b}) is essentially
based on the {\it assumption} about a purely exponential decay (\ref{1a}). 
Note, however, that
{\it a-priori} there is no reason to assume such a decay. [Just on the contrary,
it will be demonstrated below that this is
{\it not} the case for the problem in question.] The cutoff functions
(\ref{1b}), (\ref{1c}) (and many others) yield the same result (\ref{2}) as the 
function (\ref{1a}) and one can
hardly make a distinction between them from the magnetoconductance measurements (Fig. 1).

For instance, if one sticks to the choice (\ref{1b}),
one should extract $\tau_{\varphi}$ by comparing eqs. (\ref{2b}) and
(\ref{3a}). This comparison yields $\tau_{\varphi}$ independent of $\tau_H$ and
$1/\tau_{\varphi} \propto T^{2/3}$. The
latter form coincides with the well known result by Altshuler, Aronov and Khmelnitskii
(AAK) \cite{AAK} but is in an obvious
disagreement with (\ref{3b}). If, instead of (\ref{1b}), one uses the trial function
(\ref{1c}) and compares (\ref{2c}) and (\ref{3b}), one finds
$\tau_{\varphi} \propto (4-b)/(T\tau_H^{1/2})$, i.e. positive,
zero and even negative (!) dephasing times respectively for $b<4$, $b=4$ and $b>4$.
However, all these dramatic differences in the first order results have no effect both on 
the form of the magnetoconductance  $\delta \sigma_d (H)$ (Fig. 1) and on the value 
$\tau_{\varphi}$ extracted from it. 

Thus, whatever result is obtained in the first order perturbation theory in the interaction,
it is yet insufficient to draw any definite conclusion about the dephasing time 
$\tau_{\varphi}$. The problem
is essentially nonperturbative and should be treated as such. The corresponding analysis
was developed in our paper \cite{GZ2} and will be extended further in the next section.

\begin{figure}[t]
\centerline{\psfig{file=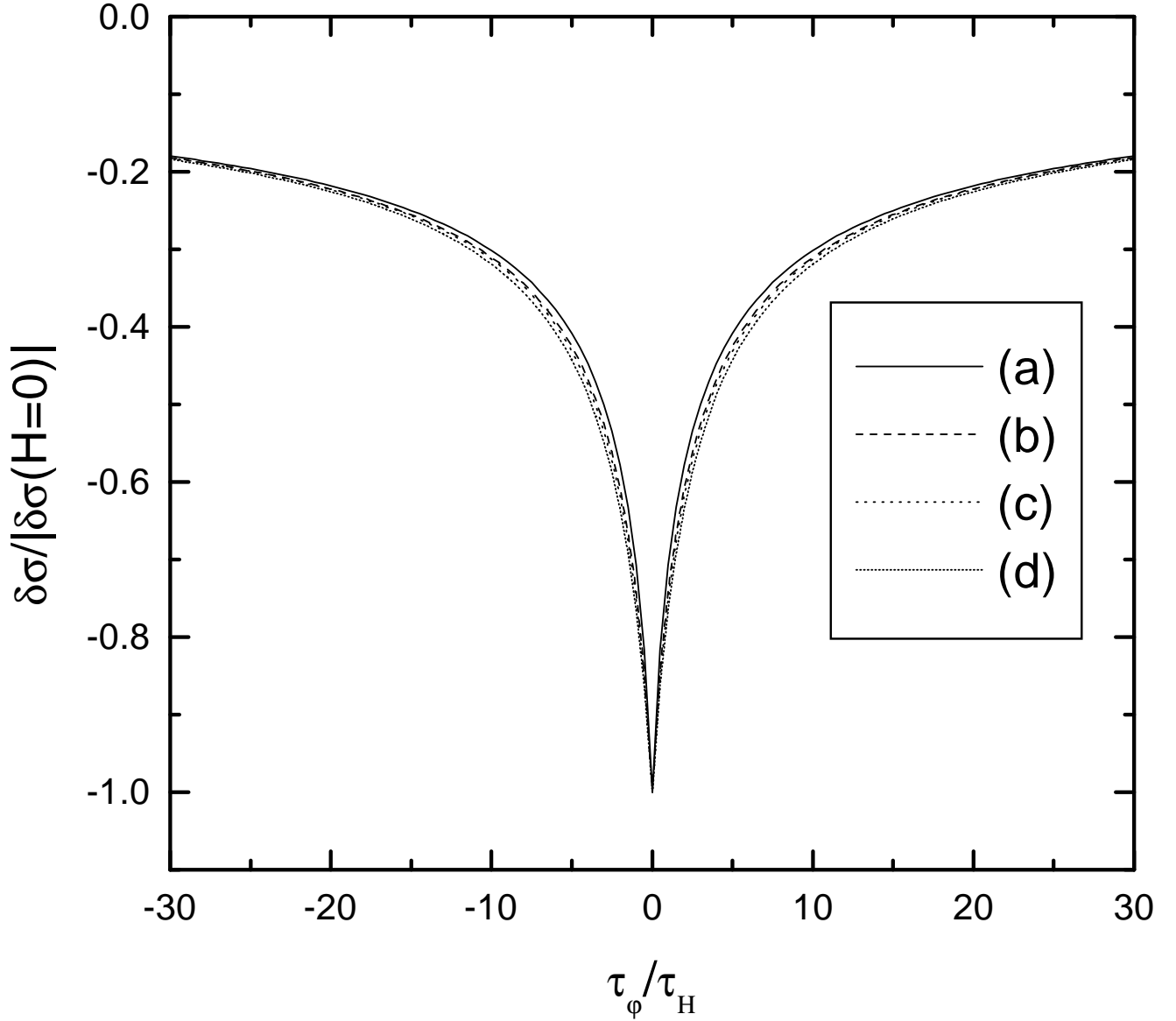,height=16cm}}
\caption{Magnetoconductance curves corresponding to 
different cutoff procedures:
(a) purely exponential cutoff (\protect\ref{1a});
(b) Airy function of eq. (\protect\ref{Airy}));
(c) the cutoff function $F$ is given by the eq. (\protect\ref{1c}) with $b=4$
($a\simeq 1.479$)
corresponding to ``zero dephasing'' in the first order perturbation theory;
(d) the cutoff function (\protect\ref{1c}) with $b=5$
($a\simeq 1.582$) which yields ``negative dephasing'' in the first order 
perturbation theory.
Here $\tau_\varphi$ is rescaled to absorb the factor $a$ in eq. (\protect\ref{2}).
}
\label{fig1}
\end{figure}

\section{Weak Localization Correction: Nonperturbative Results}

In order to
provide a complete description of the electron-electron
interaction effect on the weak localization correction (\ref{dsigWL}) it is
in general necessary to calculate both the function $f_d(t)$
in the exponent of (\ref{dsigWL}) and the pre-exponential function $A_d(t)$.
An important observation is, however, that information about the
effect of interaction on $A_d(t)$ is not needed to correctly evaluate
the dephasing time $\tau_{\varphi}$. It suffices  to find only the function
$f_d(t)$ which describes the decay of correlations in time and provides
an effective cutoff for the integral (\ref{dsigWL}) at $t \sim \tau_{\varphi}$.
The role of the pre-exponent is merely to establish an exact numerical
prefactor. Since $\tau_{\varphi}$ is defined up to a numerical prefactor
of order one anyway, it is clear that only the function $f_d(t)$ -- and not $A_d(t)$ -- 
is really important.

In the subsection A we extend our previous analysis \cite{GZ2}
and evaluate of the function $f_d(t)$ keeping all 
the subleading terms. This procedure is important in at least two aspects: 
(i) it allows to unambiguously settle the
issue of unphysical divergences which was argued by VA \cite{VA}
to be a problem in our previous calculation \cite{GZ2} and (ii) it is necessary
to establish a direct relation between our nonperturbative approach and the perturbation
theory in the interaction. In the subsection B we will perform a
semi-quantitative analysis of the effect of interactions on the pre-exponential 
function $A_d(t)$. In the subsection C we will demonstrate that the short
time perturbative expansion of both the exponent and the pre-exponent at low $T$ 
not only can (section 2B) but {\it does} lead to missing an important information
about $\tau_{\varphi}$ in disordered conductors.

\subsection{Exponent}

The function $f_d(t)$ can be evaluated by means of
the path integral formalism \cite{GZ2}. This procedure amounts to calculating the
path integral for the kernel of the evolution operator
\begin{equation}
J \sim \int {\cal D}\bbox{r}\int {\cal D}\bbox{p} \, \exp (iS_0-iS_0'-iS_R-S_I)
\label{10}
\end{equation}
within the saddle point approximation on pairs of time reversed paths and to averaging
over diffusive trajectories. Here $S_0$ and $S_0'$ represent the electron action
on the two parts of the Keldysh contour, while $iS_R+S_I$ accounts for the interaction.
The effective action (\ref{10}) was derived in our Ref. \onlinecite{GZ2}, for
the sake of convenience we reproduce the explicit expressions in Appendix A
(eqs. (\ref{J17}-\ref{SI})) together with the expression for the conductance
of a disordered metal in terms of the kernel of the evolution
operator $J$ and the electron density matrix (\ref{sigma00}).

The saddle point approximation procedure was
described in details in Ref. \onlinecite{GZ2}. One can demonstrate 
that the contribution of the real part $S_R$ of the action (\ref{SR}) vanishes on
any pair of time reversed diffusive paths. By no means this cancellation occurs by chance,
rather it is a generic property of a wide class of influence functionals describing
dissipative environments. E.g. similar cancellation is observed in the Caldeira-Leggett
model \cite{CL}, the relation to which will be discussed in Section 5.
Also we would like to
point out that for any pair of paths $S_R$ contributes only to the {\it real} part
of the effective action and, hence, can never cancel an {\it imaginary} contribution
from the term $S_I$ (\ref{SI}). Anyway, the function $f_d(t)$ in the exponent 
is determined solely by the
imaginary part of the action $S_I$ (\ref{SI}) and it is given by the following
expression \cite{GZ2}:
\begin{equation}
f_d(t)=e^2\int\limits_0^t dt_1 \int\limits_0^t dt_2
\left\langle I(t_1-t_2,\bbox{r}(t_1)-\bbox{r}(t_2))-
I(t_1+t_2-t,\bbox{r}(t_1)-\bbox{r}(t_2)) \right\rangle_{\rm diff},
\label{ft}
\end{equation}
where the function $I$ is defined in (\ref{I55}). In equilibrium it is expressed in terms of an imaginary part of the inverse effective dielectric susceptibility
$1/\epsilon (\omega ,k)$ multiplied by $\coth (\omega /2T)$.
The first term (\ref{ft}) describes the contribution of the self energy diagrams
(diagrams (a) and (b) in Fig. 2), while the second term is due to the vertex
diagrams ((c) and (d) in Fig. 2). In order to evaluate the function
$f_d(t)$ (\ref{ft}) we introduce the Fourier transform of the function $I$ and
then average over the diffusive trajectories with the aid of a standard replacement
\cite{CS}
$\langle\exp[i\bbox{k}(\bbox{r}(t_1)-\bbox{r}(t_2))]\rangle_{\rm diff}
=\exp(-Dk^2|t_1-t_2|)$. After the integration over $\bbox{k}$ we obtain
\begin{eqnarray}
f_d(t)&=&
\frac{4e^2D^{1-d/2}}{\sigma_d (2\pi)^d}
\left(\int\frac{d^dx}{1+x^4}\right)
\int\frac{d\omega\; d\omega'}{(2\pi)^2}
\left[|\omega'|^{d/2-2}(\omega-\omega')\coth\frac{\omega-\omega'}{2T}\;
\frac{1-\cos\omega t}{\omega^2}
\right.
\nonumber\\
&&
\left.
-
|\omega'|^{d/2-2}\omega\coth\frac{\omega}{2T}\;
\frac{\cos\omega t-\cos\omega' t}{\omega^{\prime 2}-\omega^2}
\right].
\label{fott}
\end{eqnarray}
Here again the first and the second terms in the square brackets are
respectively from the self energy (Fig. 2a,b) and the vertex (Fig. 2c,d)
diagrams. For 1d and 2d cases the integral of the first term over $\omega'$
diverges at $\omega'\to 0$. However, it is easy to check that this divergence
is exactly canceled by
the second term, the whole integral is finite in any dimension and does not
require artificial infrared cutoffs. Various divergences are rather
inherent to the perturbation theory in the interaction and -- at least in part -- are due
to insufficiency of the perturbative expansion in our problem,
especially at low temperatures.
It is also useful to note that at $T \to 0$ the leading contribution
to $f_d(t)$ in the long time limit is insensitive to a divergence contained
in the first term in the square brackets (\ref{fott})
and can be derived only from this term \cite{GZ2}.

The integrals in (\ref{fott}) can be handled
in a straightforward manner. Technically it is sometimes more convenient to
perform calculations in the real time rather than in the frequency representation. Here
we will present the calculation for the 1d case.

First we find the explicit expression for $I(t,x)$ (\ref{I55}):
\begin{equation}
I(t,x)=\frac{1}{\sigma_1}\int\frac{d\omega dk}{(2\pi)^2}\;\;
\frac{\omega}{k^2}\coth\frac{\omega}{2T}\;\;e^{-i\omega t+ikx}=
\frac{|x|}{2\pi\sigma_1}\left(-\frac{d}{dt}{\cal P}[\pi T\coth (\pi Tt)]\right).
\label{I1}
\end{equation}
Here ${\cal P}$ stands for the principal value, i.e.
${\cal P}[\pi T\coth (\pi Tt)]$ is a distribution rather than an ordinary
function divergent at $t\to 0$.
For a given diffusive trajectory (and on sufficiently long time scales) this function
can be replaced by the following function of time
\begin{eqnarray}
\langle I(t-0,x(t)-x(0))\rangle_{\rm diff}&=&\int dx\; I(t,x){\cal D}(|t|,x)=
\frac{1}{2\pi\sigma_1}\left(-\frac{d}{dt}{\cal P}[\pi T\coth (\pi Tt)]\right)
\int dx\; \frac{|x|\exp\left[-\frac{x^2}{4D|t|}\right]}{\sqrt{4\pi D|t|}}
\nonumber\\
&=&
\frac{1}{\pi\sigma_1}\sqrt{\frac{D|t|}{\pi}}
\left(-\frac{d}{dt}{\cal P}[\pi T\coth (\pi Tt)]\right).
\label{Idiff}
\end{eqnarray}
Substitution of this equation in eq.(\ref{ft}) yields
\begin{equation}
f_1(t)=\frac{2e^2}{\pi\sigma_1}\int\limits_0^t dt_1\int\limits_0^{t_1}dt_2
\left(-\frac{d}{dt}{\cal P}[\pi T\coth (\pi Tt)]\right)\left[
\sqrt{\frac{Dt_2}{\pi}}-\sqrt{\frac{D|2t_1-t_2-t|}{\pi}}\right].
\end{equation}
Let us first integrate this expression over $t_1$ and then integrate the result by
parts. We obtain
\begin{equation}
f_1(t)=\frac{2e^2}{\pi\sigma_1}\sqrt{\frac{D}{\pi}}
\int\limits_{2\tau_e/\pi}^t dt'\; \pi T\coth(\pi Tt')\left(
\frac{t}{2\sqrt{t'}}-\frac{3}{2}\sqrt{t'}+\sqrt{t-t'}\right).
\label{ft1}
\end{equation}
The short time cutoff in (\ref{ft1}) is equivalent to
a sharp cutoff at $\omega=1/\tau_e$ in the frequency domain.

In the quantum regime $\pi Tt\ll 1$ we find
\begin{equation}
f_1(t)=\frac{e^2}{\pi\sigma_1}\sqrt{\frac{2D}{\tau_e}}t+
\frac{2e^2}{\pi\sigma_1}\sqrt{\frac{Dt}{\pi}}\left(
\ln\frac{2\pi t}{\tau_e}-6\right), \;\;\; \pi Tt\ll 1.
\label{fquantum}
\end{equation}
Note, that apart from the leading linear in time term
there exists a a smaller term $\propto \sqrt{t}\ln(t/\tau_e)$, which also grows
in time.

In order to find the function $f_1(t)$ in the opposite thermal limit
$\pi Tt\gg 1$, let us rewrite the integral (\ref{ft1}) in the form
\begin{eqnarray}
f_1(t)&=&\frac{2e^2}{\pi\sigma_1}\sqrt{\frac{D}{\pi}}\left\{
\sqrt{\frac{\pi}{2\tau_e}}t - \pi Tt^{3/2}(\coth (\pi Tt)-1)+
t\sqrt{\pi T}\int\limits_0^{\pi Tt}\frac{dx}{\sqrt{x}}
\left(\coth x-1-\frac{x}{\sinh^2x}\right)\right.
\nonumber\\
&&\left.
-\frac{3}{2\sqrt{\pi T}}\int\limits_0^{\pi Tt}dx\sqrt{x}(\coth x-1)
+\int\limits_{2\tau_e/\pi}^t dt'\;\pi T\coth(\pi Tt')\sqrt{t-t'}\right\}.
\nonumber
\end{eqnarray}
Making use of the following integrals
\begin{equation}
\int\limits_0^\infty \frac{dx}{\sqrt{x}}
\left(\coth x -1-\frac{x}{\sinh^2 x}\right)=\sqrt{\frac{\pi}{2}}
\zeta\left(\frac{1}{2}\right), \;\;\;
\int\limits_0^\infty dx\sqrt{x}(\coth x-1)=
\frac{1}{2}\sqrt{\frac{\pi}{2}}\zeta\left(\frac{3}{2}\right),
\end{equation}
we get
\begin{eqnarray}
f_1(t)&=&\frac{2e^2}{\pi\sigma_1}\sqrt{\frac{D}{\pi}}\left\{
\sqrt{\frac{\pi}{2\tau_e}}t+\frac{2\pi}{3}Tt^{3/2}+\frac{\pi\zeta(1/2)}
{\sqrt{2}}t\sqrt{T}
-\frac{3\zeta(3/2)}{4\sqrt{2}}\frac{1}{\sqrt{T}}
\right.
\nonumber\\
&&
\left.
+
\sqrt{t}\ln\left(\frac{1}{4T\tau_e}\right)
+{\cal O}\left(\frac{1}{T\sqrt{t}}\right)+
2\pi Tt^{3/2}e^{-2\pi Tt}+
{\cal O}(\sqrt{t}e^{-2\pi Tt})\right\},\;\;\; \pi Tt\gg 1.
\label{fthermal}
\end{eqnarray}
We observe that in both cases (\ref{fquantum}) and (\ref{fthermal}) there
exists a linear in time temperature independent contribution to $f_1(t)$
which determines the dephasing time $\tau_{\varphi}$ at low temperatures \cite{GZ1,GZ2,GZ98}.
Beside that at $Tt \gg 1$ there exists another term $\propto Tt^{3/2}$
which yields dominating contribution to $\tau_{\varphi}$ at high temperatures
$T \gtrsim T_q \sim 1\sqrt{\tau_{\varphi}\tau_e}$,
where the result of AAK \cite{AAK} $\tau_{\varphi} \propto T^{-2/3}$ is recovered.

In addition to both these important contributions all four
diagrams of Fig. 2 yield  subleading terms in the expression for $f_1(t)$ which also
grow with time, albeit slower than the main terms. These subleading terms also
contribute to dephasing even at $T=0$ (cf. eq. (\ref{fquantum})), however this
contribution is always smaller than that of the leading terms, typically in the parameter
$\sqrt{\tau_e/\tau_{\varphi}}$. This result is in contrast with the statement
of Ref. \onlinecite{VA}, where it was argued that at $t\sim\tau_\varphi$ the
contribution of the vertex diagrams to $f_1(t)$ can be comparable to that of
the self-energy diagrams and the term $\propto t/\sqrt{\tau_e}$, which is the
most important at $T\to 0$, can be canceled.  A straightforward calculation
demonstrates that this is not the case.

A similar calculation can be performed in 2d and 3d dimensions. In any dimension the
result can be expressed in the form
\begin{equation}
f_d(t)=t/\tau_{\varphi}^0+\delta f_d(T,t),
\label{f}
\end{equation}
where we defined
\begin{equation}
\frac{1}{\tau_\varphi^0}=\frac{\kappa_de^2(2D)^{1-d/2}}{\pi\sigma_d\tau_e^{d/2}}.
\label{tau0}
\end{equation}
A numerical prefactor $\kappa_d$ in (\ref{tau0}) (determined for a sharp high frequency cutoff
at $\omega =1/\tau_e$) is $\kappa_1=1$ for 1d, $\kappa_2=1/4$ for 2d
and $\kappa_3=1/3\pi$ for 3d. A detailed expression for the function $\delta f_1(T,t)$ is
given in (\ref{fquantum}) and (\ref{fthermal}), where we retained also several subleading
terms needed for further comparison with perturbative results (Section 4). For 2d and 3d
systems we will present only the leading order contributions to $\delta f_d (T,t)$. In 2d
we find
\begin{eqnarray}
\delta f_2(t)&=&\frac{2\gamma_0}{\pi}\frac{\tau_e}{\tau_\varphi^0}
\ln\frac{t}{\tau_e},   \quad Tt\ll 1,
\nonumber\\
\delta f_2(t)&=&\frac{2\tau_e}{\tau_\varphi^0}Tt\ln(Tt),
\quad Tt\gg 1,
\label{deltaf2}
\end{eqnarray}
Here $\gamma_0 \simeq 0.577..$ is the Euler's constant and $t \gg \tau_e$.
Similarly, in the 3d case we obtain
\begin{eqnarray}
\delta f_3(t)&\simeq &7.8\frac{\tau_e}{\tau_\varphi^0}
\quad Tt\ll 1,
\nonumber\\
\delta f_3(t)& \simeq &3\frac{(T\tau_e)^{3/2}}{\tau_\varphi^0}t
\quad Tt\gg 1,
\label{deltaf3}
\end{eqnarray}
In 3d we
used a standard approximation and replaced
$\omega\coth(\omega/2T)-|\omega|$ by $2T\theta (T-|\omega|)$.

\subsection{Pre-exponent}

As it was already discussed above, the pre-exponential function $A_d(t)$ does
not play any significant role in our problem. Therefore its rigorous calculation
at all times (which is a separate and quite complicated problem) will not be
discussed here. Of importance is to qualitatively understand how the
function $A_d(t)$ is modified in the presence of the electron-electron
interaction. Therefore we will restrict ourselves to semi-quantitative arguments
which, however, turn out to give surprisingly good agreement with the
rigorous results obtained in Sec. 4 in the short time limit.

It is well known \cite{AAK1,CS} that without interaction the function $A_d(t)$
is related to the return probability of diffusive trajectories to the same
point after the time $t$. In the presence of dissipation (described by the term
$S_R$ in the effective action) the particle energy decreases and its diffusion
slows down. This implies that at any given time $t$ the function $A_d(t)$
should exceed the pre-exponent $A^{(0)}_d(t)$ evaluated without interaction.
On the other hand, at least if the interaction is sufficiently weak, diffusion
will still take place at all times and, hence, $A_d(t)$ will decay in time, albeit
somewhat slower than $A^{(0)}_d(t)$.

Now let us try to find a typical time scale at which the deviation of $A_d(t)$
from $A^{(0)}_d(t)$ becomes of the order of $A^{(0)}_d(t)$. For the sake of 
definiteness we restrict our analysis to the 1d case.
As we have
already discussed, the real part of the action $S_R$ vanishes on the 
time reversed diffusive paths. In order to evaluate the contribution of $S_R$ in 
the path integral (\ref{10}) or (\ref{J17}) we need to include fluctuations 
around the time reversed paths. We assume that these fluctuations are small 
and neglect them in the arguments of the functions
$R(t_1-t_2,\bbox{r}_i(t_j)-\bbox{r}_k(t_n))$ in eq. (\ref{SR}). These
fluctuations are, however, important and should be kept in the arguments 
of the functions $1-2n(\bbox{p}_i,\bbox{r}_i)$. 
In equilibrium one has $1-2n(\bbox{p}_j,\bbox{r}_j)=\tanh\frac{\xi_j}{2T}$, where 
we defined $\xi_j=\frac{\bbox{p}_j^2}{2m}+U(\bbox{r}_j)-\mu$. Within the above
approximation we get
\begin{eqnarray}
S_R & \simeq & \frac{e^2}{2}\int\limits_0^t dt_1\int\limits_0^t dt_2
\left\{\left[\left\langle
R(t_1-t_2,\bbox{r}(t_1)-\bbox{r}(t_2))\right\rangle_{\rm diff}-
\left\langle R(t_1-t_2,\bbox{r}(t_1)-\bbox{r}(t-t_2))\right\rangle_{\rm diff}
\right]\right.
\nonumber\\
&&
\left.
\times\left[\tanh(\xi_1(t_2)/2T)-\tanh(\xi_2(t_2)/2T)\right]
\right\}
\label{SR11}
\end{eqnarray}

In addition to the contribution (\ref{SR11}) one should also take care 
of the corrections to the action $S_0$ (\ref{S_0}) due to the interaction.
These corrections turn out to be of the same order as (\ref{SR11}). 
In the presence of interaction 
the classical paths change and satisfy the following Langevin equation \cite{GZ2}:
\begin{equation}
m\ddot{\bbox{r}}(t')+\nabla U(\bbox{r}(t'))+e^2
\int dt'' \nabla_{\bbox{r}(t')}R(t'-t'',\bbox{r}(t')-\bbox{r}(t''))
\tanh\frac{\xi(t'')}{2T}=
-e\bbox{E}(t',\bbox{r}(t')),
\label{lange}
\end{equation}
where $\bbox{E}(t',\bbox{r})$ is the fluctuating electric field due to the
Nyquist noise. From this equation we find
\begin{eqnarray}
S_0[\bbox{p},\bbox{r}] & \simeq &
S_0[\bbox{p}^{(0)},\bbox{r}^{(0)}]+
\int\limits_0^t dt_1 \int\limits_0^{t_1} dt_2
\left\langle e^2
\int\limits_0^t dt''\dot{\bbox{r}}(t_2)
\nabla_{\bbox{r}(t_2)}R(t_2-t'',\bbox{r}^{(0)}(t_2)-\bbox{r}^{(0)}(t''))
\tanh\frac{\xi(t'')}{2T}
\right\rangle_{\rm diff}
\nonumber\\
&\simeq&
S_0[\bbox{p}^{(0)},\bbox{r}^{(0)}]+
\int\limits_0^t dt_1
\left\langle e^2
\int\limits_0^{t} dt''\;
R(t_1-t'',\bbox{r}^{(0)}(t_1)-\bbox{r}^{(0)}(t''))
\tanh\frac{\xi(t'')}{2T}
\right\rangle_{\rm diff}
\nonumber\\
&\simeq&
S_0[\bbox{p}^{(0)},\bbox{r}^{(0)}]+
e^2\int\limits_0^t dt_1 \int\limits_0^{t} dt_2\;
\left\langle
R(t_1-t_2,\bbox{r}^{(0)}(t_1)-\bbox{r}^{(0)}(t_2))
\right\rangle_{\rm diff}
\tanh\frac{\xi(t_2)}{2T}.
\label{S011}
\end{eqnarray}
Here $\bbox{r}^{(0)}$ is a classical diffusive path without
interaction.  The energy $\xi$ is conserved along such a path.
The average energy change due the noise field $\bbox{E}(t',\bbox{r}(t'))$
(\ref{lange}) vanishes and therefore was omitted in (\ref{S011}).

Adding eqs. (\ref{SR11}) and (\ref{S011}) together we obtain
\begin{eqnarray}
S_0-S'_0-S_R
&\simeq&
S_0[\bbox{p}_1,\bbox{r}_1]-S_0[\bbox{p}_2,\bbox{r}_2]
+\frac{e^2}{2}\int\limits_0^t dt_1\int\limits_0^t dt_2 {\cal R}
\left[\tanh(\xi_1(t_2)/2T)-\tanh(\xi_2(t_2)/2T)\right],
\label{S111}
\end{eqnarray}
where we defined
\begin{equation}
{\cal R}=\left\langle
R(t_1-t_2,\bbox{r}(t_1)-\bbox{r}(t_2))\right\rangle_{\rm diff}+
\left\langle R(t_1-t_2,\bbox{r}(t_1)-\bbox{r}(t-t_2))\right\rangle_{\rm diff}.
\label{calR}
\end{equation}

Within our simple approximation the paths $\bbox{r}_1$ and $\bbox{r}_2$ in the
action $S_R$ are considered to be independent from each other. Therefore the
kernel (\ref{10}) can be expressed in the following form:
\begin{equation}
J=\tilde U(t,\bbox{r}_{1f},\bbox{r}_{1i})
\tilde U^+(t,\bbox{r}_{2f},\bbox{r}_{2i}) e^{-f_1(t)},
\label{Jpreexp}
\end{equation}
where
\begin{equation}
\tilde U(t,\bbox{r}_{1f},\bbox{r}_{1i})=\int {\cal D}\bbox{r}\int {\cal
D}\bbox{p} \exp \left[  iS_0[\bbox{p},\bbox{r}]+i \frac{e^2}{2}\int\limits_0^t
dt_1\int\limits_0^t dt_2 {\cal R}\tanh\frac{\xi_1(t_2)}{2T} \right]
\label{tildeU}
\end{equation}
It is convenient to define the following function:
\begin{eqnarray}
u(t)&=&\frac{e^2}{2}\int\limits_0^t dt_1\int\limits_0^t dt_2{\cal R}
\simeq 
\frac{e^2}{\pi\sigma_1}\sqrt{\frac{2D}{\tau_e}}t+
\frac{e^2}{2\sigma_1}\sqrt{\frac{Dt}{\pi}}
\label{u}
\end{eqnarray}
Then the operator (\ref{tildeU}) can be rewritten as follows:
\begin{equation}
\tilde U(t,\bbox{r}_{1f},\bbox{r}_{1i})=
\sum\limits_{\lambda} e^{-i\xi_\lambda t+iu(t)\tanh\frac{\xi_\lambda}{2T}}
\;\;\psi_\lambda(\bbox{r}_{1f})\psi^*_\lambda(\bbox{r}_{1i}).
\label{tildeU1}
\end{equation}
Here $\psi_\lambda$ and $\xi_\lambda$ are respectively the
eigenfunctions and the energy eigenvalues of a single electron Hamiltonian
$\hat H_0=\frac{\hat p^2}{2m}+U(\bbox{r})-\mu$.

In the absence of the
interaction the pre-exponent is given by the following expression:
$$
A_1^{(0)}(t)=\frac{1}{\sqrt{4\pi Dt}}=
\frac{1}{4\pi\sqrt{2D}}\int d\xi_1\int d\xi_2
\left(\frac{d}{d\xi_1}\tanh\frac{\xi_1}{2T}\right)
\frac{\cos[(\xi_1-\xi_2)t]}{\sqrt{|\xi_1-\xi_2|}}.
$$
According to eq. (\ref{tildeU1}) the products $\xi_j t$ should be
replaced by $\xi_jt-u(t)\tanh\frac{\xi_j}{2T}$. Thus we get
\begin{equation}
A_1(t)=
\frac{1}{4\pi\sqrt{2D}}\int d\xi_1\int d\xi_2
\left(\frac{d}{d\xi_1}\tanh\frac{\xi_1}{2T}\right)
\frac{\cos\left[(\xi_1-\xi_2)t-u(t)
\left(\tanh\frac{\xi_1}{2T}
-\tanh\frac{\xi_2}{2T}\right)\right]}
{\sqrt{|\xi_1-\xi_2|}}.
\label{A1}
\end{equation}

Let us emphasize that the estimate (\ref{A1}) was obtained with the aid of several 
crude approximations and, in particular in the long time limit, corrections to 
this simple result can easily be expected.  However, since we are not interested
in the details of the long time behavior of $A_d(t)$, the result (\ref{A1})
is already sufficient for our purposes. The main properties of $A_d(t)$ are as follows.

Firstly, eq. (\ref{A1}) determines a typical time
at which  $A_1(t)$ significantly deviates from $A_1^{(0)}(t)$. This scale
(which we will denote as $\tau_A$) is set by the function $u(t)$ and at low $T$ can be 
determined from the condition
$u(\tau_A) \sim 1$. Combining this condition with eq. (\ref{u}) and observing that
the first term in this equation equals to $t/\tau_{\varphi}^0$ and the second 
term is small for all times $t \gg \tau_e$, we
conclude that -- at least for sufficiently low temperatures -- 
the time scale $\tau_A$ is of the same order as the dephasing time 
at $T=0$ (\ref{tau0}), i.e. $\tau_A \sim \tau_{\varphi}^0$. Thus for all
$t\lesssim\tau_\varphi^0$ the effect of the interaction on the
pre-exponent is small and for such times one can safely approximate 
$A_1(t)\approx A_1^{(0)}(t)$. This approximation was already 
used within our previous analysis \cite{GZ2}.

Secondly, the estimate (\ref{A1}) illustrates again an intuitively
obvious property of the pre-exponent: in the long time limit $A_1(t)$
decays in time. Thus no compensation of the exponential decay of
correlations $\propto \exp (-f_1(t))$ can be expected from the pre-exponential
function $A_1(t)$ at long times. Hence, in our problem
the effect of interaction on the pre-exponent $A_1(t)$ can be disregarded
also in the long time limit $t > \tau_{\varphi}^0$.

The same analysis can be repeated for 2d and 3d cases
and the same conclusions will follow. 

\subsection{Discussion} 

Our consideration allows to suggest  the following transparent picture. 
The dephasing time is fully determined by the imaginary
part of the effective action $S_I$ which contains ``coth''. In other
words, the function in the exponent of (\ref{dsigWL}) is 
\begin{equation}
f_d(t)=f_d[S_I].
\label{fdsi}
\end{equation}
The real part of the effective action $S_R$ (which depends on ``tanh''
and contains information about the exclusion principle) contributes
to the (unimportant for dephasing) pre-exponent $A_d(t)$ in (\ref{dsigWL}), i.e. 
\begin{equation}
A_d(t)=A_d[S_R].
\label{adsr}
\end{equation}
The splitting between the
exponent and the pre-exponent of the type (\ref{fdsi}), (\ref{adsr}) holds 
also for the exactly solvable Caldeira-Leggett model. This will be
demonstrated in Sec. 5B.

Although the difference between $A_d(t)$ and $A_d^{(0)}(t)$
cannot have any significant impact on the dephasing time $\tau_{\varphi}$,
this difference should be taken into account if one evaluates the
weak localization correction perturbatively in the interaction. In the limit 
$\tau_H \ll \tau_{\varphi}$ a short time expansion of both the exponent and the pre-exponent
is sufficient and the weak localization correction can be
represented as a sum of three terms 
\begin{equation}
\delta \sigma_d=\delta \sigma_d^{(0)}+\delta \sigma^{\rm deph}+\delta \sigma^{\tanh},
\label{3deltas}
\end{equation}
where $\delta \sigma_d^{(0)}$ is the ``noninteracting'' correction and
\begin{equation}
\delta\sigma^{\rm deph}=
\frac{2e^2(D\tau_H)^{1-d/2}}{\pi(4\pi)^{d/2}}\int\limits_0^\infty
dx\frac{e^{-x}}{x^{d/2}}f_d(x\tau_H),
\label{corrd}
\end{equation}
\begin{equation}
\delta\sigma^{\tanh}=
\frac{2e^2(D\tau_H)^{1-d/2}}{\pi(4\pi)^{d/2}}\int\limits_0^\infty
dx\frac{e^{-x}}{x^{d/2}}[1-(4\pi Dx\tau_H)^{d/2}A_d(x\tau_H)].
\label{cortanh}
\end{equation}
Making use of the result (\ref{A1}) for the 1d case in the limit $T \to 0$
and at short times we obtain 
\begin{equation}
A_1(t) =\frac{\cos[u(t)]+\sin[u(t)]}{\sqrt{4\pi Dt}} \simeq \frac{1}{\sqrt{4\pi Dt}}
[1+u(t)+ {\cal O}( u^2(t))].
\label{correction}
\end{equation}
We will keep track only of the leading 
contribution to the function (\ref{u})
$u(t) \simeq t/\tau_{\varphi}^0$. This is sufficient within the accuracy of
our estimate (\ref{A1}). Combining (\ref{corrd}) and (\ref{cortanh}) we observe
that the sum of the last two terms in (\ref{3deltas}) depends on the combination
$$f_1(t)-u(t).$$
The term $t/\tau_{\varphi}$ drops out of this combination, it is contained both
in $f_1(t)$ and $u(t)$ and cancels out exactly. The same cancellation occurs
in 2d and 3d cases. This cancellation illustrates again the conclusion of Sec. 2:
it is impossible to obtain correct information about the dephasing time even
from the correct first order perturbative analysis.

The accuracy of our estimate of the pre-exponent at short times
$t \lesssim \tau_{\varphi}^0$ can also be checked by means of a direct perturbative
calculation. This calculation is presented in the next section. It demonstrates
that the above cancellation of the first order linear in time $T$-independent terms from
the exponent and the pre-exponent has a general origin 
and is not related to the quasiclassical approximation and/or disorder averaging
at all.
  
The final results for the weak localization correction to the conductance 
presented in the next section are mainly focused on the 1d case. 
Here we provide the results for the 2d case. In the ``perturbative'' limit
$\tau_H \ll \tau_{\varphi}$ one obtains from (\ref{deltaf2}), 
(\ref{3deltas})-(\ref{cortanh}) 
\begin{eqnarray}
\delta\sigma_2(H)-\delta \sigma_2^{(0)}(H)&=&\frac{e^2}{2\pi^2}\frac{e^2R_{\Box}}{2\pi}
T\tau_H\ln(T\tau_H) 
\label{dsigT}
\end{eqnarray}
in the thermal limit $T\tau_H \gg 1$ and
\begin{eqnarray}
\delta\sigma_{2}(H)-\delta \sigma_2^{(0)}(H)&=&\frac{2\gamma_0}{\pi}\frac{e^2}{2\pi^2}
\frac{e^2R_{\Box}}{4\pi}\left(\ln\frac{\tau_H}{\tau_e}\right)^2
\label{dsigCWL}
\end{eqnarray}
in the quantum limit $T\tau_H \ll 1$. Here $R_{\Box}$ is the sheet resistance of
a two-dimensional film. The result (\ref{dsigT}) coincides with that found by AAG
 \cite{AAG2} in the limit $T\tau_H \gg 1$. An opposite
limit of low temperatures was not considered in Ref. \onlinecite{AAG2} at all.
We will perform a detailed comparison of our results with those of AAG in the next 
section.

\section{Perturbation Expansion}

Now let us analyze the expression for the weak localization correction to the
conductivity $\delta \sigma_d$ perturbatively in the interaction. The structure 
of this section
is as follows. In the subsection A we will derive general exact results for
the system conductance in the first order in the interaction. In the subsection B we will
demonstrate that the exact first order diagrams do not cancel at $T=0$ and, moreover,
that the result cannot in general be interpreted as an effective renormalization. We will
also demonstrate that some previous statements about an exact cancellation of
the first order diagrams in the limit $T \to 0$ are nothing but artifacts of insufficient
approximations, the main of which is the golden rule approximation. A detailed calculation
of the weak localization correction in the first order in the interaction is performed
in the subsection C. There we will also identify the contributions to this correction
coming from the exponent and the pre-exponent, see section 3. In the subsection D
we will present a detailed comparison of our analysis with that developed by AAG
in Ref. \onlinecite{AAG2}.

\begin{figure}
\centerline{\psfig{file=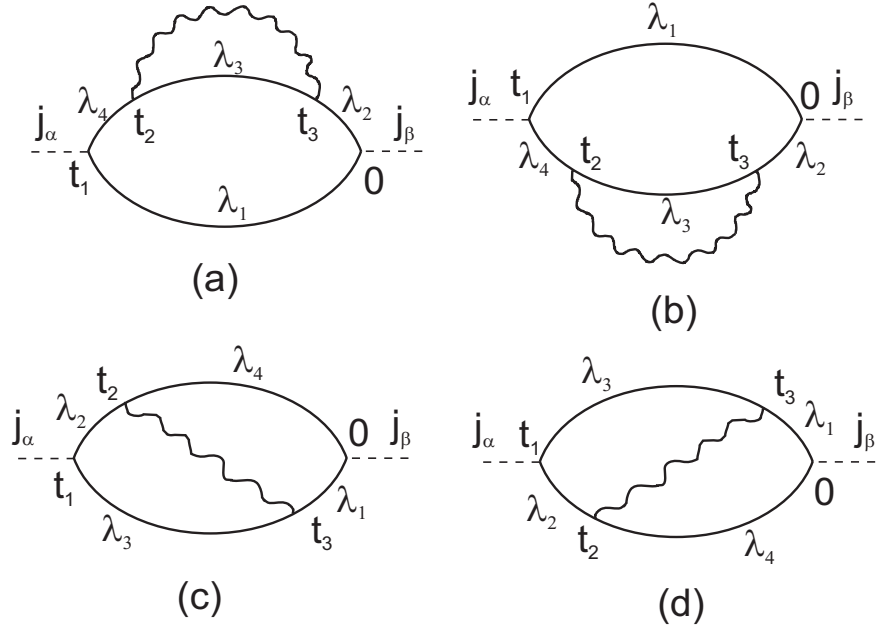,width=12cm}}
\caption{
The four first order diagrams. The time ordering is as follows: $t_1>t_2>t_3>0$.
}
\label{fig2}
\end{figure}

\subsection{General Results}

The perturbation theory can be
constructed by means of a regular expansion of the kernel of the evolution
operator $J$ (\ref{J17}) in powers of $iS_R+S_I$. In the first order one obtains
four diagrams presented in Fig. 2. The contribution of the self-energy diagrams
(Fig. 2a,b) was analyzed in details in Appendix A. It was demonstrated that this
contribution to $\delta \sigma$ can be written in terms of the
the evolution operator $U$ for noninteracting electrons. The corresponding expression is defined in (\ref{dsigma1}-\ref{I55}). It is equivalent to the
result (\ref{dsigma}) obtained diagrammatically by AAG.

Let us express the evolution operator $U$ in the basis of the exact wave
functions of noninteracting electrons:
\begin{equation}
U(t,\bbox{r}_1,\bbox{r}_2)=\langle\bbox{r}_1|\exp(-i\hat H_0t)|\bbox{r}_2\rangle=
\sum\limits_\lambda e^{-i\xi_\lambda t}\psi_\lambda(\bbox{r}_1)
\psi^*_\lambda(\bbox{r}_2).
\label{U}
\end{equation}
Obviously the representation (\ref{U}) holds
both with and without the external magnetic field with the only difference
that in the latter case the energy
levels are doubly degenerate, while in the former case this degeneracy is lifted
by the magnetic field.

The density matrix $\rho$ which enters the expression (\ref{dsigma1})
can also be expanded in the basis of the eigenfunctions $\psi_{\lambda}$. We find
\begin{eqnarray}
(1-2\rho)(\bbox{r}_1,\bbox{r}_2)=\sum\limits_\lambda
\tanh\frac{\xi_\lambda}{2T} \psi_\lambda(\bbox{r}_1) \psi^*_\lambda(\bbox{r}_2),
\label{1-2rho}
\\
\frac{\partial\rho}{\partial\mu}(\bbox{r}_1,\bbox{r}_2)=
\sum\limits_\lambda
\frac{1}{2}\left(\frac{d}{d\xi_\lambda}\tanh\frac{\xi_\lambda}{2T}\right)
\psi_\lambda(\bbox{r}_1) \psi^*_\lambda(\bbox{r}_2).
\label{drhodmu}
\end{eqnarray}

Now let us substitute
(\ref{U}-\ref{drhodmu}) into the expression (\ref{dsigma1}). Performing the two time
integrals after a straightforward algebra (see Appendix B for details) we obtain
the correction to the conductivity due to the self-energy diagrams of Fig. 2a,b
\begin{eqnarray}
\delta\sigma^{\rm se}_{\alpha\beta}&=& \delta \sigma _{\alpha \beta }^{C}
-\frac{e^2}{2\cal{V}\sigma}\int\limits_0^{+\infty} dt_1
\sum\limits_{\lambda_1..\lambda_4}
\left(\frac{d}{d\xi_{\lambda_1}}\tanh\frac{\xi_{\lambda_1}}{2T}\right)
\left(j^{\lambda_4\lambda_1}_\alpha j^{\lambda_1\lambda_2}_\beta+
j^{\lambda_4\lambda_1}_\beta j^{\lambda_1\lambda_2}_\alpha\right)
M^{\lambda_2\lambda_3;\lambda_3\lambda_4}\times
\nonumber\\
&&
\int\frac{d\omega}{2\pi}\;\;\omega
\left[\coth\frac{\omega}{2T}+\tanh\frac{\xi_{\lambda_3}}{2T}\right]
F(t_1,\omega,\xi_{\lambda_1}..\xi_{\lambda_4}),
\label{dsig11}
\end{eqnarray}
where we defined the matrix elements
\begin{equation}
j^{\lambda_1\lambda_2}_\alpha = \int d\bbox{r} \psi^*_{\lambda_1}(\bbox{r})
\hat j_\alpha \psi_{\lambda_2}(\bbox{r}), \;\;\;\;
M^{\lambda_2\lambda_3;\lambda_3\lambda_4}=
\int\frac{d^3k}{(2\pi)^3}\;\;\frac{1}{k^2}
\langle\lambda_2|e^{i\bbox{kr}}|\lambda_3\rangle
\langle\lambda_3|e^{-i\bbox{kr}}|\lambda_4\rangle
\label{M}
\end{equation}
and the function
\begin{eqnarray}
F(t_1,\omega,\xi_{\lambda_1}..\xi_{\lambda_4})&=&
\int\limits_0^{t_1}dt_2\int\limits_0^{t_2}dt_3
\cos (-\xi_{\lambda_1}t_1+\xi_{\lambda_2}t_3+
(\xi_{\lambda_3}+\omega)(t_2-t_3)+\xi_{\lambda_4}(t_1-t_2))
\label{F}\\
&=&
\cos \xi_{21}t_1\frac{\xi_{42}[\cos((\xi_{32}+\omega)t_1)-1]+(\xi_{32}+\omega )
[1-\cos \xi_{42}t_1]}{\xi_{42}(\xi_{42}-\xi_{32}-\omega)(\xi_{32}+\omega)}
\nonumber\\
&&
-\sin \xi_{21}t_1
\frac{\xi_{42}\sin ((\xi_{32}+\omega)t_1)-(\xi_{32}+\omega )\sin \xi_{42}t_1}
{\xi_{42}(\xi_{42}-\xi_{32}-\omega)(\xi_{32}+\omega)}.
\nonumber
\end{eqnarray}
Here we introduced the notation $\xi_{ij}\equiv \xi_{\lambda_i}-\xi_{\lambda_j}$.

The term $\delta \sigma _{\alpha \beta }^{C}$ in (\ref{dsig11}) describes the
correction due to the non-screened Coulomb interaction. It is defined by
the following expression:
\begin{eqnarray}
\delta \sigma _{\alpha \beta }^{C} &=&-\frac{e^2}{4{\cal {V}}}%
\int\limits_0^{+\infty }dt_1\int\limits_0^{t_1}dt_2\sum\limits_{\lambda
_1,\lambda _2,\lambda _4}\left( \frac d{d\xi _{\lambda _1}}\tanh \frac{\xi
_{\lambda _1}}{2T}\right) \left( j_\alpha ^{\lambda _4\lambda _1}j_\beta
^{\lambda _1\lambda _2}+j_\beta ^{\lambda _4\lambda _1}j_\alpha ^{\lambda
_1\lambda _2}\right) \times
\nonumber \\
&&\times \left\langle \lambda _2\right| \frac{[1-2\rho ](\bbox{r}_1,\bbox{r}%
_2)}{|\bbox{r}_1-\bbox{r}_2|}\left| \lambda _4\right\rangle \sin [-\xi
_{\lambda _1}t_1+\xi _{\lambda _2}t_2+\xi _{\lambda _4}(t_1-t_2)].
\label{Clmb}
\end{eqnarray}

The contribution to $\delta \sigma$ from the vertex diagrams of Fig. 2c,d can be found
analogously (see Appendix B). We get
\begin{eqnarray}
\delta\sigma_{\alpha\beta}^{\rm vert}&=&
-\frac{e^2}{2{\cal V}\sigma}\int\limits_0^\infty dt_1
\sum\limits_{\lambda_1..\lambda_4}
\left(\frac{d}{d\xi_{\lambda_1}}\tanh\frac{\xi_{\lambda_1}}{2T}\right)
\left(j_\alpha^{\lambda_2\lambda_3} j_\beta^{\lambda_1\lambda_4}+
j_\beta^{\lambda_2\lambda_3} j_\alpha^{\lambda_1\lambda_4}\right)
M^{\lambda_3\lambda_1;\lambda_4\lambda_2}
\nonumber\\
&&
\times \int\frac{d\omega}{2\pi}\;\;\omega
\left[-\coth\frac{\omega}{2T}+\tanh\frac{\xi_{\lambda_3}}{2T}\right]
G(t_1,\omega,\xi_{\lambda_1}..\xi_{\lambda_4}).
\label{dsigver}
\end{eqnarray}
Here we have introduced the following function:
\begin{eqnarray}
G(t_1,\omega,\xi_{\lambda_1}..\xi_{\lambda_4})&=&
\int\limits_0^{t_1}dt_2\int\limits_0^{t_2}dt_3
\cos(-\xi_{\lambda_3}(t_1-t_3)-\xi_{\lambda_1}t_3+\xi_{\lambda_4}t_2
+\xi_{\lambda_2}(t_1-t_2)+\omega(t_2-t_3))
\label{G}\\
&=&
\cos \xi_{21}t_1\frac{(\xi_{42}+\xi_{31})\cos[(\xi_{31}-\xi_{42}-\omega)t_1]
-(\xi_{42}+\omega )\cos \xi_{42}t_1-(\xi_{31}-\omega )\cos \xi_{31}t_1}
{(\xi_{42}+\xi_{31})(\xi_{31}-\omega)(\xi_{42}+\omega)}
\nonumber\\
&&
+\sin \xi_{21}t_1
\frac{(\xi_{42}+\xi_{31})\sin[(\xi_{31}-\xi_{42}-\omega)t_1]
+(\xi_{42}+\omega )\sin \xi_{42}t_1-(\xi_{31}-\omega )\sin \xi_{31}t_1}
{(\xi_{42}+\xi_{31})(\xi_{31}-\omega)(\xi_{42}+\omega)}.
\nonumber
\end{eqnarray}
Despite an obvious similarity in the structure of the self-energy (eq. (\ref{dsig11}))
and the vertex (eq. (\ref{dsigver})) corrections to the conductivity these two
expressions differ in several aspects: the terms containing $\coth (\omega /2T)$
in (\ref{dsig11}) and (\ref{dsigver}) have the opposite signs,
the functions $F$ (\ref{F}) and $G$ (\ref{G}) of the energy arguments
$\xi_{\lambda}$ are different and the matrix
elements entering (\ref{dsig11}) and (\ref{dsigver}) depend on different indices.

It is important to emphasize that the eqs. (\ref{dsig11}-\ref{G}) determine the
total correction to the conductivity tensor $\sigma_{\alpha \beta}=
\sigma^{\rm se}_{\alpha \beta}+\sigma^{\rm vert}_{\alpha \beta}$ which is
{\it identical} to the initial results (\ref{dsigma}) and (\ref{dsigma1}). In deriving
(\ref{dsig11}-\ref{G}) from (\ref{dsigma1}) no quasiclassical approximation, no
averaging over disorder and/or no other approximation of any kind has been made:
the above equations are
{\it exact} quantum mechanical results in the first order in the interaction. Therefore
these equations can be conveniently used to test the statement about the full cancellation
of the first order diagrams at $T \to 0$ which is quite frequently made in the
literature (see e.g. \cite{FA} as well as recent works \cite{IFS,RSC} and
further references therein).

\subsection{Breakdown of the Fermi Golden Rule Approximation}

Although here we are mainly
interested in the contribution of the diagrams of Fig. 2 to the
current-current correlation function,
the structure of the result is by no means specific to
this function only. The very same structure
-- perhaps apart from the matrix elements of the current operator --
is reproduced if one calculates e.g. the inelastic scattering time
\cite{FA,Blanter,RSC} and similar quantities. This is quite natural because
the results for different quantities follow
from the expansion of the same evolution operator $J$ (\ref{J17}) in the
interaction. Hence, the analysis to be presented below is general and
can be applied to various physical quantities evaluated by means of the diagrams
of Fig. 2.

Let us consider the self-energy diagrams of Fig. 2a,b. Just for the sake of
clarity let us repeat the statement we are going to test: according to
Fukuyama and Abrahams \cite{FA} and to some other authors the contribution of these diagrams
vanishes in the limit $T \to 0$ because the result contains the combination
\begin{equation}
\left(\frac{d}{d\epsilon}\tanh\frac{\epsilon}{2T}\right)
\left[\coth\frac{\omega}{2T}+\tanh\frac{\epsilon-\omega}{2T}\right]
\label{GR}
\end{equation}
under the integrals over $\epsilon$ and $\omega$. This combination restricts both
integrals to the regions $|\epsilon | \lesssim T$ and $|\omega | \lesssim T$ and
makes the result to vanish completely at $T=0$.

Already the first inspection of the expression (\ref{dsig11},\ref{F}) allows to observe
that it is the combination
\begin{equation}
\left(\frac{d}{d\xi_{\lambda_1}}\tanh\frac{\xi_{\lambda_1}}{2T}\right)
\left[\coth\frac{\omega}{2T}+\tanh\frac{\xi_{\lambda_3}}{2T}\right].
\label{exact}
\end{equation}
and {\it not} (\ref{GR}) which enters the exact quantum mechanical result. This
combination is {\it not} zero even at $T=0$ because
$\xi_{\lambda_3}\not=\xi_{\lambda_1}-\omega$, high frequencies $ |\omega | >T$
{\it do} contribute to the integral and, moreover, this integral may -- depending
on the spectrum of the fluctuation propagator -- even diverge for large $\omega$
unless one introduces an effective high frequency cutoff. We would like to
emphasize that these conclusions are general and do not depend on any
particular form of the matrix elements (\ref{M}). Thus the statement of the above
papers that the contribution of the diagrams of Fig. 2a,b vanishes in equilibrium
at $T=0$ {\it is proven to be incorrect}. Below we will demonstrate that
this poorly justified statement is a result of several rough approximations, the main of
which is the golden rule approximation. This approximation may sometimes yield correct
leading order results in the high temperature limit, but it breaks down at
sufficiently low $T$.

In order to illustrate this point let us first make a simplifying assumption.
Namely, let us for a moment restrict our attention only to
the contribution of the terms with $\xi_{\lambda_1}=\xi_{\lambda_2}=\xi_{\lambda_4}$.
Below we will see that this assumption is not sufficient to properly evaluate the
first order perturbation correction to the conductivity: in order to do that it
is important to allow for a (possibly small) difference between
$\xi_{\lambda_1}$ and $\xi_{\lambda_2}$. But such an approximation is sufficient
for calculation of some other physical quantities, like the inelastic scattering time, and
we will adopt it for a moment just in order to demonstrate the failure of
the golden-rule-type perturbation theory in the interaction.

The contribution of the terms with $\xi_{\lambda_1}=\xi_{\lambda_2}=\xi_{\lambda_4}$
to the conductivity $\delta \sigma^{\rm se}$ reads
\begin{eqnarray}
\delta \tilde \sigma_{\alpha\beta}^{\rm se}&=&
-\frac{e^2}{2\cal{V}\sigma}\int\limits_0^{+\infty} dt_1
\sum\limits_{\lambda_1..\lambda_4}^{\xi_{\lambda_1}=
\xi_{\lambda_2}=\xi_{\lambda_4}}
\left(\frac{d}{d\xi_{\lambda_1}}\tanh\frac{\xi_{\lambda_1}}{2T}\right)
\left(j^{\lambda_4\lambda_1}_\alpha j^{\lambda_1\lambda_2}_\beta+
j^{\lambda_4\lambda_1}_\beta j^{\lambda_1\lambda_2}_\alpha\right)
M^{\lambda_2\lambda_3;\lambda_3\lambda_4}\times
\nonumber\\
&&
\int\frac{d\omega}{2\pi}\;\;\omega
\left[\coth\frac{\omega}{2T}+\tanh\frac{\xi_{\lambda_3}}{2T}\right]
\frac{1-\cos((\xi_{31}+\omega)t_1)}
{(\xi_{31}+\omega)^2}.
\label{dsigdiag}
\end{eqnarray}
Let us first evaluate this expression within the Fermi golden rule approximation:
\begin{equation}
\frac{1-\cos((\xi_{31}+\omega)t_1)}
{(\xi_{31}+\omega)^2} \longrightarrow
\pi t_1 \delta(\xi_{31}+\omega).
\label{GRS}
\end{equation}
Substituting (\ref{GRS}) into (\ref{dsigdiag}) we obtain
\begin{eqnarray}
\delta \tilde \sigma_{\alpha\beta}^{\rm se,\; GR}&=&
-\frac{e^2}{4\cal{V}\sigma}\int\limits_0^{+\infty} dt_1\;\; t_1
\sum\limits_{\lambda_1..\lambda_4}^{\xi_{\lambda_1}=
\xi_{\lambda_2}=\xi_{\lambda_4}}
\left(\frac{d}{d\xi_{\lambda_1}}\tanh\frac{\xi_{\lambda_1}}{2T}\right)
\left(j^{\lambda_4\lambda_1}_\alpha j^{\lambda_1\lambda_2}_\beta+
j^{\lambda_4\lambda_1}_\beta j^{\lambda_1\lambda_2}_\alpha\right)
M^{\lambda_2\lambda_3;\lambda_3\lambda_4}\times
\nonumber\\
&&
\xi_{31}
\left[\coth\frac{\xi_{31}}{2T}
-\tanh\frac{\xi_{\lambda_3}}{2T}\right] \propto T \int\limits_0^{+\infty} dt_1\;\; t_1.
\label{dsigGR}
\end{eqnarray}
This expression implies that {\it within the Fermi
golden rule approximation} the self-energy diagrams of Fig. 2a,b
yield a linear in time decay of the initial quantum state with the corresponding
relaxation rate proportional to $T$. Obviously, {\it such} relaxation rate vanishes
at $T=0$ in agreement with Ref. \onlinecite{FA} and others.

Now let us carry out an exact frequency integration in (\ref{dsigdiag}) without
making the golden rule approximation (\ref{GRS}). It is fairly obvious that the integral
over $\omega$ is not restricted to $|\omega | \lesssim T$ and even diverges at
high frequencies. As before, in order to cure this divergence we introduce
the high frequency cutoff $\omega_c \approx 1/\tau_e$.
For simplicity we also assume that the energy difference $\xi_{31}$ is smaller than $1/\tau_e$.
Then in the limit $T \to 0$ we find
$$
\int\frac{d\omega}{2\pi}
\omega \left[\coth\frac{\omega}{2T}+\tanh\frac{\xi_{\lambda_3}}{2T}\right]_{T \to 0}
\frac{1-\cos((\xi_{31}+\omega) t_1)}{(\xi_{31}+\omega)^2}
$$
\begin{eqnarray}
=
\frac{|\xi_{31}|t_1}{2}-\frac{\xi_{31} t_1}{2}{\rm sign} \xi_{\lambda_3}
+2\int\limits_{|\xi_{31}|}^{1/\tau_e}\frac{d\omega}{2\pi}
\left(\frac{1}{\omega}-\frac{|\xi_{31}|}{\omega^2}\right)
\left(1-\cos\omega t_1\right).
\label{func}
\end{eqnarray}
The first and the third terms in the second line of this expression come from
$\coth (\omega/2T)$ while the second term originates from $\tanh(\xi_{\lambda_3}/2T)$.
We observe that the first two terms are the same as in
the golden rule approximation (\ref{dsigGR}). These terms enter with the
opposite signs and exactly cancel each other at $T=0$ because in this limit
$(d/d\xi_{\lambda_1})\tanh (\xi_{\lambda_1}/2T)$ reduces to a $\delta$-function
and therefore $\xi_{\lambda_3} \equiv \xi_{31}$. The last term does not vanish even
at zero temperature, this term
is not small and obviously contains the contribution of {\it all} frequencies up
to $1/\tau_e$. The integral over $\omega$ contained in this term can be
easily evaluated. We will do it a bit later when we fix the dependence of the
matrix elements $M^{\lambda_2,\lambda_3;\lambda_3,\lambda_4}$ on energies. Now
it is only important for us to demonstrate that
the last term in (\ref{func}) is completely missing within the golden rule approach
employed in Refs. \onlinecite{FA,IFS,RSC} and others. It is obvious, therefore,
that this approach fails to correctly describe the system behavior at sufficiently
low temperatures.

Note, that AAG \cite{AAG2} also did not observe an exact cancellation
of diagrams of the first order perturbation theory in the interaction. However, they
argued that the remaining terms provide the so-called interaction correction to
the conductance which can be viewed as an effective (temperature dependent) renormalization
of the bare parameters and has nothing to do with dephasing. Already from the form of the
third term in the right hand side of (\ref{func}) one can conclude that in general
this is not true. Indeed, if one adopts that for $\xi_{\lambda_1}=\xi_{\lambda_2}=
\xi_{\lambda_4}$ the dependence of the matrix elements
$M^{\lambda_2,\lambda_3;\lambda_3,\lambda_4}$ on the energy difference $\xi_{31}$
has the form
\begin{equation}
M^{\lambda_2,\lambda_3;\lambda_3,\lambda_4} \propto |\xi_{31}|^{d/2-2}
\label{Mxi}
\end{equation}
(cf. eq. (2.33) of Ref. \onlinecite{AAG2}), and integrates the product of
$M^{\lambda_2,\lambda_3;\lambda_3,\lambda_4}$
and the last term in (\ref{func}) over the energy $\xi_{31}$ one immediately
observes that after the cancellation of the unphysical divergence (which is also
contained in the vertex diagrams of Fig. 2c,d and enters with the opposite sign, see
also Section 3a)
one obtains the contribution
$\propto \sqrt{t}\ln(t/\tau_e)$ in 1d and $\ln(t/\tau_e)$ in 2d. This
contribution is just a part of the function $\delta f_d(T,t)$ (\ref{f}) at $T \to 0$.
It grows with time,
contributes to dephasing and obviously cannot be reduced to the renormalization
of the initial parameters which would be provided by a time-independent term.

In order to understand why AAG arrived at such a conclusion
it is appropriate to highlight the approximation employed in Ref. \onlinecite{AAG2}. 
As a first step they split the total contribution to $\delta \sigma$
into two parts, eqs. (5.12c) and (5.12d), effectively
rewriting the combination (\ref{exact}) in the following equivalent form:
\begin{equation}
\left(\frac{d}{d\xi_{\lambda_1}}\tanh\frac{\xi_{\lambda_1}}{2T}\right)
\left[\left(\coth\frac{\omega}{2T}+\tanh\frac{\xi_{\lambda_1}-\omega}{2T}\right)+
\left(\tanh\frac{\xi_{\lambda_3}}{2T}-\tanh\frac{\xi_{\lambda_1}-\omega}{2T}\right)
\right].
\label{exact2}
\end{equation}
The first two terms in the square brackets of (\ref{exact2}) were interpreted 
by AAG as a ``dephasing'' contribution (eq. (5.12c) of \cite{AAG2}) while
the last two terms are meant to be the ``interaction'' correction
(eq. (5.12d) of \cite{AAG2}). Obviously, the contribution of the first two
terms vanishes at $T \to 0$. In order to understand the behavior of the
remaining terms we make use of (\ref{dsigdiag}), (\ref{Mxi}) and observe that
the contribution of the last two terms in (\ref{exact2}) is proportional
to the following integral
\begin{equation}
\int d\omega \int d\xi_{\lambda_1} \int d\xi_{\lambda_3}
\left(\frac{d}{d\xi_{\lambda_1}}\tanh\frac{\xi_{\lambda_1}}{2T}\right)
\omega |\xi_{31}|^{d/2-2}
\left[\tanh\frac{\xi_{\lambda_3}}{2T}-\tanh\frac{\xi_{\lambda_1}-\omega}{2T}\right]
\frac{1-\cos((\xi_{31}+\omega) t)}{(\xi_{31}+\omega)^2}.
\label{nonren0}
\end{equation}
The approximation employed by AAG while evaluating such a
combination is equivalent to ignoring the oscillating $\cos$-term in (\ref{nonren0}).
After dropping this term and making the integral dimensionless one can
easily observe that the remaining integral has the form
\begin{equation}
A_1/\sqrt{T}
\label{ren1}
\end{equation}
in 1d and $A_2 \ln T$ in 2d, where $A_{1,2}$ are temperature- and time-independent
constants. AAG interpreted these contributions 
as an effective renormalization due to interaction. Note, however, that it is
correct to drop the $\cos$-term only at sufficiently long times $Tt \gg 1$, while
at smaller $Tt \lesssim 1$ this term is important. Evaluation of the integral
(\ref{nonren0}) in the latter limit yields
\begin{equation}
B_1\sqrt{t}\ln(t/\tau_e)
\label{nonren1}
\end{equation}
in 1d and $B_2 \ln(t/\tau_e)$ in 2d, where $B_{1,2}$ are again temperature- and
time-independent constants.
It is fairly obvious that the term (\ref{nonren1}) already cannot be interpreted as a
renormalization effect from an effective {\it static} potential. This term
explicitly depends on time and actually contributes to dephasing!

Now we are aware of the behavior of the integral (\ref{nonren0}) at all times: at
$t=0$ this integral is obviously zero, it grows with time as (\ref{nonren1}) for
$Tt \lesssim 1$, reaches the value (\ref{ren1}) and saturates in the long time limit
$Tt \gg 1$. Clearly, in the interesting limit $T \to 0$ the behavior (\ref{ren1})
can never be realized, the term (\ref{nonren0}) grows at all times and contributes
to dephasing. In this limit we are back to the result (\ref{func}). The perturbation
theory strongly diverges in this case. It also diverges at finite temperatures,
thus the corresponding expressions can only make
sense if one introduces a cutoff at times much smaller than the dephasing time
$\tau_{\varphi}$, because at times $t \sim \tau_{\varphi}$ {\it all} orders
of the perturbation theory should be taken into account. In Ref. \onlinecite{AAG2}
this cutoff time was chosen to be the magnetic-field-induced decoherence time
$\tau_H \ll \tau_{\varphi}$. Thus the approximation 
leading to the time-independent term (\ref{ren1}) is valid only for $T\tau_H \gg 1$,
in which case the contribution (\ref{nonren1}) is anyway much smaller than that
from the first two terms in (\ref{exact2}) and, hence, can be safely ignored
in the above limit. On the other hand, in the most interesting limit
$T\tau_H \lesssim 1$ (which is compatible with $T \tau_{\varphi} \gg 1$) the
contribution (\ref{nonren0}) dominates, its behavior is given by eq. (\ref{nonren1})
rather than by eq. (\ref{ren1}) and, consequently, {\it nonzero} low temperature
dephasing is observed already in the first order perturbation theory in
the interaction. We will come back to this discussion in Sec. 4.D and in Appendix C.

Let us emphasize again that no approximation was done during our derivation
presented in Sec. 4A.
Our main goal here was to demonstrate that the absence of the cancellation of
diagrams in the first order perturbation theory has nothing to do with 
the quasiclassical approximation and/or disorder average as it is sometimes
speculated in the literature.

\subsection{Perturbative weak localization correction}

Now let us perform a systematic evaluation of the exact expressions (\ref{dsig11})-(\ref{G})
obtained within the first order perturbation theory in the interaction. Our calculation
consists of several steps. First we notice that the expressions (\ref{dsig11})-(\ref{G})
contain the full information about contributions from all energy states. Since here
we are interested only in the weak localization correction to the conductance
we should restrict our attention to the time reversed energy states and evaluate
the matrix elements for such states. The matrix elements for the current operator
can be extracted from the expression for the weak localization correction
without interaction $\delta \sigma _{\alpha \beta }^{{\rm ni}}$. Starting from
the standard expression for this correction (see e.g. Ref. \onlinecite{AAG2}) and
rewriting it in terms of the matrix elements for the current (\ref{M}) we obtain
\begin{eqnarray}
\delta \sigma _{\alpha \beta }^{{\rm ni}} &=&\frac 18\int \frac{d\bbox{r}_1}{%
\bbox{r}_2}\int \frac{d\epsilon }{2\pi }\left( \frac d{d\epsilon }\tanh
\frac \epsilon {2T}\right) \left\{ [G_{12}^R(\epsilon )-G_{12}^A(\epsilon )]%
\hat j_\alpha [G_{21}^A(\epsilon )-G_{21}^R(\epsilon )]\hat j_\beta +\alpha
\leftrightarrow \beta \right\}  \nonumber \\
&=&\frac 1{4{\cal V}}\int\limits_0^{+\infty }dt_1\sum\limits_{\lambda
_1\lambda _2}\left( \frac d{d\xi _{\lambda _1}}\tanh \frac{\xi _{\lambda _1}%
}{2T}\right) \left( j_\alpha ^{\lambda _1\lambda _2}j_\beta ^{\lambda
_2\lambda _1}+j_\beta ^{\lambda _1\lambda _2}j_\alpha ^{\lambda _2\lambda
_1}\right) \cos \xi_{21}t_1.  \label{dsigni}
\end{eqnarray}
The expression
for the matrix elements of the currents $jj(\xi_{21})$ is readily established
by comparison (\ref{dsigni}) to the well known quasiclassical result
 in the absence of interaction (\ref{dsigWL}). We find
\begin{eqnarray}
jj(\xi_{21} ) &=& j_\alpha ^{\lambda _1\lambda _2}j_\beta
^{\lambda _2\lambda _1}+j_\beta ^{\lambda _1\lambda _2}j_\alpha ^{\lambda
_2\lambda _1} =-%
\frac{4e^2D{\cal V}}{\pi ^2}\int \frac{d^dQ}{(2\pi )^d}\;\frac{DQ^2+1/\tau _H%
}{\xi_{21}^2+(DQ^2+1/\tau _H)^2}  \nonumber \\
&=&\left\{
\begin{array}{ll}
-\frac{2e^2{\cal V}}{\pi ^2}\sqrt{D\tau _H}{\rm Re}\left( \frac 1{\sqrt{%
1+i|\xi_{21}| \tau _H}}\right) , & {\rm 1d} \\
-\frac{e^2{\cal V}}{\pi ^3}\left( \ln \frac{\tau _H}{\tau _e}-\frac 12\ln
(1+(\xi_{21} \tau _H)^2)\right) , & {\rm 2d}
\end{array}
\right.
\end{eqnarray}
As it was already pointed out above, these expressions are only valid
for the time reversed states $\lambda _1$ and $\lambda _2$ relevant for the
weak localization correction. For later purposes let us also rewrite the above
result in the 1d case in the real time representation:
\begin{equation}
\int d\xi_{21} jj(\xi_{21})\cos \xi_{21} t_1 =
-\frac{4{\cal V}e^2}{2\pi}\sqrt{\frac{D}{\pi}}\frac{e^{-|t_1|/\tau_H}}
{\sqrt{|t_1|}}.
\label{jj1}
\end{equation}

The next step in our calculation is to identify the contribution to $\delta \sigma$
responsible for dephasing. As it was already demonstrated above within the
nonperturbative analysis, this contribution is determined by the function $f_d(t)$
(\ref{fott}) in the exponent  (\ref{dsigWL}). Clearly, in the first order in
the interaction this contribution is obtained by expanding the exponent in (\ref{dsigWL}) up
to the linear term in $f_d(t)$ and ignoring the effect of interaction on the
pre-exponential function $A_d(t)$. Hence, this ``dephasing'' contribution should
have the form
\begin{equation}
\delta \sigma_{\alpha \beta }^{\rm deph}=\frac 1{4{\cal V}}\int\limits_0^{+\infty
}dt_1\;\;f_d(t_1)\sum\limits_{\lambda _1\lambda _2}\left( \frac d{d\xi
_{\lambda _1}}\tanh \frac{\xi _{\lambda _1}}{2T}\right) \left( j_\alpha
^{\lambda _1\lambda _2}j_\beta ^{\lambda _2\lambda _1}+j_\beta ^{\lambda
_1\lambda _2}j_\alpha ^{\lambda _2\lambda _1}\right) \cos \xi_{21}t_1.
\label{dsigdepha}
\end{equation}
We observe that this expression contains the function $ \cos \xi_{21}t_1$
and does not contain  $\sin \xi_{21}t_1$. Furthermore, from the above
analysis we know that the function $f_d(t)$ contains only $\coth (\omega /2T)$
and does not depend on $\tanh (\xi_{\lambda_3}/2T)$. Therefore in the general
result for the conductance correction (\ref{dsig11})-(\ref{G}) we will
first take care of all terms which contain the product
$\coth (\omega /2T)\cos \xi_{21}t_1$ leaving all the remaining terms for further
consideration.

Consider the ``coth$\times$cos'' terms originating from the self-energy diagrams
of Fig. 2a,b. For such diagrams one should put $\xi_{42}=0$,
then from (\ref{dsig11}), (\ref{F}) one will immediately observe that the contribution
of the ``coth$\times$cos'' terms can indeed be represented in the form (\ref{dsigdepha})
where
\begin{equation}
f^{\rm se }_d(t,\lambda _2)=\frac{2e^2}\sigma \sum\limits_{\lambda _3}\int
\frac{d\omega }{2\pi }\;\;M^{\lambda _2\lambda _3;\lambda _3\lambda
_2}\omega \coth \frac \omega {2T}\frac{1-\cos [(\xi _{32}+\omega )t]}{(\xi _{32}+\omega )^2}.
\label{f11}
\end{equation}
Let us replace the summation over $\lambda _3$ by the integration over
$\xi _{\lambda _3}$. Assuming that the matrix elements
$M^{\lambda _2\lambda _3;\lambda _3\lambda_2}$ depend only on the energy difference
$\xi _{32}$, making a shift $\xi_{32} +\omega \to \omega $ and
denoting $\xi_{32} \to \omega ^{\prime}$ we find
\begin{equation}
f^{\rm se }_d(t)=\frac{4\pi e^2}\sigma \int \frac{d\omega d\omega ^{\prime }}{%
(2\pi )^2}M(\omega ^{\prime })(\omega -\omega ^{\prime })\coth \frac{\omega
-\omega ^{\prime }}{2T}\;\;\frac{1-\cos \omega t}{\omega ^2}.  \label{ffin}
\end{equation}
This expression does not depend on $\lambda _2$
and exactly coincides with the first term in eq. (\ref{fott}) if we identify
the matrix element $M(\omega ^{\prime })$ as:
\begin{equation}
M(\omega ^{\prime })=\frac{D^{1-d/2}}\pi \left( \int \frac{d^dx}{(2\pi )^d}%
\frac 1{1+x^4}\right) |\omega ^{\prime }|^{d/2-2}.  \label{M1}
\end{equation}
Note, that the energy dependence of the matrix elements
$M\propto |\omega
^{\prime }|^{d/2-2}$ (\ref{M1})
determined within the above procedure is in the agreement with the conjecture
(\ref{Mxi}) as well as with eq. (2.33) of Ref. \onlinecite{AAG2}.

The contribution of the ``coth$\times$cos'' terms contained in the vertex diagrams
of Fig. 2c,d can be evaluated analogously. Again one should consider only the part
of the function $G$ (\ref{G}) which contains $\cos (\xi_{21}t_1)$.
For the contribution of the time reversed states to the vertex diagrams one
should identify $\xi_{42}=\xi_{31}$. Making use of this equation, from the corresponding
terms in (\ref{dsigver}), (\ref{G}) one finds
\begin{eqnarray}
\delta\sigma^{{\rm deph, vert}}_{\alpha\beta}&=& -\frac{1}{4{\cal V}}\int\limits_0^{+%
\infty}dt_1 f_{d}^{{\rm vert}}(t_1,\lambda_1)
\sum\limits_{\lambda_1\lambda_2} \left(\frac{d}{d\xi_{\lambda_1}}\tanh\frac{%
\xi_{\lambda_1}}{2T}\right)
\left(j_\alpha^{\lambda_2\lambda_3}j_\beta^{\lambda_1\lambda_4}+
j_\beta^{\lambda_2\lambda_3}j_\alpha^{\lambda_1\lambda_4}\right) \cos \xi_{21}t_1,
\label{vertvert}\end{eqnarray}
where
\begin{eqnarray}
f_{d}^{{\rm vert}}(t,\lambda_1)&=& -\frac{2e^2}{\sigma}\sum_{\lambda_3}\int%
\frac{d\omega}{2\pi} M^{\lambda_3\lambda_1;\lambda_4\lambda_2}\omega
\coth\frac{\omega}{2T} \frac{\cos\omega t-\cos \xi_{31}t}{\xi_{31}^2-\omega^2}.
\label{fdvert}
\end{eqnarray}
By comparing eq. (\ref{fdvert}) with the second term of the expression (\ref{fott})
we observe that they coincide provided one denotes $\xi_{31}=\xi_{42}=\omega'$
and again assumes that the matrix elements depend only on the energy difference
$M^{\lambda_3\lambda_1;\lambda_4\lambda_2}=M(\omega')$, where $M(\omega')$ is defined
in eq. (\ref{M1}). Furthermore, in order to identify eqs. (\ref{dsigdepha})
(with $f_d(t) \to f_d^{\rm vert}(t)$) and (\ref{vertvert}) we have to assume
that $j_\alpha^{\lambda_2\lambda_3}j_\beta^{\lambda_1\lambda_4}+
j_\beta^{\lambda_2\lambda_3}j_\alpha^{\lambda_1\lambda_4}=
j_\alpha^{\lambda_1\lambda_2}j_\beta^{\lambda_2\lambda_1}
+j_\beta^{\lambda_1\lambda_2}j_\alpha^{\lambda_2\lambda_1}$. This
completes the analysis of the ``coth$\times$cos''-contribution from the vertex diagrams
of Fig. 2c,d.

Thus, we have explicitly demonstrated that the perturbative ``dephasing''
contribution to the conductance obtained before from the nonperturbative
analysis can also be identified
in the first order perturbative expansion provided one infers the matrix elements
$M$ in the form (\ref{M1}). With this in mind one can immediately write down the
expression for $\delta \sigma_{\alpha \beta }^{\rm deph}$ in the form (\ref{corrd}).
For a quasi-1d case at low $T$ we find
\begin{equation}
\delta\sigma^{\rm deph}=\frac{e^2}{\pi}\frac{e^2}{\sigma_1}
\left[\frac{1}{\pi\sqrt{2}}\frac{D\tau_H^{3/2}}{\sqrt{\tau_e}}+
\frac{2D\tau_H}{\pi^2}\left(\ln\frac{2\pi\tau_H}{\tau_e}-6-\gamma_0 \right)\right],
\;\;\; \pi T\tau_H\ll 1
\label{dephquantum1}
\end{equation}
In the opposite high temperature limit we obtain
\begin{equation}
\delta\sigma^{\rm deph}=\frac{e^2}{\pi}\frac{e^2}{\sigma_1}
\left[\frac{4}{3\pi}DT\tau_H^2+\frac{1}{\pi\sqrt{2}}
\frac{D\tau_H^{3/2}}{\sqrt{\tau_e}}+\frac{\zeta(1/2)}{\sqrt{2\pi}}
D\sqrt{T}\tau_H^{3/2}+
\frac{2D\tau_H}{\pi^2}\ln\left(\frac{1}{4T\tau_e}\right)
-\frac{3\zeta(3/2)}{2\pi}\sqrt{\frac{D^2\tau_H}{2\pi T}}\right],
\;\;\; \pi T\tau_H\gg 1
\label{dephthermal1}
\end{equation}

Now let us come to the final step of our calculation
and evaluate the remaining terms in the general result
(\ref{dsig11})-(\ref{G}). We notice that the contribution of all terms containing
the combination $\coth (\omega /2T)\sin \xi_{21}t_1$ vanish after the integration over
the energy $\xi_{21}$. The same is true for the terms containing $\tanh (\xi_{\lambda_3}/2T)$
in the contribution of the vertex diagrams (\ref{dsigver}). These observations imply
that all the remaining nonvanishing terms come from the self-energy diagrams of Fig. 2a,b
and contain $1-2\rho$ or $\tanh$. We will denote their total contribution as
$\delta \sigma^{\tanh}$. We already know from the above analysis that this
contribution comes from the expansion of the pre-exponent $A_1(t)$ to the first order in the
interaction. Collecting all such terms from (\ref{dsig11}), (\ref{F}) and
(\ref{Clmb}), we obtain
\begin{equation}
\delta \sigma ^{\tanh }=\delta \sigma _{\cos}^{\tanh }+\delta \sigma _{\sin }^{\tanh }
+\delta \sigma ^{C},
\label{dsigtanh}
\end{equation}
where
\begin{eqnarray}
\delta\sigma^{\tanh}_{\cos}&=&
-\frac{e^2}{2{\cal V}\sigma_1}\int\limits_0^{+\infty} dt_1\int d\xi_{\lambda_1}
\int d\xi_{\lambda_2}\int d\xi_{\lambda_3}
\int\frac{d\omega}{2\pi}
\left(\frac{d}{d\xi_{\lambda_1}}\tanh\frac{\xi_{\lambda_1}}{2T}\right) jj(\xi_{21})
\cos \xi_{21}t_1
\nonumber\\
&&\times
M(\xi_{32}) \omega\tanh\frac{\xi_{\lambda_3}}{2T}
\frac{1-\cos[(\xi_{32}+\omega)t_1]}{(\xi_{32}+\omega)^2}.
\label{dephtanh}
\end{eqnarray}
\begin{eqnarray}
\delta \sigma _{\sin }^{\tanh } &=&\frac{e^2}{2{\cal V}\sigma _1}%
\int\limits_0^{+\infty }dt_1\int d\xi_{\lambda_1}\int d\xi_{\lambda_2}
\int d\xi{\lambda_3}\int \frac{d\omega }{%
2\pi }\left( \frac d{d\xi_{\lambda_1}}\tanh \frac{\xi_{\lambda_1}}{2T}\right) jj(\xi_{21})\sin \xi_{21}t_1  \nonumber \\
&&\ \times M(\xi_{32})\omega \tanh \frac{\xi_{\lambda_3}}{2T}
\frac{(\xi_{32}+\omega )t_1-\sin (\xi_{32}+\omega )t_1}{(\xi_{32}+\omega )^2},
\label{dephtanhsin}
\end{eqnarray}
\begin{equation}
\delta \sigma ^{C} =-\frac{e^2}{4{\cal {V}}}%
\int\limits_0^{+\infty }dt_1\ t_1\ \left( \frac d{d\xi_{\lambda_1}}
\tanh \frac{\xi_{\lambda_1}}{2T}\right) jj(\xi_{21})
\left\langle \lambda _2\right| \frac{[1-2\rho ](\bbox{r}_1,%
\bbox{r}_2)}{|\bbox{r}_1-\bbox{r}_2|}\left| \lambda _2\right\rangle \sin \xi_{21}t_1,
\label{dsigcoulomb}
\end{equation}
As before, the above equations were obtained from the exact ones by imposing $\xi_{42}=0$.
In order to establish a somewhat closer relation to the approach developed by AAG 
we also note,
that it is the contribution $\delta \sigma^{\tanh}$ (\ref{dsigtanh}) which contains the
so-called Hikami boxes within the diagrammatic analysis of Ref. \onlinecite{AAG2}. AAG argued
that (partial) cancellation of the first order diagrams can only be observed
if one takes the Hikami boxes into account. Below we will demonstrate that this is not the case.
Actually we have already shown in Sec. 4B (eqs. (\ref{dsigGR}), (\ref{func})) that
this cancellation (of the
linear in time ``golden rule'' terms only!) in the first order at $T=0$ occurs already before
disorder averaging and thus has nothing to do with the Hikami boxes. Now we will
illustrate this fact again by means of a direct calculation.

Let us first consider the term $\delta\sigma^{\tanh}_{\cos}$ (\ref{dephtanh}).
The integral over $\omega$ can be evaluated exactly and we get
\begin{equation}
\delta \sigma _{\cos }^{\tanh }=\frac{e^2}{4{\cal V}\sigma _1}
\int\limits_0^{+\infty }dt_1\int d\xi_{\lambda_1}\int d\xi_{21}
\int d\xi_{32}\left(
\frac d{d\xi_{\lambda_1}}\tanh \frac{\xi_{\lambda_1}}{2T}\right) jj(\xi_{21})
\cos \xi_{21}t_1M(\xi_{32})\xi_{32}t_1\tanh \frac{\xi_{\lambda_1}+\xi_{21}+\xi_{32}}{2T}.
\label{dephtanh1}
\end{equation}
Further calculation will be performed for a quasi-1d case. We also make use of
the real time representation of our integrals, as it was already done before.
For 1d systems we obtain from (\ref{M1})
\begin{equation}
\xi_{32} M(\xi_{32})=\frac{\sqrt{D}}{2\sqrt{2}\pi}
\frac{\xi_{32}}{|\xi_{32}|^{3/2}}
=\frac{-i}{4\pi}\int dt''
e^{i\xi_{32} t''}\left(\frac{d}{dt}\sqrt{\frac{4D|t''|}{\pi}}\right).
\label{M2}
\end{equation}
Also we will use the following relation
\begin{equation}
\int d\xi_{\lambda_1}
\left(\frac{d}{d\xi_{\lambda_1}}\tanh\frac{\xi_{\lambda_1}}{2T}\right)
\tanh\frac{\xi_{\lambda_1}+\xi_{21}+\xi_{32}}{2T}=
2\pi\int dt' e^{i(\xi_{21}+\xi_{32})t'}
\frac{Tt'}{\sinh\pi Tt'}{\cal P}\frac{-iT}{\sinh\pi Tt'}.
\label{tanhabc}
\end{equation}
Substituting (\ref{jj1}), (\ref{M2}) and (\ref{tanhabc}) into eq.(\ref{dephtanh1}),
we find
\begin{equation}
\delta \sigma _{\cos }^{\tanh }=-\frac{e^2}{4\pi ^3}\frac{e^2D}{\sigma _1}%
\int\limits_0^{+\infty }dt_1\int\limits_{-\infty }^{+\infty }dt^{\prime
}\left( {\cal P}\frac{\pi T}{\sinh \pi Tt^{\prime }}\right) \frac{\pi T}{%
\sinh \pi Tt^{\prime }}\;t_1\sqrt{|t^{\prime }|}\left[ \frac{%
e^{-|t_1+t^{\prime }|/\tau _H}}{\sqrt{|t_1+t^{\prime }|}}+\frac{%
e^{-|t_1-t^{\prime }|/\tau _H}}{\sqrt{|t_1-t^{\prime }|}}\right] .
\end{equation}
After simple algebra this equation can be converted into the following integral:
\begin{equation}
\delta \sigma _{\cos }^{\tanh }=-\frac{e^2}{\pi ^3}\frac{e^2D}{\sigma _1}%
\int\limits_0^{+\infty }dt\frac{e^{-t/\tau _H}}{\sqrt{t}}\left\{
t\int\limits_{2\tau _e/\pi }^tdt^{\prime }\left( \frac{\pi T}{\sinh \pi
Tt^{\prime }}\right) ^2\sqrt{t^{\prime }}+\int\limits_t^{+\infty }dt^{\prime
}\left( \frac{\pi T}{\sinh \pi Tt^{\prime }}\right) ^2{t^{\prime }}%
^{3/2}\right\}
\label{dephtanh2}
\end{equation}
In the quantum limit $\pi T\tau_H\ll 1$ we get
\begin{eqnarray}
\delta\sigma^{\tanh}_{\cos}&=&
-\frac{e^2}{\pi^3}\frac{e^2D}{\sigma_1}
\int\limits_0^{+\infty} dt\frac{e^{-t/\tau_H}}{\sqrt{t}}
\left\{2\sqrt{\frac{\pi}{2\tau_e}}t-2\sqrt{t}+
\frac{3\zeta(3/2)}{4\sqrt{2}}\frac{1}{\sqrt{T}}\right\}
\nonumber\\
&=& -\frac{e^2}{\pi}\frac{e^2}{\sigma_1}\left[\frac{1}{\pi\sqrt{2}}
\frac{D\tau_H^{3/2}}{\sqrt{\tau_e}}-\frac{2D\tau_H}{\pi^2}+
\frac{3\zeta(3/2)}{4\pi}\sqrt{\frac{D^2\tau_H}{2\pi T}}\right],\;\;\;
\pi T\tau_H\ll 1.
\label{tanhquantum}
\end{eqnarray}
To consider the opposite thermal limit $\pi T\tau_H\gg 1$ it is
convenient to rewrite this equation in the following form:
\begin{eqnarray}
\delta\sigma^{\tanh}_{\cos}&=&
-\frac{e^2}{\pi^3}\frac{e^2D}{\sigma_1}
\int\limits_0^{+\infty} dt\frac{e^{-t/\tau_H}}{\sqrt{t}}
\left\{2t\sqrt{\frac{\pi}{2\tau_e}}-2\sqrt{t}\frac{\pi Tt}{\sinh\pi Tt}-
4t\sqrt{\pi T}\int\limits_{0}^{\pi Tt} dx \frac{\sqrt{x}}{\sinh^2x}
(x\coth x-1)\right.
\nonumber\\
&&
\left.
+\frac{1}{\sqrt{\pi T}}\int\limits_{\pi Tt}^{+\infty} dx
\frac{x^{3/2}}{\sinh^2x} \right\}.
\end{eqnarray}
This equation yields
\begin{eqnarray}
\delta\sigma^{\tanh}_{\cos}&=&
-\frac{e^2}{\pi^3}\frac{e^2D}{\sigma_1}
\int\limits_0^{+\infty} dt\frac{e^{-t/\tau_H}}{\sqrt{t}}
\left\{
2t\sqrt{\frac{\pi}{2\tau_e}}+\frac{\pi\zeta(1/2)}{\sqrt{2}}t\sqrt{T}\right\}
\nonumber\\
&=&
-\frac{e^2}{\pi}\frac{e^2}{\sigma_1}\left[
\frac{1}{\pi\sqrt{2}}\frac{D\tau_H^{3/2}}{\sqrt{\tau_e}}+
\frac{\zeta(1/2)}{2\sqrt{2\pi}}D\sqrt{T}\tau_H^{3/2}\right],\;\;\;
\pi T\tau_H\gg 1.
\label{tanhthermal}
\end{eqnarray}
Here we have used the following integrals:
\begin{equation}
\int\limits_0^{+\infty} dx\frac{\sqrt{x}}{\sinh^2x}(x\coth x-1)=
-\sqrt{\frac{\pi}{2}}\frac{\zeta(1/2)}{4},\;\;\;\;
\int\limits_0^{+\infty} dx\frac{x^{3/2}}{\sinh^2x}=
\sqrt{\frac{\pi}{2}}\frac{3\zeta(3/2)}{4}.
\end{equation}

Now we turn to the correction (\ref{dephtanhsin}). To begin with, we should
handle a divergence  which appears in the integral over $\omega $ for the term
linear in $t_1$. It is easy to demonstrate, however, that this divergence is
fictitious. It disappears completely if a more accurate expression for the matrix
elements is used. This expression reads:
$$
\frac{M(\xi_{32})\omega }{\sigma _1}=\int \frac{dk}{2\pi }\ \frac 1{k^2}%
\left\langle \lambda _2\right| e^{i\bbox{kr}}\left| \lambda _3\right\rangle
\left\langle \lambda _3\right| e^{-i\bbox{kr}}\left| \lambda _2\right\rangle
\text{Im}\left( \frac{-4\pi }{\epsilon (\omega ,k)}\right) .
$$
Now we can use the analytical properties of the function $1/\epsilon $ and write
$$
{\cal P}\int \frac{d\omega }{2\pi }\ \text{Im}\left( \frac 1{\epsilon
(\omega ,k)}\right) \frac 1{\xi_{32}+\omega }=\frac 12\text{Re}\left(
\frac 1{\epsilon (\xi_{23},k)}-1\right) .
$$
Substitution this identity into eq. (\ref{dephtanhsin}) we immediately
observe that the term containing $-1$ is exactly canceled by the
correction $\delta \sigma^C.$ Thus the result is finite and has the form
\begin{equation}
\delta \sigma ^{C}+\delta \sigma _{\sin }^{\tanh }=\delta
\sigma _1^{\tanh }+\delta \sigma _2^{\tanh },
\label{manysigmas}
\end{equation}
where
\begin{equation}
\delta \sigma _1^{\tanh }=-\frac{e^2}{4{\cal V}\sigma _1}\int\limits_0^{+%
\infty }dt_1\int d\xi_{\lambda_1}\int d\xi_{\lambda_2}
\int d\xi_{\lambda_3}\left( \frac d{d\xi_{\lambda_1}}\tanh
\frac{\xi_{\lambda_1}}{2T}\right) jj(\xi_{21})\sin \xi_{21}t_1
M(\xi_{32})\tanh \frac{\xi_{\lambda_3}}{2T}\ .
\label{dsig022}
\end{equation}
and
\begin{eqnarray}
\delta \sigma _2^{\tanh } &=&-\frac{e^2D}{4{\cal V}\sigma _1}%
\int\limits_0^{+\infty }dt_1\ t_1\int d\xi_{\lambda_1} \int d\xi_{\lambda_2}
\int d\xi_{\lambda_3}\left( \frac d{%
d\xi_{\lambda_1}}\tanh \frac{\xi_{\lambda_1}}{2T}\right) jj(\xi_{21})\sin \xi_{21}t_1\times   \nonumber \\
&&\left\langle \lambda _2\right| _{\bbox{r}_1}\left\langle \lambda _3\right|
_{\bbox{r}_2}\delta (\bbox{r}_1-\bbox{r}_2)\left| \lambda _3\right\rangle _{%
\bbox{r}_1}\left| \lambda _4\right\rangle _{\bbox{r}_2}\tanh \frac{\xi_{\lambda_3}}{2T%
}.  \label{dsig021}
\end{eqnarray}
Here we have used the formula
$$
{\rm Re}\left( \frac 1{\epsilon (\xi_{23},k)}%
\right) =\frac{Dk^2}{4\pi \sigma _1}.
$$

The contribution (\ref{dsig022}) can be transformed and evaluated analogously to the
term $\delta \sigma _{\cos}^{\tanh }$. We find
\begin{equation}
\delta \sigma _1^{\tanh }=\frac{e^2}{2\pi ^3}\frac{e^2D}{\sigma _1}%
\int\limits_0^{+\infty }dt_1\int\limits_{-\infty }^{+\infty }dt^{\prime
}\left( {\cal P}\frac{\pi T}{\sinh \pi Tt^{\prime }}\right) \frac{\pi T}{%
\sinh \pi Tt^{\prime }}\;t^{\prime }\sqrt{|t^{\prime }|}\left[ \frac{%
e^{-|t_1+t^{\prime }|/\tau _H}}{\sqrt{|t_1+t^{\prime }|}}-\frac{%
e^{-|t_1-t^{\prime }|/\tau _H}}{\sqrt{|t_1-t^{\prime }|}}\right] .
\end{equation}
After simple transformations we obtain
\begin{equation}
\delta \sigma _1^{\tanh }=-\frac{2e^2}{\pi ^3}\frac{e^2D}{\sigma _1}%
\int\limits_0^{+\infty }dt\ \frac{e^{-t/\tau _H}}{\sqrt{t}}\
\int\limits_t^{+\infty }dt^{\prime }\ \left( \frac{\pi T}{\sinh \pi
Tt^{\prime }}\right) ^2t^{^{\prime }3/2}=\left\{
\begin{array}{cc}
-\frac{e^2}\pi \frac{e^2}{\sigma _1}\frac{3\zeta (3/2)}{2\pi }\sqrt{\frac{%
D^2\tau _H}{2\pi T}} & \pi T\tau _H\ll 1 \\
0 & \pi T\tau _H\gg 1
\end{array}
\right. .
\label{dsigmasin2}
\end{equation}
In order to evaluate the term $\delta \sigma _2^{\tanh }$ we use the
following expression
\begin{equation}
\left\langle \lambda _2\right| _{\bbox{r}_1}\left\langle \lambda _3\right| _{%
\bbox{r}_2}\delta (\bbox{r}_1-\bbox{r}_2)\left| \lambda _3\right\rangle _{%
\bbox{r}_1}\left| \lambda _4\right\rangle _{\bbox{r}_2}=\frac 1\pi \int
dt^{\prime \prime }\ \frac{e^{i\xi_{32}t^{\prime \prime }}}{\sqrt{%
4\pi D|t^{\prime \prime }|}}.
\label{delt}
\end{equation}
It can be obtained by comparing the two expressions for the
matrix element $M(\xi_{32})$. The first expression,
$$
M(\xi_{32})=\frac{1}{\pi}\int dt\, e^{-i\xi_{32}t} \int dx
\frac{-|x|}{2}{\cal D}(|t|,x),
$$
follows directly from the eqs. (\ref{ft},\ref{Idiff}) and (\ref{f11}),
and the second one,
$$
M(\xi_{32})=
\left\langle \lambda _2\right| _{\bbox{r}_1}\left\langle \lambda _3\right| _{
\bbox{r}_2}\frac{-|\bbox{r}_1-\bbox{r}_2|}{2}\left| \lambda _3\right\rangle _{
\bbox{r}_1}\left| \lambda _4\right\rangle _{\bbox{r}_2}
$$
can be derived from (\ref{M}). Thus we obtain
$$
\left\langle \lambda _2\right| _{\bbox{r}_1}\left\langle \lambda _3\right| _{
\bbox{r}_2} f(\bbox{r}_1-\bbox{r}_2)\left| \lambda _3\right\rangle _{
\bbox{r}_1}\left| \lambda _4\right\rangle _{\bbox{r}_2}=
\frac{1}{\pi}\int dt e^{-i\xi_{32}t} \int dx\, f(x){\cal D}(|t|,x)
$$
and arrive at eq. (\ref{delt}).
With the aid of this formula we find
\[
\delta \sigma _2^{\tanh }=-\frac{e^2}{2\pi ^3}\frac{e^2D}{\sigma _1}%
\int\limits_0^{+\infty }dt_1\int\limits_{-\infty }^{+\infty }dt^{\prime
}\left( {\cal P}\frac{\pi T}{\sinh \pi Tt^{\prime }}\right) \frac{\pi T}{%
\sinh \pi Tt^{\prime }}\;t_1\frac{t^{\prime }}{\sqrt{|t^{\prime }|}}\left[
\frac{e^{-|t_1+t^{\prime }|/\tau _H}}{\sqrt{|t_1+t^{\prime }|}}-\frac{%
e^{-|t_1-t^{\prime }|/\tau _H}}{\sqrt{|t_1-t^{\prime }|}}\right] ,
\]
which yields
\begin{equation}
\delta \sigma _2^{\tanh }=\frac{2e^2}{\pi ^3}\frac{e^2D}{\sigma _1}%
\int\limits_0^{+\infty }dt\ \frac{e^{-t/\tau _H}}{\sqrt{t}}\
\int\limits_0^{+\infty }dt^{\prime }\ \left( \frac{\pi T}{\sinh \pi
Tt^{\prime }}\right) ^2t^{^{\prime }3/2}=\frac{e^2}\pi \frac{e^2}{\sigma _1}%
\frac{3\zeta (3/2)}{2\pi }\sqrt{\frac{D^2\tau _H}{2\pi T}}.
\label{dsigmasin1}
\end{equation}
With the aid of eqs. (\ref{dsigmasin2}) and (\ref{dsigmasin1}) we observe that 
the result (\ref{manysigmas})
is zero at $T \to 0$ and it is equal to $\delta \sigma _2^{\tanh }$ (\ref{dsigmasin1})
for $\pi Tt \gg 1$. Combining eqs. (\ref{dsigtanh}), (\ref{tanhquantum}), (\ref{tanhthermal}),
(\ref{manysigmas}), (\ref{dsigmasin2}) and (\ref{dsigmasin1})
we arrive at the final results for $\delta \sigma^{\tanh}$:
\begin{equation}
\delta\sigma^{\tanh}=
-\frac{e^2}{\pi}\frac{e^2}{\sigma_1}\left[\frac{1}{\pi\sqrt{2}}
\frac{D\tau_H^{3/2}}{\sqrt{\tau_e}}-\frac{2D\tau_H}{\pi^2}+
\frac{3\zeta(3/2)}{4\pi}\sqrt{\frac{D^2\tau_H}{2\pi T}}\right],\;\;\;
\pi T\tau_H\ll 1,
\label{tanhquantum02}
\end{equation}
\begin{equation}
\delta\sigma^{\tanh}=
-\frac{e^2}{\pi}\frac{e^2}{\sigma_1}\left[
\frac{1}{\pi\sqrt{2}}\frac{D\tau_H^{3/2}}{\sqrt{\tau_e}}+
\frac{\zeta(1/2)}{2\sqrt{2\pi}}D\sqrt{T}\tau_H^{3/2}-
\frac{3\zeta (3/2)}{2\pi }\sqrt{\frac{D^2\tau _H}{2\pi T}}\right],\;\;\;
\pi T\tau_H\gg 1.
\label{tanhthermal02}
\end{equation}

In order to find the total expression for the weak localization correction one should
simply add the two contributions $\delta \sigma^{\rm deph}$ (\ref{dephquantum1}),
(\ref{dephthermal1}) and $\delta \sigma^{\tanh}$  (\ref{tanhquantum02}), (\ref{tanhthermal02})
together. We observe that the temperature independent terms $\propto \tau_H^{3/2}$ are equal
in these two expressions, they enter with the opposite signs and cancel each other exactly
in the sum $\delta \sigma^{\rm deph}+\delta \sigma^{\tanh}$ in both limits $\pi Tt \ll 1$
and $\pi Tt \gg 1$. As we have already discussed, these are just the linear in time
``golden rule'' terms coming from the exponent  ($\delta \sigma^{\rm deph}$) and
the pre-exponent ($\delta \sigma^{\tanh}$). Their cancellation occurs in no relation to
(and due to much more general reasons than) averaging over disorder. Other (``non-golden-rule'')
terms do not cancel and combine in the final result which we will present below.

\subsection{Discussion}

Although the main differences between our approach and that of AAG \cite{AAG2}
can already be understood from the above analysis, we will briefly summarize them again for 
the sake of clarity.

\begin{enumerate}

\item
The first crucial difference to be emphasized here is that our method \cite{GZ2} is 
essentially nonperturbative in the interaction while
the approach \cite{AAG2} is only the first order perturbation theory. In
the most interesting limiting case $\tau_H \gtrsim \tau_{\varphi}$ (which was
only considered in our Refs. \onlinecite{GZ1,GZ2,GZ98}) one {\it cannot} proceed
perturbatively in the interaction at any temperature including $T=0$.
This is precisely what AAG do: it is demonstrated in Appendix A
that the general result for the conductivity \cite{AAG2} is identical to the
first order expansion of (\ref{J17}) in the interacting terms $iS_R+S_I$
while all higher order terms (which are larger than the first order term for
$\tau_H \gtrsim \tau_{\varphi}$) were not taken into account in \cite{AAG2}. In contrast,
our path integral approach is equivalent to an effective summation of diagrams
in {\it all} orders with the exponential accuracy. This is sufficient
for correct evaluation of $\tau_{\varphi}$. Within our analysis only the action in the
exponent (rather than the whole expression
for $\sigma$) is expanded in the interaction (this is correct as long as $p_Fl \gg 1$).
Our method also allows for a clear distinction between the exponent and
the pre-exponential contribution to $\delta \sigma$.

It remains unclear to us why AAG repeatedly stated that
our procedure ``is nothing but a perturbative expansion'' \cite{AAG1} and our
results are ``purely perturbative'' \cite{AAG2}. The only justification of the
above statements which we could extract from the above papers is that
our result for the dephasing rate $1/\tau_{\varphi}$ ``is proportional to
the first power of the fluctuation propagator'' \cite{AAG2}. Although the 
latter is true in some limits, it is
hard to understand how this could help to turn a nonperturbative problem into
a perturbative one. Indeed, if one formally multiplies the photon propagator by
a constant $\lambda$ everywhere in our calculation, one would obtain
$1/\tau_{\varphi} \propto \lambda$. The same holds for the
calculation \cite{AAG2}. Note, however, that it is not the decoherence
rate $1/\tau_{\varphi}$ (we are not aware of a quantum mechanical operator
which expectation value would correspond to such a quantity) but rather the
expectation value for the current operator which is calculated theoretically and
measured in experiments. For $\tau_H \gtrsim \tau_{\varphi}$ the result for the weak
localization correction $\delta \sigma $ depends on $\lambda$ as
$$
\delta \sigma_1 \propto -1/\sqrt{\lambda}
$$
in 1d (cf. eq. (\ref{2})) and $ \delta \sigma_2 \propto \ln \lambda$ in 2d. Obviously,
these results are purely nonperturbative in the ``interaction strength'' $\lambda$.
Any attempt to calculate the expectation value of the current operator perturbatively
may only yield to divergences in all orders of the expansion in powers of $\lambda$.
As to the decoherence rate $1/\tau_{\varphi}$, it is only {\it extracted} from the
nonperturbative results for the conductance correction. Hence, the relation
$1/\tau_{\varphi} \propto \lambda$ cannot by itself tell anything
about the perturbative or nonperturbative character of the calculation.
In the limit $\tau_H \ll \tau_{\varphi}$ the conductance correction $\delta \sigma (H)$
can be evaluated perturbatively in $\lambda$. However, as it was explained in Sec. 2,
even in this limit $\tau_{\varphi}$ can be unambiguously determined only within
the nonperturbative procedure, while any perturbative expansion yields
ambiguous results for $\tau_{\varphi}$ which fully depend on the assumption
about the decay of correlations in time.

\item
Another crucial difference is that AAG essentially use
the {\it assumption} about a purely exponential decay of the phase correlations
in time while no such assumption was used within our analysis. Specifically,
eq. (3.2) of Ref. \onlinecite{AAG2} is equivalent to our eq. (\ref{2}) only provided
one assumes that $f_d(t)$ is a linear function of time $f_d(t)=t/\tau_{\varphi}$
and ignores the effect of the interaction on the pre-exponent, i.e. puts
$A_d(t)=1/(4\pi Dt)^{d/2}$. This assumption cannot be checked within the
perturbation theory in the interaction and, as it was already explained above,
in general it can only be valid within the golden rule approximation. The whole
comparison between ours and AAG's results carried out by the authors 
\cite{AAG2} is essentially based on their eq. (3.2) which was neither used
nor even written down in our paper \cite{GZ2}.  

Let us emphasize that AAG (unlike many others) {\it do not} use the golden rule
approximation in their perturbative calculation of the weak localization correction
$\delta \sigma (H)$ in the limit $\tau_H \ll \tau_{\varphi}$. However, they explicitly
use this approximation while extracting $\tau_{\varphi}$ from $\delta \sigma (H)$:
eq. (3) of Ref. \onlinecite{AAG1} and eq. (4.3) of Ref. \onlinecite{AAG2} are valid
{\it only} within the golden rule approximation. As it was
demonstrated above, $f_d$ is not a linear function of time
(cf. eqs. (\ref{fott})-(\ref{deltaf2})) and, moreover,
in the presence of interaction the pre-exponent in (\ref{2}) deviates from its
``noninteracting'' form $A^{(0)}_d(t)=1/(4\pi Dt)^{d/2}$. As a result, the relation
between $\delta \sigma (H)$ and $\tau_{\varphi}$ depends on
temperature and is different from eq. (3) of Ref. \onlinecite{AAG1} (or eq. (4.3) of 
Ref. \onlinecite{AAG2})
at any $T$ even in the limit $\tau_H \ll \tau_{\varphi}$. Since the linear in time
$T$-independent contributions from the exponent and the pre-exponent exactly cancel
each other in the first order perturbation theory, the golden-rule-type assumption
\cite{AAG1,AAG2} about purely exponential decay of correlations in time inevitably
yields to missing of the $T$-independent contribution (\ref{tau0}) to $\tau_{\varphi}$.

\item
Let us compare all the approximations used by AAG \cite{AAG2} and in our paper \cite{GZ2}.
In both papers the same quasiclassical condition $p_Fl \gg 1$ was assumed and the
expressions for the photon propagators were defined within RPA. 
In order to perform the perturbative expansion in
the interaction AAG considered the limit of strong
magnetic fields $\tau_H \ll \tau_{\varphi}$ (in Ref. \onlinecite{AAG1} this
condition was not quoted). AAG also performed the expansion in
the inverse dimensionless conductance $1/g(L_H)$, i.e. they assumed that
$g \gg 1$ on the scale of the magnetic length $L_H=\sqrt{D\tau_H}$. Although we
do not need these approximations within our nonperturbative analysis \cite{GZ2}, their
appearance in the perturbative treatment \cite{AAG2} is understandable.

As to an additional condition $T \tau_H \gg 1$, in our opinion it is not needed
even within the perturbative procedure of AAG. Indeed, the
condition $g(L_H) \gg 1$ does not depend on temperature at all, and the
inequality  $\tau_H \ll \tau_{\varphi}$ can only become stronger at lower $T$
provided it is already satisfied at higher temperatures. Therefore under the
two latter conditions the perturbative expansion \cite{AAG2} should be justified
down to $T \to 0$ and the condition $T \tau_H \gg 1$ is not needed at all. This
condition should also be irrelevant for eqs. (2.42) of Ref. \onlinecite{AAG2}.
According to AAG ``all the corrections to these formulas
are small as $1/(T\tau_{\varphi})$''. Combining $\tau_H \ll \tau_{\varphi}$
and $g(L_H) \gg 1$ with eq. (2) of Ref. \onlinecite{AAG1} (or eq. (4.9)
of Ref. \onlinecite{AAG2}) $1/\tau_{\varphi}^{AAG}=T/g(L_H)$ we observe that the inequality
$T\tau_{\varphi}^{AAG} \gg 1$ is satisfied at all temperatures including $T \to 0$.

Thus the perturbative results \cite{AAG2} can be analyzed
in both limits $T \tau_H >1$ and $T\tau_H <1$. Since the latter limit of lower
temperatures was not discussed by AAG we carried out
the corresponding analysis in Appendix C.
Combining eqs. (\ref{lTtH}), (\ref{sTtH}) with eq. (4.3b) of Ref. \onlinecite{AAG2} we find
$1/\tau_{\varphi}^{AAG}=Te^2\sqrt{D\tau_H}/(4\sigma_1)$ for $T\tau_H \gg 1$
(just like in Ref. \onlinecite{AAG2}) and $1/\tau_{\varphi}^{AAG}=
3e^2\sqrt{D\tau_H}/(2\pi\sigma_1\tau_H)$ for $T\tau_H \ll 1$. The latter result
(which was not presented by AAG) demonstrates that a nonzero dephasing time
at $T=0$ is obtained even if one explicitly follows the procedure of
Ref. \onlinecite{AAG2}. Although due to the reasons explained above
this result differs from the correct one (\ref{tau0})  it is interesting to
observe that a nonzero dephasing rate at $T=0$ is already contained in
the formulas derived by AAG.

\item Subtle details of disorder averaging
do not play any significant role in the problem in question and can merely
influence some numerical prefactors of order one.
As it was demonstrated above without making {\it any} approximation,
no exact cancellation of the first order diagrams occurs even at $T=0$.
The ``non-canceled'' $T$-independent terms describe not only renormalization due to
interaction but also contribute to dephasing. These conclusions are
general and hold both before and after averaging.

As to the (partial) cancellation, it indeed occurs at $T \to 0$, but only for the linear
in time ``golden rule'' terms coming both from the exponent and the pre-exponent.
This partial cancellation is also due to very general reasons, it occurs already
in the exact (non-averaged) perturbative expression and has no relation
to the quasiclassical approximation and/or disorder averaging. AAG \cite{AAG2}
argued that this cancellation can not be reproduced if averaging over disorder
does not involve the so-called Hikami boxes. Our analysis demonstrates that this
is not true.

Let us compare the perturbative results for the weak localization correction
obtained in Ref. \onlinecite{AAG2} and within our analysis. In the limit $T\tau_H \gg 1$
for the 1d case AAG get (cf. eq. (4) of Ref. \onlinecite{AAG1}
or eq. (4.13a) of Ref. \onlinecite{AAG2}):
\begin{equation}
\delta\sigma_{WL}^{AAG}\simeq \frac{e^2}{\pi}\frac{e^2}{\sigma_1}
DT\tau_H^2\left[\frac14+\frac{\zeta(1/2)}{2\sqrt{2\pi T\tau_H}}
+{\cal O}\left(\frac{1}{(T\tau_H)^{3/2}}\right)\right].
\label{AAG155}
\end{equation}
In the same limit with the aid of our eqs. (\ref{dephthermal1}) and (\ref{tanhthermal02})
for the weak localization correction $\delta\sigma_{WL}=
\delta\sigma^{\rm deph}+\delta\sigma^{\tanh}$ we find
\begin{equation}
\delta\sigma_{WL}=\frac{e^2}{\pi}\frac{e^2}{\sigma_1}DT\tau_H^2
\left[\frac{4}{3\pi}+\frac{\zeta(1/2)}{2\sqrt{2\pi T\tau_H}}+
{\cal O}\left(\frac{1}{T\tau_H}\right) \right].
\label{GZ155}
\end{equation}
In the opposite limit $T\tau_H \ll 1$ {\it our} calculation of the integrals \cite{AAG2}
(see Appendix C) yields
\begin{equation}
\delta\sigma_{WL}^{AAG} \simeq
\frac{e^2}{\pi}\frac{e^2}{\sigma_1}
\frac{3D\tau_H}{2\pi}(1+{\cal O} (\sqrt{T\tau_H})).
\label{AAG156}
\end{equation}
Combining our eqs. (\ref{dephquantum1}) and (\ref{tanhquantum02}) in the same limit
$T \tau_H \ll 1$ we obtain
\begin{equation}
\delta\sigma_{WL}\simeq \frac{e^2}{\pi}\frac{e^2}{\sigma_1}
\frac{2D\tau_H}{\pi^2}\left(\ln\frac{2\pi\tau_H}{\tau_e}-5-\gamma_0+
{\cal O} (\sqrt{T\tau_H}) \right),
\label{GZ156}
\end{equation}
Note, that the ``renormalization'' terms $\propto \sqrt{\tau_H/T}$ (which are
irrelevant for dephasing and can be added to the interaction correction) are dropped
in eqs. (\ref{AAG155})-(\ref{GZ156}) for the sake of simplicity.

We observe that in both limits $T\tau_H \gg 1$ and $T\tau_H \ll 1$ the $T$-independent
terms $\propto \tau_H^{3/2}$ (see eqs. (\ref{dephquantum1}), (\ref{dephthermal1}),
(\ref{tanhquantum02}), (\ref{tanhthermal02})) exactly cancel each
other and do not contribute to the results (\ref{GZ155}), (\ref{GZ156}) at all.
The same cancellation occurs in the expressions \cite{AAG2} (\ref{AAG155}), (\ref{AAG156}).
The latter equations were derived within the averaging procedure involving
the Hikami boxes. In order to obtain (\ref{GZ155}), (\ref{GZ156}) we used a somewhat
different averaging procedure which amounts to deriving the matrix elements
$M(\omega')$ from the general properties of diffusive trajectories. Since in
both cases exactly the same cancellation occurs in both limits of high and low temperatures,
we conclude that the issue of the Hikami boxes raised in Ref. \onlinecite{AAG2} is
completely unimportant for this cancellation.

We can also add that the averaging procedure employed by AAG is
efficient within the perturbation theory while our procedure is developed to average the
nonperturbative results obtained within the path integral technique. The perturbative
results obtained within both methods are essentially the same in 2d (see Sec. 3C)
and practically the same in 1d apart from some unimportant details. Both procedures
yield nonzero ``dephasing'' terms even at $T=0$.

\end{enumerate}

\section{Caldeira-Leggett model}

As we have already discussed before \cite{GZ2,GZ98,GZr1,GZr2}, the physical
nature of the interaction-induced decoherence can be understood with the
aid of a simple model of a quantum particle interacting with a bath
of harmonic oscillators \cite{FV,FH}. By a proper choice of both the interaction
term and the frequency spectrum of the bath oscillators one can easily realize
the important limit of Ohmic dissipation and arrive at the Caldeira-Leggett
(CL) model \cite{CL}. Some rigorous results obtained within this exactly solvable
model are presented in Appendix D for the sake of completeness.

An important advantage of the CL model is that the density matrix and the expectation
values of the quantum mechanical operators can be calculated exactly. This enables one
not only to avoid worries concerning the validity range of various approximations, but also
to test these approximations employed in some other models which cannot be
solved exactly. In particular, here we are interested in checking the approximations
which have led various authors \cite{AGA,AAG1,AAG2,CI,IFS,RSC,FA}
to the conclusion about the absence of interaction-induced decoherence in 
disordered metals at $T=0$, or to the conclusion \cite{VA} that the quantum correction
to the classical decoherence rate is small and decreases this rate below its
classical value. Since it is well known that the off-diagonal elements of the
particle density matrix $\rho (x,x')$ are suppressed due to interaction with the CL bath
even at $T=0$ in equilibrium (this effect is nothing but nonzero decoherence at $T=0$),
it is interesting to test if it is possible to reproduce this result within the
approximations employed in the above papers.

Also, it is sometimes speculated that the results derived within the CL model
cannot be compared to ones obtained for electrons in a disordered metal because
of different statistics. One could conjecture that electrons
in a disordered metal should have zero decoherence rate at $T=0$ predominantly due to
the Pauli principle, while in the CL model nonzero decoherence at $T=0$ is allowed
because no exclusion principle exists for bosons. The role of the
Pauli principle can also be clarified by performing a direct comparison of the results
obtained within the CL model with ones for electrons in a disordered metal.

On a perturbative level this program will be carried out in the subsection A. In the
subsection B we will discuss the relation between the exponent and the pre-exponent
for the CL model and illustrate the analogy between the results of this
subsection and those of Sec. 3. We will develop this comparison further in the
subsection C where we analyze the properties of the ``Cooperon'' in the CL model.
The validity range of various approximations is discussed in the subsection D.

\subsection{Perturbation theory}

Since in practically all cases the conclusion about the zero decoherence
in the interacting systems at $T=0$ in equilibrium was reached only within
the first order perturbation theory in the interaction, it is instructive to
examine the structure of the first order perturbative terms in the CL model.

Let us expand the kernel of the evolution operator (\ref{J15}) in the interaction part
of the action $iS_R+S_I$. In the zeroth order we get a simple result
\begin{equation}
J_0(t,x_{1f},x_{2f},x_{1i},x_{2i})=
U(t,x_{1f},x_{1i})U^+(t,x_{2f},x_{2i}),
\label{J0}
\end{equation}
where $U(t,x_f,x_i)=\langle x_f|\exp(-i\hat p^2 t/2m)|x_i\rangle$ is a free
particle evolution operator. Investigating the transport properties of disordered conductors
one usually expresses the results in terms of advanced and retarded Green functions
$G^{R,A}$. In order to emphasize the
analogy with the CL model, we note that the expression (\ref{J0}) can be
rewritten as
\begin{equation}
J_0(t,x_{1f},x_{2f},x_{1i},x_{2i})=
G^R(t,x_{1f},x_{1i})G^A(-t,x_{2f},x_{2i}),
\label{J0RA}
\end{equation}
where $G^R(t,x_f,x_i)=-i\theta(t)U(t,x_f,x_i)$,
$G^A(t,x_f,x_i)=i\theta(-t)U(t,x_f,x_i)$. Comparing this expression to that
for the conductivity of a disordered metal (\ref{sigma00}), we note that
the latter contains an additional time integral,
$\sigma\propto\int dt\; J_0(t)$. This difference is not important though, in order to
simplify the comparison of the corresponding perturbative results one can always keep the
time $t$ finite (exactly as it was done in the preceding section) and perform the
time integration only at the last stage of the calculation.

Let us consider the first order correction to the kernel $J$ due to the interaction. This
correction is again given by the sum of the four diagrams of Fig. 2.
The current operators $\hat j_\alpha$ are, however, not applied. Also
the ``photon propagators'' are now different. Namely, instead of the function
$R(t_3-t_4,\bbox{r}_3-\bbox{r}_4)$ one should substitute the function
$\alpha_R(t_3-t_4)x_3x_4$, while instead of $I(t_3-t_4,\bbox{r}_3-\bbox{r}_4)$
one should use $\alpha_I(t_3-t_4)x_3x_4$ (see eqs. (\ref{alphaR}), (\ref{alphaI})).
In contrast to the case of an electron propagating in a disordered metal
(\ref{J17}-\ref{SI}) the action in the exponent of (\ref{J15}) does not
contain the factor $1-2n(\bbox{p},\bbox{r})$. Therefore the operator $1-2\rho$,
related to the Fermi statistics,
does not appear in the perturbation theory. The free particle states are
labeled by its momentum, therefore the indices $\lambda_j$ in the diagrams
of Fig. 2 should be understood as the momentum values.

For the sake of brevity we will omit the general result for the
first order correction $\delta J^{(1)}$ to the operator $J$ which is
expressed in terms of the same functions (\ref{F}) and (\ref{G}) with
$\xi_{\lambda_j} \to E_p$. Rather we immediately
go over to the part of the kernel $J$ describing the evolution of the
diagonal elements of the
density matrix, which corresponds to the "diagonal" part of the conductivity
(\ref{dsigdiag}). This is sufficient for our illustration purposes.
For the probability of the transition from the state with the
momentum $q$ to the state with the momentum $p$ after the time $t$ we find
\begin{eqnarray}
\delta J^{(1)}_{pp,qq}(t)&=&
\langle p|_{x_{1f}}\langle q|_{x_{2i}}
\delta J^{(1)}(t,x_{1f},x_{1i},x_{2f},x_{2i})
|p\rangle_{x_{2f}} |q\rangle_{x_{1i}}=
\nonumber\\
&&
-2\eta\delta_{pq}\sum\limits_k |x_{pk}|^2\int\frac{d\omega}{2\pi}
\omega\left(\coth\frac{\omega}{2T}+1\right)
\frac{1-\cos\left[(E_q-E_k-\omega)t\right]}{(E_q-E_k-\omega)^2}
\nonumber\\
&&
+2\eta |x_{pq}|^2\int\frac{d\omega}{2\pi}
\omega\left(\coth\frac{\omega}{2T}-1\right)
\frac{1-\cos\left[(E_p-E_q-\omega)t\right]}{(E_p-E_q-\omega)^2},
\label{dJ}
\end{eqnarray}
where $x_{pk}=\langle p|x|k\rangle$ is the matrix element of the operator $x$
and $E_p=p^2/2m \geq 0$ is the energy of the free particle with the momentum $p$.

Let us first evaluate the above expression within the standard golden rule
approximation performed e.g. in Refs. \onlinecite{FA,IFS,RSC} and others.
As it was already discussed in Section 4.2, this approximation is equivalent to
the following replacement
\begin{equation}
\frac{1-\cos\left[(E_p-E_q-\omega)t\right]}{(E_p-E_q-\omega)^2}
\longrightarrow
\pi t\delta(E_p-E_q-\omega).
\label{GRO}
\end{equation}
Performing this replacement in (\ref{dJ}) we get
\begin{eqnarray}
\delta J^{(1)}_{ppqq}(t)&=&
-\eta t\delta_{pq}\sum\limits_k |x_{pk}|^2 (E_q-E_k)\left(
\coth\frac{E_q-E_k}{2T}+1\right)
\nonumber\\
&&
+\eta t|x_{pq}|^2(E_p-E_q)\left(\coth\frac{E_p-E_q}{2T}-1\right).
\label{dJGR}
\end{eqnarray}
The first term in this expression, as well as in the expression (\ref{dJ}),
originates from the self-energy diagrams (a) and (b) in Fig. 2. This term
describes the out-scattering rate. It is equal to the sum over all possible transitions
from the initial state $q$ to all finite states $k$. It is mostly
important as far as dephasing is concerned. The second term in the eq.
(\ref{dJ},\ref{dJGR}) just gives the transition rate from a given initial
state $q$ to a given final state $p$. It comes from the vertex diagrams
(c) and (d) in Fig. 2.

Now let us investigate the time evolution of the density matrix
provided the initial density matrix is just the equilibrium
one for a free particle. In the momentum representation it has the form:
$$
\rho^{\rm eq}_{pq}=\delta_{pq}\frac{1}{L}\sqrt{\frac{2\pi}{mT}}
\exp\left(-\frac{p^2}{2mT}\right),
$$
where $L$ is the length of the system.
Substituting all these results into
eq. (\ref{rho}) we obtain the following expression for the occupation
probability of the state $p$ within the golden rule approximation:
\begin{eqnarray}
\delta \rho^{(1)}_{pp}(t)&=&
-\eta t\frac{1}{L}\sqrt{\frac{2\pi}{mT}}
\sum\limits_k |x_{pk}|^2 (E_p-E_k)\left(
\coth\frac{E_p-E_k}{2T}+1\right)\exp\left(-\frac{E_p}{T}\right)
\nonumber\\
&&
+\eta t\frac{1}{L}\sqrt{\frac{2\pi}{mT}}
\sum\limits_q
|x_{pq}|^2(E_p-E_q)\left(\coth\frac{E_p-E_q}{2T}-1\right)
\exp\left(-\frac{E_q}{T}\right).
\label{dwGR}
\end{eqnarray}
We observe that the combination
\begin{equation}
\left(\coth\frac{E_p-E_k}{2T}+1\right)\exp\left(-\frac{E_p}{T}\right).
\label{combCL}
\end{equation}
appeared in the result (\ref{dwGR}). It is very similar to the combination
(\ref{GR}) in the case of a disordered metal. Keeping in mind the condition $E_p>0$,
one can easily see that this combination again yields zero result at $T=0$, i.e.
within the golden rule approximation relaxation processes are forbidden
in the zero temperature limit. This result is fully equivalent
to one presented e.g. in eqs. (1,2) of Ref. \onlinecite{IFS}, and it is in an obvious
contradiction to the exact solution of the CL model.

Now let us perturbatively find the occupation probabilities {\it without}
making the golden rule approximation. We get
\begin{eqnarray}
\delta \rho^{(1)}_{pp}(t)&=&
-2\eta \frac{1}{L}\sqrt{\frac{2\pi}{mT}}\exp\left(-\frac{E_p}{T}\right)
\sum\limits_k |x_{pk}|^2\int\frac{d\omega}{2\pi}
\omega\left(\coth\frac{\omega}{2T}+1\right)
\frac{1-\cos\left[(E_p-E_k-\omega)t\right]}{(E_p-E_k-\omega)^2}
\nonumber\\
&&
+2\eta\frac{1}{L}\sqrt{\frac{2\pi}{mT}}
\sum\limits_q
|x_{pq}|^2\int\frac{d\omega}{2\pi}
\omega\left(\coth\frac{\omega}{2T}-1\right)
\frac{1-\cos\left[(E_p-E_q-\omega)t\right]}{(E_p-E_q-\omega)^2}
\exp\left(-\frac{E_q}{T}\right).
\label{dw}
\end{eqnarray}
As it was already done for the case of a disordered metal (Sec. 4B), let us
consider the first part of this expression determined by the self-energy
diagrams of Fig. 2a,b. In the zero temperature limit we find
\begin{equation}
\int\frac{d\omega}{2\pi}\omega \left[ \coth\frac{\omega}{2T}+1\right]_{T \to 0}
\frac{1-\cos((E_{pk}-\omega)t)}{(E_{pk}-\omega)^2}=
\frac{|E_{pk}|t}{2}+\frac{E_{pk} t}{2}+
2\int\limits_{|E_{pk}|}^{\omega_c}\frac{d\omega}{2\pi}
\left(\frac{1}{\omega}-\frac{|E_{pk}|}{\omega^2}\right)
\left(1-\cos\omega t_1\right),
\label{funcCL}
\end{equation}
where we defined $E_{pk}=E_p-E_k$. We observe a close similarity between the
results (\ref{funcCL}) and (\ref{func}).
In both cases the first two terms in the right hand side are the same as in the golden
rule approximation, they cancel each other at $T=0$. In both cases the third term survives
even at $T=0$, it is due to quantum noise and originates only from the $\coth$-part
of the effective action.
Moreover, we observe that the last terms in the equations (\ref{func}) and (\ref{funcCL})
are {\it exactly} the same, one should only identify the energy difference
$\xi_{31}$ in (\ref{func}) with $E_{pk}$ in (\ref{funcCL}). Thus we conclude that at $T=0$
the only difference between the two problems considered here lies in the
matrix elements of the interaction. Everything else is the same and,
hence, we explicitly demonstrated that the Pauli principle cannot cause
any important distinction between the problems in question at $T=0$. The above 
difference in the matrix
elements results only in some quantitatively different features, like e.g. different
functional dependences of the density matrix on time, however the decay of the
off-diagonal elements of $\rho$ (and thus decoherence) is present in both
models at any temperature including $T=0$. In both cases at low temperatures this decay
cannot be correctly described within the golden rule approximation. This approximation
fails completely at $T \to 0$.
Dropping the cos-term while evaluating the result (\ref{dw}) (this approximation
is be equivalent to one we discussed in the end of Sec. 4B) is clearly insufficient
at low temperatures (see below).

Finally, a close similarity between the perturbative results obtained here and in Sec. 4B
demonstrates again that averaging over disorder is absolutely irrelevant for the
issue of cancellation (or non-cancellation) of the diagrams in the first order
of the perturbation theory. In the CL model no such average exists at all, however
at $T=0$ diagrams cancel or do not cancel depending on whether to employ or not to
employ the golden rule approximation. This property is completely general, and it
does not depend on the form of the matrix elements, Fermi or Bose statistics, averaging
over disorder and other details of the model.

\subsection{Exponent and pre-exponent}

As the structure of the first order result in the perturbation theory is clear from
the above consideration one can try to proceed further and calculate higher order
terms. Then one can try to sum up diagrams in all orders in order to recover
the nonperturbative result. This program appears to be quite involved from
a technical point of view and, to the best of our knowledge, was not yet
carried out. Fortunately in the case of the Caldeira-Leggett model
one does not need to sum up diagrams, the exact result can be obtained much easier,
just by performing several Gaussian integrations. In this way one arrives at the
equations (\ref{Jexp}-\ref{Gt}) which are, of course, equivalent
to the result of {\it exact} summation of {\it all} diagrams of the perturbation theory.
The result (\ref{Jexp}) can be expressed in the following form:
\begin{equation}
J(t)=
\frac{\eta}{2\pi(1-e^{-\gamma t})} \exp [i\tilde R (t)-\tilde I(t)],
\label{Jexp1}
\end{equation}
where $\gamma =\eta /m$, the function $\tilde R =\tilde R(t,x_i,x_f)$ does not
depend on temperature, while the function $\tilde I(t)$ is proportional to
the frequency integral of the combination $\omega \coth (\omega /2T)$ (see
eqs. (\ref{Jexp})-(\ref{Gt})) rather than the combination $\omega (\coth (\omega /2T)+1)$
( see eqs. (\ref{dwGR})-(\ref{dw})).

By looking at the above formula (or at (\ref{Jexp})) one immediately observes that
this result cannot be simply guessed from the first order perturbation
theory, e.g. just by exponentiating the first order results or by a similar
procedure. As it was demonstrated above, the combination ``coth + 1'' (\ref{combCL})
appears in the perturbation theory, while the time dependence of a real part $\tilde I(t)$
of the exponent (\ref{Jexp1}) (or (\ref{Jexp})) is governed only by ``coth'' and
not by ``coth + 1''. This implies that in the course of the exact summation
of all the diagrams terms combine in a nontrivial way, so that ``coth'' gets split
from ``1'',
this combination does not appear in higher orders in the same form as in the
first order result. From the exact result (\ref{Jexp})-(\ref{Gt}) one can
immediately draw a conclusion on {\it how} ``coth'' gets split from ``1'': all
terms of the diagrammatic expansion containing ``coth'' gather in the exponent.
Moreover, {\it only} such terms contribute to the real part of
the exponent $\tilde I$ (\ref{Jexp1}). In other words, since ``coth'' is contained only
in the imaginary part of the effective action $S_I$ (\ref{SI15}), we conclude that
the real part of the exponent $\tilde I$ (\ref{Jexp1}) (or (\ref{Jexp}) is determined
only by $S_I$.
This result could, of course, be expected in advance because it
is the imaginary part of the action $S_I$ which should be responsible for the
decay of the kernel $J$ (\ref{Jexp}) in time.

The real part of the action $S_R$ (\ref{SR15}), in contrast, does not (and cannot)
contribute to the real part of the exponent $\tilde I$. Now we will demonstrate
that $S_R$ (and only $S_R$) contributes to the pre-exponential function in (\ref{Jexp1}),
(\ref{Jexp}).
The pre-exponent is determined by the path integral (\ref{J15}) with
zero boundary conditions, $x_{1i}=x_{2i}=x_{1f}=x_{2f}=0$. From the structure of the
action (\ref{J15}-\ref{SI15}) one easily observes that its real part can be represented
in the form
\begin{equation}
iS_0[x_1]-iS_0[x_2]-iS_R[x_1,x_2]=ix^-\hat L x^+,
\end{equation}
while its imaginary part has a different structure:
\begin{equation}
S_I[x_1,x_2]=x^-\hat A x^-,
\end{equation}
where $\hat L$ and $\hat A$  are two different operators. The Gaussian integrals
can be easily performed and we get
\begin{equation}
\int {\cal D} x^-\int {\cal D} x^+
e^{ix^-\hat L x^+ - x^-\hat Ax^-}=\int {\cal D} x^- \delta(\hat L^T x^-)
e^{-x^-\hat Ax^-}=\frac{1}{\det \hat L}=\frac{\eta}{2\pi(1-e^{-\gamma t})}.
\label{preexp}
\end{equation}
This relation proves that the pre-exponent in (\ref{Jexp1}) (or (\ref{Jexp})) is
determined solely
by the real part of the action $S_R$ and does not depend on its imaginary part
$S_I$ at all.

Thus the above analysis allows for a clear distinction between the two parts of the effective
action: $S_I$ determines the real part of the exponent $\tilde I (t)$ and governs the
decay of the
off-diagonal elements of the density matrix (hence, playing a crucial role for
dephasing), while $S_R$ determines the time dependence of the pre-exponent
which is only relevant for the kinematics of classical trajectories and has nothing
to do with the issue of decoherence. In the first order perturbation theory
the terms from the exponent and the pre-exponent mix and partially cancel each other, thus
making any clear distinction between them impossible. This situation is fully
analogous to one encountered in the preceding sections for the problem
of electron dephasing in disordered conductors.

Depending on
the boundary conditions $S_R$ can also contribute to the imaginary part
of the exponent $\tilde R$ (cf. (\ref{Jexp}). However, this {\it imaginary} part does not
determine the decay of the off-diagonal elements and in no way can it cancel
(or contribute to) the {\it real} part of the exponent $\tilde I(t)$ determined
exclusively by $S_I$.
On top of that, for the boundary conditions corresponding to the ``Cooperon'' this
imaginary part of the exponent vanish $\tilde R \equiv 0$. This particular case will be
considered below for illustration.

\subsection{``Cooperon'' in the Caldeira-Leggett model}

Let us define the ``Cooperon'' configuration as
\begin{equation}
C(t,x)=J(t,x,0;0,x)
\label{cooperondef}
\end{equation}
That means the path integral (\ref{J15}) is evaluated on the trajectories with
the time reversed boundary conditions: $x_{1i}=x_{2f}=0$, $x_{1f}=x_{2i}=x$ or,
equivalently, $x^+_i=x^+_f=x/2$, $x^-_i=-x^-_f=-x$.
Substituting these boundary conditions into (\ref{Jexp}) one easily finds
\begin{equation}
C(t,x)=\frac{\eta}{2\pi(1-e^{-\gamma t})}
\exp\left[ -\eta \big\{ g_1(t)+4g_2(t)-2g_3(t)\big\}x^2\right]
\label{cooperonexact}
\end{equation}
This is the exact result. In the long time limit $\gamma t\gg 1$ we obtain
\begin{equation}
C\simeq \frac{\eta}{2\pi}\; e^{-\eta g_1(t) x^2}, \;\;\;
g_1(t)\simeq Tt+\ln\frac{1-e^{-2\pi Tt}}{2\pi(T/\omega_c)} .
\label{cooplong}
\end{equation}
This expression decays in time, and the analog of the dephasing time
(which we will also denote as $\tau_\varphi$ here) can be defined from
the equation
\begin{equation}
\eta g_1(\tau_\varphi) x^2 \approx 1.
\label{tau1}
\end{equation}
We observe that this time is determined only by the
exponent of the equation (\ref{cooperonexact}).

Let us now expand separately the combination of $g$-functions in the exponent
and the pre-exponent in the exact result (\ref{cooperonexact}) to the first order in
the interaction parameter $\eta$ each. We get
\begin{equation}
C(t,x)=\left(\frac{m}{2\pi t}+\frac{\eta}{4\pi}\right)
\exp\left[ -\eta g_{\rm pert}(t) x^2  \right]
\label{cooperonapp}
\end{equation}
where
\begin{equation}
g_{\rm pert}(t)=\frac{1}{2}\int\limits_0^t ds \int\limits ds'
\left(1+4\frac{ss'}{t^2}-2\frac{s+s'}{t} \right) G(s-s')
\simeq \left\{\begin{array}{ll}
              (\ln(\omega_c t)+\gamma -1)/\pi, & Tt\ll 1 \\
              Tt/3, &  Tt\gg 1
              \end{array}\right.
\label{gpert}
\end{equation}
The function $g_{\rm pert}$ is just the imaginary part of the action $S_I(t)$
(\ref{SI15}) evaluated on the two time reversed paths: $x_1(s)=xs/t$ and
$x_2(s')=x(t-s')/t$. These are the saddle point paths for the noninteracting
part of the action $S_0$. Comparing the functions $g_{\rm pert}$ (\ref{gpert}) and
$g_1(t)$ (\ref{cooplong}) we observe that with a sufficient accuracy
one has  $g_{\rm pert}(t)\simeq g_1(t)$ for $Tt\ll 1$ and
$g_{\rm pert}(t)\simeq g_1(t)/3$ for $Tt\gg 1$.
Hence, if the dephasing time
$\tau_{\varphi}$ is determined from the perturbative result (\ref{cooperonapp}) as
\begin{equation}
\eta g_{\rm pert}(\tau_\varphi)x^2 \approx 1,
\end{equation}
the result will differ from the exact one only by a numerical
factor of order one which anyway can be absorbed in the definition of $\tau_{\varphi}$.
As before (cf. (\ref{tau1})) the dephasing time extracted from this equation
will depend only on the exponent (i.e. on $S_I$), while the pre-exponent in the
eq.(\ref{cooperonapp}) (defined by $S_R$) must be ignored again.

Now let us expand the whole expression (\ref{cooperonexact}) in powers of
$\eta$. The first order correction for the Cooperon has the form
\begin{equation}
\delta C_1(t,x)=C_0(t,x)\left(\frac{\eta t}{2m}-\eta g_{\rm pert}(t) x^2\right),
\label{cooperonpert}
\end{equation}
where $C_0(t)=m/2\pi t$ is the Cooperon in the absence of the interaction defined
e.g. by (\ref{J0RA}). The equation (\ref{cooperonpert}) is nothing but the short
time expansion of the exact result because the interaction parameter $\eta$ always
enters into this result being multiplied by $t$. This expansion is fully equivalent
to the perturbation theory in the interaction in the weak localization theory.

As it was already discussed in Sec. 2, it is impossible to unambiguously define
the dephasing time $\tau_{\varphi}$ within the frames of the perturbation theory
only, and an additional assumption about the decay of the Cooperon in time should
necessarily be made. For instance, AAG assumed that the
Cooperon decay is purely exponential: $C(t)=C_0(t)e^{-t/\tau_\varphi}$
(cf. e.g. eqs. (2.45) and (3.2) of Ref. \onlinecite{AAG2}). Assuming
such a form here and combining it with the perturbative result (\ref{cooperonpert}),
we find
\begin{equation}
\frac{1}{\tau_{\varphi}^{\rm pert}}=\eta x^2\frac{g_{\rm pert}(t)}{t}-\frac{\eta}{2\pi}
\label{tau3}
\end{equation}
Comparing this equation with eq. (\ref{tau1}) which follows from the exact result
(\ref{cooperonexact}) we observe the presence of an additional term $-\eta/2\pi$
in eq. (\ref{tau3}). This term originates from the expansion of the
pre-exponent and has nothing to do with dephasing. However, it is not small
and can strongly influence the result for $\tau_{\varphi}$, provided
the latter is determined from the perturbative expansion (\ref{cooperonpert}).
Depending on the choice of $x$ one can obtain positive, almost zero and
even negative values of the dephasing time (cf. Sec. 2), which is an obvious nonsense.
This simple example demonstrates again that it is impossible to make any conclusion
about the long time behavior of the system (and, hence, about the
dephasing time $\tau_{\varphi}$) from the first order
expansion in the interaction, since the latter is valid in the short time limit only.

\subsection{Other approximations}

Let us check some other approximations which are sometimes employed in the
literature. A deficiency of the golden rule approximation applied within
the perturbation theory has already been illustrated above. In certain situations
one proceeds beyond the perturbation theory, correctly get ``coth'' in the exponent,
and only then apply the golden rule approximation. Also in this case the true
low temperature behavior will be missing completely. In order to observe
this property let us evaluate the function $g_1(t)$ (\ref{g1}) within
the golden rule approximation (\ref{GRO}). Extending the integral over $s-s'$ in
(\ref{g1}) to infinite limits and performing this integration first we obtain
the delta-function $\delta (\omega)$. This is just the golden rule approximation,
cf. eq. (\ref{GRO}). After that the remaining integrals trivially yield
$$
g_1(t) \simeq Tt,
$$
i.e. only the first term in (\ref{g1}) (or (\ref{cooplong})) is reproduced,
while the second term is missing. After that one could
incorrectly conclude that no quantum decoherence occurs
in the CL model at $T=0$. An obvious mistake here is
to extend the integral over $s-s'$ to infinite limits. An exact calculation
of the function $g_1(t)$ allows to recover both the zero-frequency contribution
$Tt$ as well as an additional term (see eq. (\ref{cooplong})) which originates
from frequencies $\omega > T$ and does not vanish at $T=0$.

The above approximation was applied e.g. in a recent preprint by Levinson \cite{Lev}.
In this paper a transparent formulation of the problem of quantum decoherence in
quantum dots is derived. Within this formulation Levinson arrived at the result for 
the equilibrium dephasing rate which contains only ``coth'', while ``tanh'' or ``1'' do not
appear in the exponent at all (cf. eqs. (4), (5), (7) and (14) of Ref.
\onlinecite{Lev}). It is interesting that a nonzero dephasing rate at $T=0$ is contained 
in these formulas before the golden rule approximation is made.
In order to see that it is sufficient to combine eqs. (5), (7) and (14)
of Ref. \onlinecite{Lev}, substitute them into eq. (4) of the same paper and perform the
time integration. A finite result for the decoherence rate will follow immediately at all
temperatures including $T=0$. High frequencies will contribute to this result
which will actually depend on a physical high frequency cutoff. However, if one
applies the golden rule approximation, only the zero frequency contribution will
be recovered (eq. (6) of Ref. \onlinecite{Lev}) and the whole decoherence effect
at low $T$ will be missing. At low $T$ the decay of
correlations in the problem \cite{Lev} is not exponential
in time. But also in this situation the dephasing time
$\tau_{\varphi}$ can be easily defined. This time just sets a scale on which the
quantum coherence is sufficiently suppressed and the integral over time
(eq. (4) of Ref. \onlinecite{Lev}) becomes convergent.

For nonzero (but possibly relatively small) values of $T$ one can try to argue
that at long times it is sufficient to consider the limit $Tt \gg 1$
and expand the function $g_1(t)$ (\ref{g1}) (or (\ref{cooplong})) in powers of $1/Tt$.
Then one gets
\begin{equation}
g_1(t) \approx Tt + \ln (\omega_c/T),
\label{nonanaly}
\end{equation}
while all higher order terms of this expansion will be equal to zero because
$g_1(t)$ is a non-analytic function of $1/Tt$. From (\ref{nonanaly}) one could
conclude that the first term in this equation describes dephasing due to thermal
fluctuations while the second term is the ``interaction correction'' which does
not depend on time and has nothing to do with dephasing. Since the first term
vanishes in the limit $T \to 0$, one could again arrive at the conclusion that
no decoherence
is present in the CL model at $T \to 0$. As it is clear from the exact
solution (\ref{Jexp}-\ref{Gt}), this conclusion is not correct.

The expansion in $1/Tt$
(or in $1/T\tau_H$) is just the expansion performed in Ref. \onlinecite{AAG2} within
the first order perturbation theory in the interaction (see also Sec. 4D and
Appendix C). As it was already discussed
in Sec. 4B, this expansion is equivalent to dropping the oscillating cos-term
e.g. in the expression (\ref{nonren0}) (or, equivalently, in the expression (\ref{dw})
for the CL model).
Obviously, this approximation has nothing to do with averaging over disorder.
The above example also invalidates
any attempt to approach the correct low temperature behavior by means of
a high temperature expansion, like it was suggested e.g. by Vavilov and
Ambegaokar in Ref. \onlinecite{VA}.
For instance, the terms $\sim \exp (-2\pi Tt)$ cannot be recovered in any order in $1/Tt$.
We will come back to the analysis of the paper \cite{VA} in Appendix E.

Finally, let us briefly discuss the role of the low frequency modes of the
effective environment. One could conjecture that, since high frequencies up
to $\omega_c$ contribute to the dephasing time, the result will not change
even if one introduces the low frequency cutoff $\omega_{c0}$.
In order to test this conjecture it is again sufficient to consider the
behavior of the function $g_1(t)$.
Introducing the low frequency cutoff $\omega_{c0}$ in the integral (\ref{g1}),
at $T=0$ one readily finds
\begin{equation}
g_1(t) \simeq \ln (\omega_c/\omega_{c0}), \;\;\;\; \omega_{c0}t \gg 1,
\label{g1const}
\end{equation}
while in the opposite limit $\omega_{c0}t \lesssim 1$ the result is the
same as without the low frequency cutoff, i.e. $g_1(t) \simeq \ln \omega_ct$.
Thus the cutoff at $\omega \sim \omega_{c0}$ leads to a different long
time behavior of $g_1(t)$. It increases at short times
but then saturates at a value $\sim \ln (\omega_c/\omega_{c0})$. In this case
no time decay of the off-diagonal elements of the density matrix occurs at long
times and therefore the coherence is not fully suppressed. However,
it is not the long time limit which is interesting in the dephasing problem,
but rather the system behavior at $t \sim \tau_{\varphi}$. If $\omega_{c0}\tau_{\varphi}
\ll 1$ ($\tau_{\varphi}$ was defined e.g. in (\ref{tau1})), by the time the behavior
(\ref{g1const}) is reached the coherence will already be very strongly suppressed.
Thus from a practical point of view there is no substantial difference between
the cases $\omega_{c0}=0$ and $\omega_{c0}\tau_{\varphi} \ll 1$. In the
opposite limit $\omega_{c0}\tau_{\varphi} \gg 1$ the function $g_1(t)$ saturates
earlier than quantum coherence gets suppressed. In this case cutting out the
low frequency oscillators changes the result significantly.

This simple consideration clarifies the role of the low frequency modes
in the dephasing problem.
At low $T$ these modes do not really affect the expression for $\tau_{\varphi}$
which depends on the high frequency cutoff $\omega_c$. However, if the low frequency
cutoff is chosen such that $\omega_{c0}\tau_{\varphi} \gg 1$, the dephasing
time $\tau_{\varphi}$ simply looses its meaning because $g_1(t)$ saturates
already at much shorter times $t \sim 1/\omega_{c0} \ll \tau_{\varphi}$.
The same conclusion
applies to disordered metals in which case the function $f_d(t)$ (\ref{fott})
should be considered instead of $g_1(t)$.

For an extended discussion of various approximations analyzed for the exactly
solvable CL model we refer the reader to the paper \cite{Al}.

\section{Concluding Remarks}

Let us briefly summarize our main observations.

\begin{enumerate}

\item We explicitly demonstrated that the perturbation theory in the interaction
is principally insufficient in the problem of electron dephasing in disordered conductors.
The decoherence time $\tau_{\varphi}$ cannot be extracted even from a correct
perturbative calculation unless one {\it assumes} some particular form of the
decay of correlations in time. However, this form should be calculated rather than
assumed. This task can be accomplished only by means of a nonperturbative calculation.

\item A nonperturbative analysis shows that the dephasing time $\tau_{\varphi}$
is determined only by the function $f_d(t)$ in eq. (\ref{dsigWL}), while the effect
of interaction on the pre-exponential function $A_d(t)$ is not important for
the calculation of $\tau_{\varphi}$. Therefore, in order to evaluate the
dephasing time it is sufficient to perform a nonperturbative analysis with 
the exponential accuracy.

\item The effect of interaction on the pre-exponent $A_d(t)$ {\it is} important
if one calculates the interaction-dependent part of the weak localization 
correction in the limit of strong magnetic fields $\tau_H \ll \tau_{\varphi}$.
The zero temperature dephasing time $\tau_{\varphi}^0$ drops out of this
correction in the first order due to the exact cancellation of the linear in 
time $T$-independent contributions from the exponent $\exp (-f_d(t))$ and the
pre-exponent $A_d(t)$.

\item Nonlinear in time $T$-independent contributions do not cancel out already in
the first order of the perturbation theory.
In general these terms not only account for the renormalization effects but also
contribute to dephasing at all temperatures including $T=0$.
   
\item We demonstrated that there exists a close formal similarity between
the problem of electron dephasing in disordered conductors and the exactly
solvable Caldeira-Leggett model for a particle interacting with a
bath of harmonic oscillators.

\end{enumerate}

Our analysis allows to establish a simple correspondence between the results of 
Refs. \onlinecite{AAK,GZ2} and \onlinecite{AAG2}. 
The effect of the interaction in the expression for the magnetoconductance
(\ref{dsigWL}) is described by the function 
$$
A_d[\tanh ]\exp (-f_d [\coth ]).
$$ 
At high temperatures the function $f_d$ decays 
on a short time scale. For such times the effect 
of the interaction (related to ``tanh'') on the pre-exponent is negligible. In this
case the nonperturbative analysis of AAK \cite{AAK} applies. At low temperatures
the interaction effect on both the exponent and the pre-exponent becomes of
order one on the same time scale $\sim \tau_{\varphi}^0$. Since the effective
cutoff in the integral (\ref{dsigWL}) is determined by the function $f_d(t)$ in
the exponent, ``tanh'' can be neglected again. We arrive at our nonperturbative 
results \cite{GZ2}. Finally, in the limit $\tau_H \ll \tau_{\varphi}$ for
the relevant times $t \lesssim \tau_H$ one has $f_d(t) \ll 1$ and, hence,
$\exp (-f_d [\coth ]) \simeq 1-f_d[\coth ]$. Performing also a short 
time expansion of $A_d [\tanh ]$ we obtain the combination ``coth - tanh''
in the first order and reproduce the AAG's perturbative results \cite{AAG2}.

These observations conclude our analysis.

We acknowledge stimulating discussions with B.L. Altshuler, C. Bruder, M. B\"uttiker, 
K. Eriksen, P. Hedegard, D. Khveshenko, 
V.E. Kravtsov, D. Loss, Yu. Makhlin, A. Mirlin, D. Polyakov, G. Sch\"on, 
S. Sharov, E. Sukhorukov and J. von Delft.
One of us (D.S.G.) also acknowledges the support from the INTAS-RFBR Grant N 95-1305
and the hospitality of the Karlsruhe University and the Forschungszentrum Karlsruhe 
where a part of this work was performed.

\appendix

\section{}

Here we will compare the expressions for the
weak localization correction to the conductivity in the presence
of interaction obtained by means of our path integral technique \cite{GZ2} and
within  the diagrammatic approach of AAG.
We will demonstrate that both results are identical if analyzed
on a perturbative level before disorder averaging.

We proceed in two steps. We first transform the result \cite{AAG2} for the conductance
and demonstrate that by virtue of the causality principle one
can completely remove the terms of the type $G^RG^AG^RG^A$ emphasized by AAG. 
We will arrive at the Eqs. (\ref{dsig}, \ref{dsigma1}) which are exactly equivalent to the
result \cite{AAG2}. Our second step is to expand our expression for the
conductance (\ref{sigma00},\ref{J17}) in the interaction terms $iS_R+S_I$.
This will lead us to the Eq. (\ref{dsigma00}) which is identical to (\ref{dsigma1}).

We start from reproducing the
expression for the correction to the conductivity due to electron-electron
interaction obtained in Ref. \onlinecite{AAG2}
\begin{eqnarray}
\delta \sigma _{\alpha \beta } &=&-\frac i{16}\int \frac{d\bbox{r}_1d\bbox{r}
_2d\bbox{r}_3d\bbox{r}_4}{{\cal V}}\int \frac{d\omega }{2\pi }\frac{
d\epsilon }{2\pi }\left( \frac d{d\epsilon }\tanh \frac \epsilon {2T}\right)
\coth \frac \omega {2T}\left[ {\cal L}_{34}^R(\omega )-{\cal L}
_{34}^A(\omega )\right] \times   \nonumber \\
&&\left\{ 2\hat j_\alpha \left[ G_{12}^R(\epsilon )-G_{12}^A(\epsilon
)\right] \hat j_\beta \left[ G_{23}^A(\epsilon )G_{34}^A(\epsilon -\omega
)G_{41}^A(\epsilon )-G_{23}^R(\epsilon )G_{34}^R(\epsilon -\omega
)G_{41}^R(\epsilon )\right] +...+\alpha \leftrightarrow \beta \right\}
\nonumber \\
&&-\frac i{16}\int \frac{d\bbox{r}_1d\bbox{r}_2d\bbox{r}_3d\bbox{r}_4}{{\cal
V}}\int \frac{d\omega }{2\pi }\frac{d\epsilon }{2\pi }\left( \frac d{
d\epsilon }\tanh \frac \epsilon {2T}\right) \tanh \frac{\epsilon -\omega }{2T
}\times   \nonumber \\
&&\big\{ 2\hat j_\alpha \left[ G_{12}^R(\epsilon )-G_{12}^A(\epsilon
)\right] \hat j_\beta \left[ G_{23}^A(\epsilon )G_{41}^A(\epsilon ){\cal L}
_{34}^A(\omega )-G_{23}^R(\epsilon )G_{41}^R(\epsilon ){\cal L}
_{34}^R(\omega )\right] \left[ G_{34}^R(\epsilon -\omega )-G_{34}^A(\epsilon
-\omega )\right]
\nonumber\\
&& +...+\alpha \leftrightarrow \beta \big\} .
\label{dsigma}
\end{eqnarray}
For simplicity we keep the same notations as in Ref. \onlinecite{AAG2}: $G^{R(A)}$
are the retarded (advanced) Green functions for noninteracting electrons
and ${\cal L}^{R(A)}$ are photon propagators. The coordinate
dependence of the propagators is indicated by the subscripts, e.g.
$G^R_{12}(\epsilon)=G^R(\epsilon,\bbox{r}_1,\bbox{r}_2)$.
Note that in (\ref{dsigma}) only the contribution of the two
self-energy diagrams (Fig. 2a,b) was reproduced, while the remaining
contribution from the vertex diagrams of Fig. 2c,d which contains
the terms with two $\omega $-dependent Green
functions is denoted by ... . We will not consider
them in this Appendix for the sake of simplicity.

We observe that the factor $\tanh \frac{\epsilon -\omega }{2T}$
enters in this expression together with the difference $\left[
G_{34}^R(\epsilon -\omega )-G_{34}^A(\epsilon -\omega )\right] $. This
combination is just the Keldysh function
\begin{equation}
G^K(\epsilon ,\bbox{r}_1,\bbox{r}_2)=\tanh \frac \epsilon {2T}\left[
G^R(\epsilon ,\bbox{r}_1,\bbox{r}_2)-G^A(\epsilon ,\bbox{r}_1,\bbox{r}
_2)\right] =\tanh \frac \epsilon {2T}\left[ \frac 1{\epsilon +\mu -\hat H+i0}
-\frac 1{\epsilon +\mu -\hat H-i0}\right] .  \label{GK1}
\end{equation}
This function can also be rewritten as follows
\begin{eqnarray}
G^K(\epsilon ,\bbox{r}_1,\bbox{r}_2) &=&\tanh \frac \epsilon {2T}
\sum\limits_\lambda \left[ \frac 1{\epsilon -\xi _\lambda +i0}-\frac 1{
\epsilon -\xi _\lambda -i0}\right] \Psi _\lambda (\bbox{r}_1)\Psi _\lambda
^{*}(\bbox{r}_2)  \nonumber \\
&=&\sum\limits_\lambda (-2\pi i)\left( \tanh \frac{\xi _\lambda }{2T}\right)
\delta (\epsilon -\xi _\lambda )\Psi _\lambda (\bbox{r}_1)\Psi _\lambda ^{*}(
\bbox{r}_2)  \nonumber   \\
&=&\sum\limits_\lambda \left( \tanh \frac{\xi _\lambda }{2T}\right) \left[
\frac 1{\epsilon -\xi _\lambda +i0}-\frac 1{\epsilon -\xi _\lambda -i0}
\right] \Psi _\lambda (\bbox{r}_1)\Psi _\lambda ^{*}(\bbox{r}_2)  \nonumber
\\
&=&\int d\bbox{r}^{\prime }\left[ G^R(\epsilon ,\bbox{r}_1,\bbox{r}^{\prime
})-G^A(\epsilon ,\bbox{r}_1,\bbox{r}^{\prime })\right] (\delta (\bbox{r}
^{\prime }-\bbox{r}_2)-2\rho (\bbox{r}^{\prime },\bbox{r}_2)),
\label{trans1}
\end{eqnarray}
where $\xi _\lambda $, $\Psi _\lambda $ are respectively the eigenvalues
and the eigenfunctions of the Hamiltonian $\hat H-\mu $;
$\rho (\bbox{r}^{\prime },
\bbox{r}_2)$ is the equilibrium single electron density matrix, $\hat \rho
=1/(\exp ((\hat H-\mu )/T)+1)$. In a similar manner one obtains
\begin{equation}
\left( \frac d{d\epsilon }\tanh \frac \epsilon {2T}\right) \left[
G^R(\epsilon ,\bbox{r}_1,\bbox{r}_2)-G^A(\epsilon ,\bbox{r}_1,\bbox{r}
_2)\right] =2\int d\bbox{r}^{\prime }\frac{\partial \rho (\bbox{r}_1,\bbox{r}
^{\prime })}{\partial \mu }\left[ G^R(\epsilon ,\bbox{r}^{\prime },\bbox{r}
_2)-G^A(\epsilon ,\bbox{r}^{\prime },\bbox{r}_2)\right] .  \label{trans2}
\end{equation}
We also introduce the evolution operator $\hat U
(t)=\exp (-i(\hat H-\mu)t)$ which is defined both for positive and
negative times. The functions $G^R$ and $G^A$ are related to this
operator by means of the following equations:
\begin{equation}
G^R(t,\bbox{r}_1,\bbox{r}_2)=-i\theta (t)U(t,\bbox{r}_1,\bbox{r}_2);\quad
G^A(t,\bbox{r}_1,\bbox{r}_2)=i\theta (-t)U(t,\bbox{r}_1,\bbox{r}_2).
\label{GU}
\end{equation}
Now let us write down the two equivalent forms of the Keldysh Green
function in the real time representation. We find from (\ref{GK1}):
\begin{eqnarray}
G^K(t,\bbox{r}_1,\bbox{r}_2) &=&\int\limits_{-\infty }^{+\infty }dt^{\prime }
\frac{-iT}{\sinh (\pi T(t-t^{\prime }))}\left[ G^R(t^{\prime },\bbox{r}_1,
\bbox{r}_2)-G^A(t^{\prime },\bbox{r}_1,\bbox{r}_2)\right]   \nonumber \\
&=&-\int\limits_{-\infty }^{+\infty }dt^{\prime }\frac T{\sinh (\pi
T(t-t^{\prime }))}U(t^{\prime },\bbox{r}_1,\bbox{r}_2),  \label{GKA}
\end{eqnarray}
and from (\ref{trans1}) we get
\begin{eqnarray}
G^K(t,\bbox{r}_1,\bbox{r}_2) &=&\int d\bbox{r}^{\prime }\left[ G^R(t,\bbox{r}
_1,\bbox{r}^{\prime })-G^A(t,\bbox{r}_1,\bbox{r}^{\prime })\right] (\delta (
\bbox{r}^{\prime }-\bbox{r}_2)-2\rho (\bbox{r}^{\prime },\bbox{r}_2))
\nonumber \\
&=&-i\int d\bbox{r}^{\prime }U(t,\bbox{r}_1,\bbox{r}^{\prime})
(\delta (\bbox{r}
^{\prime }-\bbox{r}_2)-2\rho (\bbox{r}^{\prime },\bbox{r}_2)).  \label{GKour}
\end{eqnarray}
Analogously we obtain
\begin{eqnarray}
\left( \frac d{d\epsilon }\tanh \frac \epsilon {2T}\right) \left[
G^R(\epsilon ,\bbox{r}_1,\bbox{r}_2)-G^A(\epsilon ,\bbox{r}_1,\bbox{r}
_2)\right]  &\Rightarrow &\int\limits_{-\infty }^{+\infty }dt^{\prime }\frac{
T(t-t^{\prime })}{\sinh (\pi T(t-t^{\prime }))}\left[ G^R(t^{\prime },
\bbox{r}_1,\bbox{r}_2)-G^A(t^{\prime },\bbox{r}_1,\bbox{r}_2)\right]
\nonumber \\
&=&\int\limits_{-\infty }^{+\infty }dt^{\prime }\frac{-iT(t-t^{\prime })}{
\sinh (\pi T(t-t^{\prime }))}U(t^{\prime },\bbox{r}_1,\bbox{r}_2);
\label{dtanhA}
\end{eqnarray}
and
\begin{eqnarray}
2\int d\bbox{r}^{\prime }\frac{\partial \rho (\bbox{r}_1,\bbox{r}^{\prime })
}{\partial \mu }\left[ G^R(\epsilon ,\bbox{r}^{\prime },\bbox{r}
_2)-G^A(\epsilon ,\bbox{r}^{\prime },\bbox{r}_2)\right]  &\Rightarrow &2\int
d\bbox{r}^{\prime }\frac{\partial \rho (\bbox{r}_1,\bbox{r}^{\prime })}{
\partial \mu }\left[ G^R(t,\bbox{r}^{\prime },\bbox{r}_2)-G^A(t,\bbox{r}
^{\prime },\bbox{r}_2)\right]   \nonumber  \\
&=&-2i\int d\bbox{r}^{\prime }\frac{\partial \rho (\bbox{r}_1,\bbox{r}
^{\prime })}{\partial \mu }U(t,\bbox{r}^{\prime },\bbox{r}_2).
\label{dtanhour}
\end{eqnarray}
It is easy to observe that the eqs. (\ref{GKA},\ref{dtanhA}) contain
the integral over time which does not enter the eqs.
(\ref{GKour},\ref{dtanhour}). It is this additional time integration
that leads to violation of the normal time ordering at the level of the
perturbation theory and is responsible for the appearance of the diagrams
$G^RG^AG^RG^A$. The interpretation of such diagrams in terms of the path
integral is not possible. However, if one uses the other form of the same
functions (\ref{GKour},\ref{dtanhour}) the normal time ordering is
automatically restored, the combinations $G^RG^AG^RG^A$ disappear
due to the causality principle
and the path integral interpretation of the remaining terms of the
perturbation theory can be made.

We emphasize that all the above transformations are exact and have the
advantage that in the final expressions only the propagators depend
on the frequencies $\epsilon $ and $\omega $ (except for the factor $\coth \frac
\omega {2T}$ in $\delta \sigma _{\alpha \beta }$). This allows one to use
the analytical properties of the propagators related to the causality principle.
Namely, $G^R(\epsilon )$ and ${\cal L}^R(\omega )$ have no singularities in
the upper half-plane, while $G^A(\epsilon )$ and ${\cal L}^A(\omega )$ are
analytic functions in the lower half-plane. Making use of these properties
one can easily prove the identities
\begin{eqnarray}
\int d\omega {\cal L}^R(\omega )G^A(\epsilon -\omega ) &\equiv &0,\int
d\epsilon G_{12}^A(\epsilon )G_{23}^A(\epsilon )G_{34}^A(\epsilon -\omega
)G_{41}^A(\epsilon )\equiv 0,  \nonumber \\
\int d\omega {\cal L}^A(\omega )G^R(\epsilon -\omega ) &\equiv &0,\int
d\epsilon G_{12}^R(\epsilon )G_{23}^R(\epsilon )G_{34}^R(\epsilon -\omega
)G_{41}^R(\epsilon )\equiv 0.  \label{causality}
\end{eqnarray}
Consider e.g. the integral $\int d\omega {\cal L}^R(\omega
)G^A(\epsilon -\omega )$. Since both functions ${\cal L}
^R(\omega )$ and $G^A(\epsilon -\omega )$ are regular in the upper
half-plane, the integral vanishes. Alternatively, we can write $\int d\omega
{\cal L}^R(\omega )G^A(\epsilon -\omega )=\int dt\exp (i\epsilon t){\cal L}
^R(t)G^A(t)$ and note that ${\cal L}^R(t)\equiv 0$ for $t<0$ due to
the causality principle, while $G^A(t)\equiv 0$ for $t>0$ and
the integral is identically equal to zero. Analogously
one can prove all the other identities (\ref{causality}).

The corrections to the conductivity can now be considerably simplified:
\begin{eqnarray}
\delta \sigma _{\alpha \beta } &=&-\frac i4\int \frac{d\bbox{r}_1d\bbox{r}_2d
\bbox{r}_3d\bbox{r}_4d\bbox{r}_5}{{\cal V}}\int \frac{d\omega }{2\pi }\frac{
d\epsilon }{2\pi }\coth \frac \omega {2T}\left[ {\cal L}_{34}^R(\omega )-
{\cal L}_{34}^A(\omega )\right] \times   \nonumber \\
&&\left\{ \hat j_\alpha \left[ G_{15}^R(\epsilon )\frac{\partial \rho _{52}}{
\partial \mu }\right] \hat j_\beta G_{23}^A(\epsilon )G_{34}^A(\epsilon
-\omega )G_{41}^A(\epsilon )+\hat j_\alpha \left[ G_{15}^A(\epsilon )\frac{
\partial \rho _{52}}{\partial \mu }\right] \hat j_\beta G_{23}^R(\epsilon
)G_{34}^R(\epsilon -\omega )G_{41}^R(\epsilon )+..+\alpha \leftrightarrow
\beta \right\}   \nonumber \\
&&-\frac i4\int \frac{d\bbox{r}_1d\bbox{r}_2d\bbox{r}_3d\bbox{r}_4d\bbox{r}
_5d\bbox{r}_6}{{\cal V}}\int \frac{d\omega }{2\pi }\frac{d\epsilon }{2\pi }
\left\{ -\hat j_\alpha G_{15}^R(\epsilon )\frac{\partial \rho _{52}}{
\partial \mu }\hat j_\beta G_{23}^A(\epsilon )G_{36}^A(\epsilon -\omega
)\left( 1-2\rho \right) _{64}G_{41}^A(\epsilon ){\cal L}_{34}^A(\omega
)\right.   \nonumber \\
&&+\left. \hat j_\alpha G_{15}^A(\epsilon )\frac{\partial \rho _{52}}{
\partial \mu }\hat j_\beta G_{23}^R(\epsilon )G_{36}^R(\epsilon -\omega
)\left( 1-2\rho \right) _{64}G_{41}^R(\epsilon ){\cal L}_{34}^R(\omega
)+...+\alpha \leftrightarrow \beta \right\} .  \label{dsig}
\end{eqnarray}
We observe that the terms of the type $\hat j
_\alpha G_{12}^R(\epsilon )\hat j_\beta G_{23}^A(\epsilon )G_{34}^R(\epsilon
-\omega )G_{41}^A(\epsilon )$ do not enter the expression (\ref{dsig}) at all.
For later purposes it will be useful to rewrite the above expression in the form
of the time integral:
\begin{eqnarray}
\delta \sigma _{\alpha \beta } &=&-\frac{e^2}2\int \frac{d\bbox{r}
_1d\bbox{r}_2d\bbox{r}_3d\bbox{r}_4d\bbox{r}_5}{{\cal V}}\int\limits_0^{+
\infty }dt_1\int\limits_0^{t_1}dt_2\int\limits_0^{t_2}dt_3\times   \nonumber
\\
&&\left\{ \hat j_\alpha U_{15}(t_1)\frac{\partial \rho _{52}}{\partial \mu }
\hat j_\beta U_{23}^{+}(t_3)\left[ I_{34}(t_2-t_3)U_{34}^{+}(t_2-t_3)\right]
U_{41}^{+}(t_1-t_2)\right.   \nonumber \\
&&+\left. \hat j_\alpha U_{15}^{+}(t_1)\frac{\partial \rho _{52}}{\partial
\mu }\hat j_\beta U_{23}(t_3)\left[ I_{34}(t_2-t_3)U_{34}(t_2-t_3)\right]
U_{41}(t_1-t_2)+...+\alpha \leftrightarrow \beta \right\}   \nonumber \\
&&-\frac{ie^2}4\int \frac{d\bbox{r}_1d\bbox{r}_2d\bbox{r}_3d\bbox{r}_4d
\bbox{r}_5d\bbox{r}_6}{{\cal V}}\int\limits_0^{+\infty
}dt_1\int\limits_0^{t_1}dt_2\int\limits_0^{t_2}dt_3\times   \nonumber \\
&&\left\{ -\hat j_\alpha U_{15}(t_1)\frac{\partial \rho _{52}}{\partial \mu }
\hat j_\beta U_{23}^{+}(t_3)\left[ R_{34}(t_2-t_3)U_{36}^{+}(t_2-t_3)\left(
1-2\rho \right) _{64}\right] U_{41}^{+}(t_1-t_2)\right.   \nonumber \\
&&+\left. \hat j_\alpha U_{15}^{+}(t_1)\frac{\partial \rho _{52}}{\partial
\mu }\hat j_\beta U_{23}(t_3)\left[ R_{34}(t_2-t_3)U_{36}(t_2-t_3)\left(
1-2\rho \right) _{64}\right] U_{41}(t_1-t_2)+...+\alpha \leftrightarrow
\beta \right\} ;  \label{dsigma1}
\end{eqnarray}
where
\begin{equation}
R(t,\bbox{r}) = \int \frac{d\omega d^3k}{(2\pi )^4}\frac{4\pi }{k^2\epsilon
(\omega ,k)}e^{-i\omega t+i\bbox{kr}}=-\frac 1{e^2}{\cal L}^R(t,\bbox{r})=-\frac 1{
e^2}{\cal L}^A(-t,\bbox{r}),
\label{R55}
\end{equation}
\begin{eqnarray}
I(t,\bbox{r}) = \int \frac{d\omega d^3k}{(2\pi )^4}\text{Im}\left( \frac{
-4\pi }{k^2\epsilon (\omega ,k)}\right) \coth \left( \frac \omega {2T}
\right) e^{-i\omega t+i\bbox{kr}} \nonumber \\
=\frac 1{2e^2i}\int \frac{d\omega d^3k}{
(2\pi )^4}\coth \left( \frac \omega {2T}\right) \left[ {\cal L}^R(\omega ,k)-
{\cal L}^A(\omega ,k)\right] e^{-i\omega t+i\bbox{kr}}.
\label{I55}
\end{eqnarray}

Now we will demonstrate that the equation (\ref{dsigma1}) can be obtained
within the path integral formalism.
The formal expression for the conductivity has the form \cite{GZ2}
\begin{equation}
\sigma=
\frac{e^2}{3m}
\int\limits_{-\infty}^t dt'\int d\bbox{r}_{i1}d\bbox{r}_{i2}
\left.\left(\nabla_{r_{1f}}-\nabla{r_{2f}}\right)\right|_{\bbox{r}_{1f}
=\bbox{r_{2f}}}J(t,t';\bbox{r}_{1f},\bbox{r}_{2f};\bbox{r}_{1i},\bbox{r}_{2i})
(\bbox{r}_{1i}-\bbox{r}_{2i})\rho_0(\bbox{r}_{1i},\bbox{r}_{2i}).
\label{sigma00}
\end{equation}
The kernel $J$ is given by the path integral over electron coordinates
and momentums $\bbox{r}_1(t),\bbox{p}_1(t)$ and $\bbox{r}_2(t),\bbox{p}_2(t)$
corresponding respectively to the forward and backward parts of the Keldysh
contour. The explicit expression for this kernel reads \cite{GZ2}:
\begin{eqnarray}
J(t,t';\bbox{r}_{1f},\bbox{r}_{2f};\bbox{r}_{1i},\bbox{r}_{2i})&=&
\int\limits_{\bbox{r}_1(t')=\bbox{r}_{1i}}^{\bbox{r}_1(t)=\bbox{r}_{1f}}
{\cal D}\bbox{r}_1\int\limits_{\bbox{r}_2(t')=\bbox{r}_{2i}}^{\bbox{r}_2(t)
=\bbox{r}_{2f}}{\cal D}\bbox{r}_2\int{\cal D}\bbox{p}_1\int{\cal D}
\bbox{p}_2\times
\nonumber \\
&&
\times \exp\big\{iS_0[\bbox{r}_1,\bbox{p}_1]-iS_0[\bbox{r}_2,\bbox{p}_2]-
iS_R[\bbox{r}_1,\bbox{p}_1,\bbox{r}_2,\bbox{p}_2]-
S_I[\bbox{r}_1,\bbox{r}_2]\big\};
\label{J17}
\end{eqnarray}
where
\begin{equation}
S_0[\bbox{r},\bbox{p}]=\int\limits_{t'}^t dt''
\bigg(\bbox{p\dot r} - \frac{\bbox{p}^2}{2m} - U(\bbox{r})\bigg);
\label{S_0}
\end{equation}
\begin{eqnarray}
S_R[\bbox{r}_1,\bbox{p}_1,\bbox{r}_2,\bbox{p}_2]&=&
\frac{e^2}{2}\int\limits_{t'}^t dt_1 \int\limits_{t'}^t dt_2
\big\{R(t_1-t_2,\bbox{r}_1(t_1)-\bbox{r}_1(t_2))
\big[1-2n\big(\bbox{p}_1(t_2),\bbox{r}_1(t_2)\big)\big]
\nonumber \\
&&
-R(t_1-t_2,\bbox{r}_2(t_1)-\bbox{r}_2(t_2))
\big[1-2n\big(\bbox{p}_2(t_2),\bbox{r}_2(t_2)\big)\big]
\nonumber \\
&&
+R(t_1-t_2,\bbox{r}_1(t_1)-\bbox{r}_2(t_2))
\big[1-2n\big(\bbox{p}_2(t_2),\bbox{r}_2(t_2)\big)\big]
\nonumber \\
&&
-R(t_1-t_2,\bbox{r}_2(t_1)-\bbox{r}_1(t_2))
\big[1-2n\big(\bbox{p}_1(t_2),\bbox{r}_1(t_2)\big)\big]
\big\};
\label{SR}
\end{eqnarray}
and
\begin{eqnarray}
S_I[\bbox{r}_1,\bbox{r}_2]&=&
\frac{e^2}{2}\int\limits_{t'}^t dt_1 \int\limits_{t'}^t dt_2
\bigg\{I(t_1-t_2,\bbox{r}_1(t_1)-\bbox{r}_1(t_2))+
I(t_1-t_2,\bbox{r}_2(t_1)-\bbox{r}_2(t_2))
\nonumber \\
&&
-I(t_1-t_2,\bbox{r}_1(t_1)-\bbox{r}_2(t_2))-
I(t_1-t_2,\bbox{r}_2(t_1)-\bbox{r}_1(t_2))\bigg\}.
\label{SI}
\end{eqnarray}
The functions $R(t,\bbox{r})$ and $I(t,\bbox{r})$ are defined in (\ref{R55},\ref{I55}).

In order to obtain the perturbative result (\ref{dsigma1}) from the
formally exact
expression (\ref{sigma00}) one needs to expand the kernel $J$ (\ref{J17})
in $iS_R+S_I$. In the first
order one obtains eight different terms. Again we will consider only the terms
originating from the self-energy diagrams of Fig. 2a,b, i.e. the terms containing
$R(t_1-t_2,\bbox{r}_1(t_1)-\bbox{r}_1(t_2)),$
$R(t_1-t_2,\bbox{r}_2(t_1)-\bbox{r}_2(t_2)),$
$I(t_1-t_2,\bbox{r}_1(t_1)-\bbox{r}_1(t_2))$ and
$I(t_1-t_2,\bbox{r}_2(t_1)-\bbox{r}_2(t_2).)$
Four other terms which relate two different branches of the Keldysh contour
and contain both $\bbox{r}_1$ and $\bbox{r}_2,$ come from the vertex diagrams
of Fig. 2c,d. As it was already discussed before these terms
determine only a part of the function $\delta f_d(t)$ (\ref{f}) and -- although
they do not vanish even at $T=0$ -- always yield only
a subleading contribution to $f_d(t)$. Therefore we will not consider these
terms here for the sake of simplicity.

The correction to the kernel $J$ due to the term
$I(t_1-t_2,\bbox{r}_1(t_1)-\bbox{r}_1(t_2))$ has the form
\begin{eqnarray}
\delta
J_I^{11}(t,t';\bbox{r}_{1f},\bbox{r}_{2f};\bbox{r}_{1f},\bbox{r}_{2f})
&=&-e^2\int\limits_{t'}^{t} dt_3 \int\limits_{t'}^{t_3} dt_2
\int\limits_{\bbox{r}_1(t')=\bbox{r}_{1i}}^{\bbox{r}_1(t)=\bbox{r}_{1f}}
{\cal D}\bbox{r}_1\int\limits_{\bbox{r}_2(t')=\bbox{r}_{2i}}^{\bbox{r}_2(t)
=\bbox{r}_{2f}}{\cal D}\bbox{r}_2\int{\cal D}\bbox{p}_1\int{\cal D}
\bbox{p}_2\times
\nonumber \\
&&\times I(t_3-t_2,\bbox{r}_1(t_3)-\bbox{r}_1(t_2))
\exp\big\{ iS_0[\bbox{r}_1,\bbox{p}_1]-iS_0[\bbox{r}_2,\bbox{p}_2]\big\}
\nonumber \\
&=&-e^2\int d\bbox{r}_3 \int d\bbox{r}_4
\int\limits_{t'}^{t} dt_3 \int\limits_{t'}^{t_3} dt_2\times
\nonumber\\
&&U^{+}_{\bbox{r}_{2f},\bbox{r}_{2i}}(t-t')
U_{\bbox{r}_{1f},3}(t-t_3)I_{34}(t_3-t_2)U_{34}(t_3-t_2)
U_{4,\bbox{r}_{1i}}(t_2-t').
\label{dJI11}
\end{eqnarray}
Here we made use of a simple property of a path integral:
\begin{equation}
\int\limits_{\bbox{r}(t')=\bbox{r}_{i}}^{\bbox{r}(t)=\bbox{r}_{f}}
{\cal D}\bbox{r}\int{\cal D}\bbox{p} f(t'',\bbox{r}(t''))
\exp\big\{iS_0[\bbox{r},\bbox{p}]\big\}=
\int d\bbox{r}'' U(t-t'';\bbox{r}_f,\bbox{r}'')f(t'',\bbox{r}'')
U(t''-t;\bbox{r}'',\bbox{r}_i),
\label{svojstvo}
\end{equation}
which holds for an arbitrary function $f(t'',\bbox{r}(t''))$. Actually in
deriving (\ref{dJI11}) the property (\ref{svojstvo}) was used twice because the
function of two arguments $I(t_2-t_3,\bbox{r}_1(t_2)-\bbox{r}_1(t_3))$ enters
under the integral (\ref{dJI11}).
Already at this stage one can observe the similarity between the expression
(\ref{dJI11}) and the second term in the expression (\ref{dsigma1}). To
establish the equivalence between these two expressions the
following steps are in order:
i) after substituting the result (\ref{dJI11}) into the expression for
the conductivity (\ref{sigma00}) and applying the current
operator $\bbox{j}=(ie/m)(\nabla_{{r}_{1f}}-\nabla_{r_{2f}})$ one puts
$\bbox{r}_{1f}=\bbox{r}_{2f}=\bbox{r}_2$, $\bbox{r}_{1i}=\bbox{r}_1$,
$\bbox{r}_{2i}=\bbox{r}_5$; ii) one denotes $t-t'\to t_1$,
$t-t_2\to t_2$, $t-t_3\to t_3$; iii) one introduces an additional integration
$\int d\bbox{r}_2/{\cal V}$ which is just averaging of
the expression (\ref{sigma00}) over the sample volume and iv) one
transforms the effective initial density matrix as follows
\begin{equation}
(\bbox{r}_{1i}-\bbox{r}_{2i})\rho_0(\bbox{r}_{1i},\bbox{r}_{2i})=
i\sum\limits_{\lambda_1\lambda_2}
\frac{\langle\Psi_{\lambda_1}|\bbox{p}|\Psi_{\lambda_2}\rangle}{m}
\frac{n(\xi_{\lambda_1})-n(\xi_{\lambda_2})}{\xi_{\lambda_1}-\xi_{\lambda_2}}
\Psi_{\lambda_1}(\bbox{r}_{1i})\Psi_{\lambda_2}^*(\bbox{r}_{2i})
\simeq -i\frac{\bbox{\hat p}}{m}
\frac{\partial\rho(\bbox{r}_{1i},\bbox{r}_{2i})}{\partial\mu}.
\label{transform}
\end{equation}

After these transformations one can immediately observe the equivalence
of the results obtained by means of two methods \cite{GZ2,AAG2}
on the level of the perturbation theory. The terms arising from the
real part of the action $S_R$
can be transformed analogously, the only difference in this case is the
presence of an additional factor $(1-2\rho)_{34}$ related to the term
$1-2n(\bbox{p},\bbox{r})$ in the expression (\ref{SR}). Finally we get
\begin{eqnarray}
\delta \sigma &=&-\frac{e^3}{3}\int \frac{d\bbox{r}
_1d\bbox{r}_2d\bbox{r}_3d\bbox{r}_4d\bbox{r}_5}{{\cal V}}\int\limits_0^{+
\infty }dt_1\int\limits_0^{t_1}dt_2\int\limits_0^{t_2}dt_3\times   \nonumber
\\
&&\left\{ \frac{\bbox{\hat p}}{m} \frac{\partial \rho _{15}}{\partial \mu
}U_{52}(t_1) \bbox{\hat j} U_{23}^{+}(t_3)\left[
I_{34}(t_2-t_3)U_{34}^{+}(t_2-t_3)\right] U_{41}^{+}(t_1-t_2)\right.
\nonumber  \\
&&+\left. \frac{\bbox{\hat p}}{m} \frac{\partial \rho _{15}}{\partial
\mu }U_{52}^{+}(t_1)
\bbox{\hat j} U_{23}(t_3)\left[ I_{34}(t_2-t_3)U_{34}(t_2-t_3)\right]
U_{41}(t_1-t_2)+... \right\}
\nonumber \\
&&-\frac{ie^3}{6}\int \frac{d\bbox{r}_1d\bbox{r}_2d\bbox{r}_3d\bbox{r}_4d
\bbox{r}_5d\bbox{r}_6}{{\cal V}}\int\limits_0^{+\infty
}dt_1\int\limits_0^{t_1}dt_2\int\limits_0^{t_2}dt_3\times
\nonumber \\
&&\left\{ -\frac{\bbox{\hat p}}{m}
\frac{\partial \rho _{15}}{\partial \mu }U_{52}(t_1)
\bbox{\hat j} U_{23}^{+}(t_3)\left[ R_{34}(t_2-t_3)U_{36}^{+}(t_2-t_3)\left(
1-2\rho \right)_{64}\right] U_{41}^{+}(t_1-t_2)\right.
\nonumber \\
&&+\left. \frac{\bbox{\hat p}}{m}
\frac{\partial \rho_{15}}{\partial
\mu } U_{52}^{+}(t_1)
\bbox{\hat j} U_{23}(t_3)\left[ R_{34}(t_2-t_3)U_{36}(t_2-t_3)\left(
1-2\rho \right)_{64}\right] U_{41}(t_1-t_2)+... \right\}.
\label{dsigma00}
\end{eqnarray}
In order to verify complete equivalence of (\ref{dsigma1}) and (\ref{dsigma00})
one should a) replace the operator $e\bbox{\hat p}/m$ by
$\bbox{\hat j}$; b) adjust the factor 3 by observing that
(\ref{dsigma00}) and (\ref{dsigma1}) are the corrections respectively to
to the scalar conductivity and the conductivity tensor (in the isotropic case
one has $\delta\sigma=(\delta\sigma_{xx}+\delta\sigma_{yy}+\delta\sigma_{zz})/3$)
and c) adjust another factor 2 having in mind symmetrisation of (\ref{dsigma1})
with respect to indices $\alpha$ and $\beta$.
Also, one should keep in mind that the operator
$\partial\rho_{15}/\partial\mu$ commutes with the evolution operator
$U_{52}$.
This completes the
proof of equivalence of the results (\ref{dsigma1}) and (\ref{dsigma00}).

The difference between the diagrammatic representations
of the AAG's perturbation theory and in our approach
is illustrated in Fig. 3. We have demonstrated that
the initial diagram (Fig. 3a) can be expressed in two equivalent ways:
by means of the diagram of Fig. 3b or as a sum of the two diagrams of Fig. 3c.
In the first case causality is explicitly
maintained and an additional coordinate integration appeared together
with the density matrix $1-2\rho$. In the second case only the
sum of the diagrams of Fig. 3c is meaningful. Causality is violated in
each of this diagrams if one considers them separately. This is
particularly clear from Fig. 3d which shows the classical paths corresponding 
to the second diagram of Fig. 3(c) (cf. Fig. 10c of Ref. \onlinecite{AAG2}) 
According to Ref. \onlinecite{AAG2} it is this path configuration
which was ``mistreated'' in our analysis \cite{GZ2}. In Fig. 3d we observe that 
electrons move backward in time between the moments $t'$ and
$t_1$. Such paths cannot appear within our path integral formalism, they are forbidden by the 
causality principle. Hence, their ``mistreatment'' could not occur within our analysis either.

\begin{figure}[t]
\centerline{\psfig{file=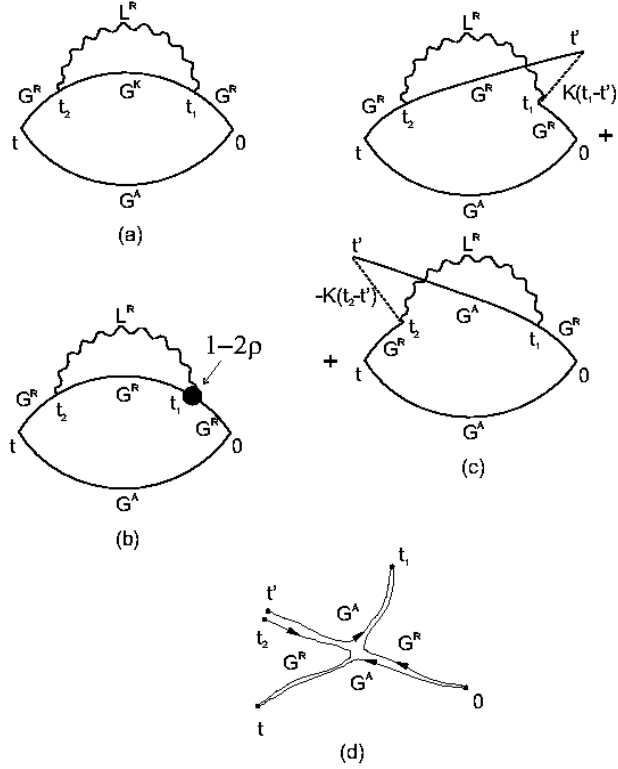,height=12cm}}
\caption{ The first order diagrams which contain the Keldysh function
$G^K$ and $\tanh\frac{\epsilon-\omega}{2T}$. (a) Initial diagram.
(b) $G^K$ is written in the form (\protect\ref{GKour}). 
(c) $G^K$ is written in the form (\protect\ref{GKA}). Two diagrams appear.
The second diagram $G^RG^AG^RG^A$ contains Hikami boxes.
Here we have defined $K(t)=\frac{-iT}{\sinh(\pi Tt)}$.
(d) The classical paths which correspond to the second diagram of Fig. 3(c).
Such paths violate causality and, hence, do not appear in the path integral.}
\label{fig3}
\end{figure}

\section{}
In this Appendix we will present some details of the derivation of eqs.
(\ref{dsig11}-\ref{G}). Let us first consider the contribution of the
self-energy diagrams of Fig. 2a,b and
consider only the terms in (\ref{dsigma1}) which contain the function $I(t,\bbox{r})$,
or $\coth(\omega/2T)$.
Substituting eqs. (\ref{U}-\ref{drhodmu}) into these terms of (\ref{dsigma1}) we find:
\begin{eqnarray}
\delta\sigma_{\alpha\beta}^{\rm coth, se}&=&
-\left[\sum\limits_{\lambda_1..\lambda_4}\frac{e^2}{2}
\int\frac{d\bbox{r}_1 d\bbox{r}_2 d\bbox{r}_3 d\bbox{r}_4}{{\cal V}}
\int\limits_0^\infty dt_1 \int\limits_0^{t_1} dt_2 \int\limits_0^{t_2} dt_3
\right.
\nonumber\\
&&
\left\{\frac{1}{2}\left(\frac{d}{d\xi_{\lambda_1}}
\tanh\frac{\xi_{\lambda_1}}{2T}\right)\hat j_\alpha
\psi_{\lambda_1}(\bbox{r}_1)\psi^*_{\lambda_1}(\bbox{r}_2)
e^{-i\xi_{\lambda_1}t_1}
\hat j_\beta \psi_{\lambda_2}(\bbox{r}_2)\psi^*_{\lambda_2}(\bbox{r}_3)
e^{i\xi_{\lambda_2}t_3}
\right.
\nonumber\\
&&
\left.
\left.
\big[I(t_2-t_3,\bbox{r}_3-\bbox{r}_4)
\psi_{\lambda_3}(\bbox{r}_3)\psi^*_{\lambda_3}(\bbox{r}_4)
e^{i\xi_{\lambda_3}(t_2-t_3)}\big]
\psi_{\lambda_4}(\bbox{r}_4)\psi^*_{\lambda_4}(\bbox{r}_1)
e^{i\xi_{\lambda_4}(t_1-t_2)}
+\alpha\leftrightarrow\beta\right\}+{\rm c.c.}\right]
\end{eqnarray}
This complicated expression can be rewritten in a simpler form if we
introduce the matrix elements
\begin{eqnarray}
I^{\lambda_1\lambda_3;\lambda_3\lambda_2}(t)&=&
\int d\bbox{r}_1 d\bbox{r}_2
\psi^*_{\lambda_1}(\bbox{r}_1)\psi_{\lambda_3}(\bbox{r}_1)
I(t,\bbox{r}_1-\bbox{r}_2)
\psi^*_{\lambda_3}(\bbox{r}_2)\psi_{\lambda_2}(\bbox{r}_2)
\label{I15}
\end{eqnarray}
and $j^{\lambda_1\lambda_2}_\alpha$ (\ref{M}). Then we get
\begin{eqnarray}
\delta\sigma_{\alpha\beta}^{\rm coth, se}&=&
-\frac{e^2}{2{\cal V}}\sum\limits_{\lambda_1..\lambda_4}
\int\limits_0^\infty dt_1 \int\limits_0^{t_1} dt_2 \int\limits_0^{t_2} dt_3
\left(\frac{d}{d\xi_{\lambda_1}}\tanh\frac{\xi_{\lambda_1}}{2T}\right)
\left(j^{\lambda_4\lambda_1}_\alpha j^{\lambda_1\lambda_2}_\beta+
j^{\lambda_4\lambda_1}_\beta j^{\lambda_1\lambda_2}_\alpha\right)
I^{\lambda_2\lambda_3;\lambda_3\lambda_4}(t_2-t_3)
\nonumber\\
&&
\cos\left(-\xi_{\lambda_1}t_1+\xi_{\lambda_2}t_3+\xi_{\lambda_3}(t_2-t_3)
+\xi_{\lambda_4}(t_1-t_2)\right).
\label{dsigma10}
\end{eqnarray}

Analogously, one can find the contribution of the remaining terms
in (\ref{dsigma1}) containing the function $R(t,\bbox{r})$, or $\tanh(\xi_\lambda/2T)$:
\begin{eqnarray}
\delta\sigma_{\alpha\beta}^{\rm tanh, se}&=&
-\frac{e^2}{4{\cal V}}\sum\limits_{\lambda_1..\lambda_4}
\int\limits_0^\infty dt_1 \int\limits_0^{t_1} dt_2 \int\limits_0^{t_2} dt_3
\left(\frac{d}{d\xi_{\lambda_1}}\tanh\frac{\xi_{\lambda_1}}{2T}\right)
\left(j^{\lambda_4\lambda_1}_\alpha j^{\lambda_1\lambda_2}_\beta+
j^{\lambda_4\lambda_1}_\beta j^{\lambda_1\lambda_2}_\alpha\right)
R^{\lambda_2\lambda_3;\lambda_3\lambda_4}(t_2-t_3)
\nonumber\\
&&
\tanh\frac{\xi_{\lambda_3}}{2T}
\;\;\sin\left(-\xi_{\lambda_1}t_1+\xi_{\lambda_2}t_3+\xi_{\lambda_3}(t_2-t_3)
+\xi_{\lambda_4}(t_1-t_2)\right).
\label{dsigma2}
\end{eqnarray}

Now we rewrite the functions
$I^{\lambda_2\lambda_3;\lambda_3\lambda_4}(t_2-t_3)$ and
$R^{\lambda_2\lambda_3;\lambda_3\lambda_4}(t_2-t_3)$ as follows:
\begin{equation}
I^{\lambda _2\lambda _3;\lambda _3\lambda _4}(t_2-t_3)=\int \frac{d\omega }{%
2\pi }\frac{d^3k}{(2\pi )^3}\text{Im}\left( \frac{-4\pi }{k^2\epsilon
(\omega ,k)}\right) \coth \frac \omega {2T}\left\langle \lambda _2\right|
e^{i\bbox{kr}}\left| \lambda _3\right\rangle \left\langle \lambda _3\right|
e^{-i\bbox{kr}}\left| \lambda _4\right\rangle \cos (\omega (t_2-t_3)),
\label{Iexact}
\end{equation}
\begin{equation}
R^{\lambda _2\lambda _3;\lambda _3\lambda _4}(t_2-t_3)=\int \frac{d\omega }{%
2\pi }\frac{d^3k}{(2\pi )^3}\frac{4\pi }{k^2\epsilon (\omega ,k)}%
\left\langle \lambda _2\right| e^{i\bbox{kr}}\left| \lambda _3\right\rangle
\left\langle \lambda _3\right| e^{-i\bbox{kr}}\left| \lambda _4\right\rangle
e^{-i\omega (t_2-t_3)}.  \label{Rexact}
\end{equation}
Now let us express the
kernel $R$ (\ref{Rexact}) in terms of the imaginary part of the inverse dielectric
susceptibility. Firstly we write
$1/\epsilon =1-(1-1/\epsilon )$. The function $1-1/\epsilon $ is regular in the
upper half-plane of $\omega$ and it tends to zero if
$\omega \rightarrow \infty $. Therefore for $t_2>t_3$ one has
$$
\int \frac{d\omega }{2\pi }\left( \frac 1{%
\epsilon (\omega ,k)}-1\right) e^{i\omega (t_2-t_3)}\equiv 0.
$$
Since in the integral (\ref{dsigma2}) the time $t_2$ indeed exceeds
$t_3$, we can replace $e^{-i\omega (t_2-t_3)}$ by $%
e^{-i\omega (t_2-t_3)}-e^{i\omega (t_2-t_3)}$ in the integral (\ref{Rexact}%
). Then we arrive at the following result:
\begin{eqnarray}
R^{\lambda _2\lambda _3;\lambda _3\lambda _4}(t_2-t_3) &=&\delta
(t_2-t_3-0)\left\langle \lambda _2\right| _{\bbox{r}_1}\left\langle \lambda
_3\right| _{\bbox{r}_2}\frac 1{|\bbox{r}_1-\bbox{r}_2|}\left| \lambda
_3\right\rangle _{\bbox{r}_1}\left| \lambda _4\right\rangle _{\bbox{r}_2}-
\nonumber \\
&&-2\int \frac{d\omega }{2\pi }\frac{d^3k}{(2\pi )^3}\text{Im}\left( \frac{%
-4\pi }{k^2\epsilon (\omega ,k)}\right) \left\langle \lambda _2\right| e^{i%
\bbox{kr}}\left| \lambda _3\right\rangle \left\langle \lambda _3\right| e^{-i%
\bbox{kr}}\left| \lambda _4\right\rangle \sin (\omega (t_2-t_3)).
\label{Rexact1}
\end{eqnarray}
The correction to the conductivity due to the self-energy diagrams
can now be written as
\begin{eqnarray}
\delta \sigma^{\rm se} _{\alpha \beta } &=&\delta \sigma _{\alpha \beta }^{C}
-\frac{e^2}{2{\cal {V}}}\int\limits_0^{+\infty
}dt_1\sum\limits_{\lambda _1..\lambda _4}\left( \frac d{d\xi _{\lambda _1}}%
\tanh \frac{\xi _{\lambda _1}}{2T}\right) \left( j_\alpha ^{\lambda
_4\lambda _1}j_\beta ^{\lambda _1\lambda _2}+j_\beta ^{\lambda _4\lambda
_1}j_\alpha ^{\lambda _1\lambda _2}\right) \times  \nonumber \\
&&\ \times \int \frac{d\omega }{2\pi }\frac{d^3k}{(2\pi )^3}\;\;\text{Im}%
\left( \frac{-4\pi }{k^2\epsilon (\omega ,k)}\right) \left\langle \lambda
_2\right| e^{i\bbox{kr}}\left| \lambda _3\right\rangle \left\langle \lambda
_3\right| e^{-i\bbox{kr}}\left| \lambda _4\right\rangle \left[ \coth \frac
\omega {2T}+\tanh \frac{\xi _{\lambda _3}}{2T}\right] F(t_1,\omega ,\xi
_{\lambda _1}..\xi _{\lambda _4}),  \label{dsig01}
\end{eqnarray}
where $\delta \sigma _{\alpha \beta }^{C}$ (\ref{Clmb}) is the correction due
to the non-screened Coulomb interaction and the function $F$ is defined in eq. (\ref{F}).

Now let us make use of the Drude approximation for the dielectric susceptibility
of a disordered metal
$$
\epsilon(\omega,k)=\frac{4\pi\sigma}{-i\omega+Dk^2}.
$$
In this case it is convenient to introduce the matrix elements
$M^{\lambda_2\lambda_3;\lambda_3\lambda_4}$ as defined in eq. (\ref{M}).
Combining the above expressions we immediately arrive at the final result
for the first order correction to the conductivity from self-energy diagrams of Fig. 2a,b
(\ref{dsig11}), (\ref{F}).

The conductivity correction from the vertex diagrams of Fig. 2c,d  is evaluated
analogously. After a straightforward algebra we obtain
\begin{eqnarray}
\delta\sigma_{\alpha\beta}^{\rm coth, vert}&=&
\frac{e^2}{2{\cal V}}\sum\limits_{\lambda_1..\lambda_4}
\int\limits_0^\infty dt_1\int\limits_0^{t_1}dt_2\int\limits_0^{t_2}dt_3
\left(\frac{d}{d\xi_{\lambda_1}}\tanh\frac{\xi_{\lambda_1}}{2T}\right)
\left(j_\alpha^{\lambda_2\lambda_3} j_\beta^{\lambda_1\lambda_4}+
j_\beta^{\lambda_2\lambda_3}j_\alpha^{\lambda_1\lambda_4}\right)
I^{\lambda_3\lambda_1;\lambda_4\lambda_2}(t_2-t_3)
\nonumber\\
&&
\cos(-\xi_{\lambda_3}(t_1-t_3)-\xi_{\lambda_1}t_3+\xi_{\lambda_4}t_2
+\xi_{\lambda_2}(t_1-t_2)),
\label{dsigver1}
\end{eqnarray}
\begin{eqnarray}
\delta\sigma_{\alpha\beta}^{\rm tanh, vert}&=&
-\frac{e^2}{4{\cal V}}\sum\limits_{\lambda_1..\lambda_4}
\int\limits_0^\infty dt_1\int\limits_0^{t_1}dt_2\int\limits_0^{t_2}dt_3
\left(\frac{d}{d\xi_{\lambda_1}}\tanh\frac{\xi_{\lambda_1}}{2T}\right)
\left(j_\alpha^{\lambda_2\lambda_3} j_\beta^{\lambda_1\lambda_4}+
j_\beta^{\lambda_2\lambda_3}j_\alpha^{\lambda_1\lambda_4}\right)
R^{\lambda_3\lambda_1;\lambda_4\lambda_2}(t_2-t_3)
\nonumber\\
&&
\tanh\frac{\xi_{\lambda_3}}{2T}
\sin(-\xi_{\lambda_3}(t_1-t_3)-\xi_{\lambda_1}t_3+\xi_{\lambda_2}t_2
+\xi_{\lambda_4}(t_1-t_2)).
\label{dsigver2}
\end{eqnarray}
Introducing again the matrix elements (\ref{M}) and combining
(\ref{dsigver1}) and (\ref{dsigver2}) we arrive at the result (\ref{dsigver}), (\ref{G}).

\section{}

In this Appendix we will analyze the expressions for the weak localization
correction to the conductance obtained by AAG
in the limit $\tau_H \ll \tau_{\varphi}$ perturbatively in the interaction.
This additional analysis is necessary because the final results
for the most interesting low temperature limit $T\tau_H \lesssim 1$ were
not presented in Ref. \onlinecite{AAG2}. For the sake of definiteness
we will consider only the 1d case which is sufficient for our purposes.

AAG split the total expression for the weak localization
correction to the conductance into two terms:
\begin{equation}
\delta \sigma_{WL}^{AAG}=\delta \sigma_{\rm deph}^{AAG}+
\delta \sigma_{CWL}^{AAG},
\label{sumoftwo}
\end{equation}
where -- according to Ref. \onlinecite{AAG2} -- the first term describes
dephasing while the second term accounts for the renormalization due to
interaction and has nothing to do with dephasing. Below we will demonstrate
that such a separation of the weak localization correction into
``dephasing'' and ``interaction'' terms is not possible even within the
perturbation theory employed by AAG.

Let us consider the first term in (\ref{sumoftwo}) which was defined in
eqs. (4.5), (4.6) of Ref. \onlinecite{AAG2} and has the form
\begin{equation}
\delta\sigma_{\rm deph}^{AAG}=\frac{\sigma}{\pi{\cal V}}
\int\frac{dQdq}{(2\pi)^2}\int\frac{d\omega}{2\pi}
\left(\frac{\omega}{2T}\right)^2\frac{1}{\sinh^2\frac{\omega}{2T}}
\left(\frac{4Te^2}{\sigma_1q^2}\right)
\left(\left[{\cal C}(Q,0)\right]^2{\cal C}(Q+q,\omega)-
\left|{\cal C}(Q,\omega)\right|^2{\cal C}(Q+q,0)\right),
\label{deph33}
\end{equation}
where the Cooperon and the diffuson are defined as follows:
\begin{equation}
{\cal C}(\omega,q)=\frac{1}{-i\omega+Dq^2+1/\tau_H}, \;\;\;
{\cal D}(\omega,q)=\frac{1}{-i\omega+Dq^2}.
\label{CD}
\end{equation}
The integrals over $Q$ and $q$ in (\ref{deph33}) can be evaluated exactly
and we find
\begin{eqnarray}
\delta\sigma_{\rm deph}^{AAG}&=&\frac{\sigma}{2\pi{\cal V}^2}\frac{\tau_H^2}{D}
{\rm Im}\int\frac{d\omega}{2\pi}\frac{\omega}{2T\sinh^2\frac{\omega}{2T}}
\left[\frac{i\omega\tau_H}{2\sqrt{1+i\omega\tau_H}(\sqrt{1+i\omega\tau_H}-1)^2}
\left(\frac{1}{\sqrt{1+i\omega\tau_H}-1}-\frac{1}{2}\right)-\right.
\nonumber\\
&&\left.
-\frac{4}{(i\omega\tau_H)^2}-
\frac{(\sqrt{1+i\omega\tau_H}-1)^2}{2(i\omega\tau_H)^2\sqrt{1+i\omega\tau_H}}
\right]
\label{sigdeph}
\end{eqnarray}
Making use of this equation we get
\begin{eqnarray}
\delta\sigma_{\rm deph}^{AAG}&\simeq &\frac{e^2}{\pi}\frac{e^2}{\sigma_1}
\left\{\frac{DT\tau_H^2}{4}\left(1+\zeta(1/2)\sqrt{\frac{2}{\pi T\tau_H}}\right)
+\frac{5}{8}\frac{\zeta(3/2)}{\pi}\sqrt{\frac{D^2\tau_H}{2\pi T}}\right\},
\;\;\; T\tau_H\gg 1,
\nonumber\\
\delta\sigma_{\rm deph}^{AAG}&\simeq & \frac{e^2}{\pi}\frac{e^2}{\sigma_1}
\frac{\pi^4}{30}D\tau_H(T\tau_H)^4,\;\;\; T\tau_H\ll 1
\label{deph}
\end{eqnarray}
In the limit $T\tau_H \gg 1$ the expression (\ref{deph}) practically coincides
with the analogous expression obtained in Ref. \onlinecite{AAG2} (see
eq. (4.11) of that paper) with the only difference in a numerical
coefficient in front of the last term in the first equation (\ref{deph})
($5/8$ in our calculation and $1/2$ in Ref. \onlinecite{AAG2}). In the
opposite limit $T\tau_H \ll 1$ the expression (\ref{deph33}) was not
evaluated by AAG at all.

Now let us consider the second contribution (\ref{sumoftwo}) which
was denoted by AAG as a cross term of weak localization and interaction.
According to Ref. \onlinecite{AAG2} this term has the following structure
(see eq. (5.23) of that paper)
\begin{equation}
\delta\sigma^{\rm CWL}=\frac{\sigma}{2\pi {\cal V}^2}{\rm Im}
\int\frac{d\omega}{2\pi}\left[\frac{d}{d\omega}\left(\omega\coth
\frac{\omega}{2T}\right)\right]\left[
I_1(\omega)+2I_2(\omega)-4I_3(\omega)+4I_4(\omega)+I_5(\omega)+8I_6(\omega)
\right],
\label{CWL}
\end{equation}
where the terms in the square brackets are defined in eqs. (5.25) of \cite{AAG2}:
\begin{eqnarray}
I_1(\omega)&=&\int\frac{dQdq}{(2\pi)^2}\frac{1}{Dq^2}\left\{
{\cal C}^2(0,Q)\left[\frac{{\cal C}(-\omega,Q+q)}{{\cal D}(-\omega,q)}-1 \right]
+\left[{\cal C}(-\omega,Q)-{\cal C}(0,Q)\right]{\cal D}(-\omega,q)+
2{\cal C}(-\omega,Q+q){\cal C}(0,Q)\right\},
\nonumber\\
I_2(\omega)&=&\int\frac{dQdq}{(2\pi)^2}\frac{q}{Dq^2}
\frac{\partial}{\partial q}{\cal C}(-\omega,Q+q){\cal C}(0,Q),
\nonumber\\
I_3(\omega)&=&\int\frac{dQdq}{(2\pi)^2} {\cal D}(-\omega,q)
{\cal C}(-\omega,Q+q){\cal C}(0,Q),
\nonumber\\
I_4(\omega)&=&\int\frac{dQdq}{(2\pi)^2} \frac{Q}{q} {\cal D}(-\omega,q)
{\cal C}(0,Q-q){\cal C}(-\omega,Q),
\nonumber\\
I_5(\omega)&=&\int\frac{dQdq}{(2\pi)^2} {\cal D}(-\omega,q){\cal C}^2(-\omega,Q),
\nonumber\\
I_6(\omega)&=&\int\frac{dQdq}{(2\pi)^2} Dq^2{\cal D}^3(-\omega,q){\cal
C}(-\omega, Q).
\label{I16}
\end{eqnarray}

These integrals can also be evaluated exactly and we arrive at the following
results:
\begin{eqnarray}
I_1(\omega)&=&\frac{\tau_H^2}{4D}
\frac{\sqrt{1+i\omega\tau_H}-1}{(i\omega\tau_H)^{3/2}\sqrt{1+i\omega\tau_H}},
\nonumber\\
I_2(\omega)&=&\frac{\tau_H^2}{4D}\frac{1}{(i\omega\tau_H)^2}
\left(2-\sqrt{1+i\omega\tau_H}-\frac{1}{\sqrt{1+i\omega\tau_H}}\right),
\nonumber\\
I_3(\omega)&=&\frac{\tau_H^2}{8D}
\left(\frac{1}{\sqrt{i\omega\tau_H}\sqrt{1+i\omega\tau_H}}-
\frac{\sqrt{1+i\omega\tau_H}-1}{i\omega\tau_H\sqrt{1+i\omega\tau_H}}\right),
\nonumber\\
I_4(\omega)&=&\frac{\tau_H^2}{4D}\left(
\frac{\sqrt{1+i\omega\tau_H}-1}{(i\omega\tau_H)^2}+
\frac{\sqrt{1+i\omega\tau_H}-1}{2(i\omega\tau_H)^{3/2}}-
\frac{1}{2(i\omega\tau_H)}\right),
\nonumber\\
I_5(\omega)&=&\frac{\tau_H^2}{8D}
\frac{1}{\sqrt{i\omega\tau_H}(1+i\omega\tau_H)^{3/2}},
\nonumber\\
I_6(\omega)&=&\frac{\tau_H^2}{32D}
\frac{1}{(i\omega\tau_H)^{3/2}\sqrt{1+i\omega\tau_H}}.
\label{Iexac}
\end{eqnarray}

Let us compare these exact expressions with those obtained in Ref.
\onlinecite{AAG2}. Unfortunately AAG
calculated the integrals approximately only in the limit of high
frequencies $\omega\tau_H\gg 1$. In this limit they found
\begin{eqnarray}
I_1^{AAG}(\omega)&\simeq & \frac{\tau_H^2}{4D}\left(\frac{1}{(i\omega\tau_H)^{3/2}}-
\frac{1}{(i\omega\tau_H)^2}\right),
\nonumber\\
I_2^{AAG}(\omega)&\simeq &\frac{\tau_H^2}{4D}\left(-\frac{1}{(i\omega\tau_H)^{3/2}}+
\frac{2}{(i\omega\tau_H)^2}\right),
\nonumber\\
I_3^{AAG}(\omega)&\simeq &\frac{\tau_H^2}{4D}\left(\frac{1}{2(i\omega\tau_H)^{3/2}}-
\frac{1}{4(i\omega\tau_H)^2}\right),
\nonumber\\
I_4^{AAG}(\omega)&\simeq &\frac{\tau_H^2}{8D}
\left(\frac{1}{(i\omega\tau_H)^{3/2}}-
\frac{3}{2(i\omega\tau_H)^2}\right),
\nonumber\\
I_5^{AAG}(\omega)&\simeq &\frac{\tau_H^2}{8D}\frac{1}{(i\omega\tau_H)^2},
\nonumber\\
I_6^{AAG}(\omega)&\simeq &\frac{\tau_H^2}{64D}\frac{1}{(i\omega\tau_H)^2}.
\label{IAleiner}
\end{eqnarray}
In order to find a numerical coefficient in front of the term $\sim 1/(i\omega\tau_H)^2$
in the expression for $I_4^{AAG}$ (\ref{IAleiner}) we exactly evaluated
the integral
$$
\int_0^{\infty} dxdydz \frac{y \exp(-x-y)}{(y+z)\sqrt{xy+xz+yz}}=
\frac{\pi}{2}
$$
(see eq. (5.25d) of Ref. \onlinecite{AAG2}) which was not calculated by AAG.

The high frequency asymptotics of the functions $I_1,I_2,I_3,I_4$
and $I_5$ coincide with the results (\ref{IAleiner}), while the asymptotic
results for  $I_6$ derived from the exact expressions (\ref{Iexac}) in the
limit $\omega \tau_H \gg 1$ takes the form
$$
I_6(\omega)\simeq \frac{\tau_H^2}{32D}\frac{1}{(i\omega\tau_H)^2}.
$$
We observe a difference in the numerical prefactor
in the exact expression and in the term
$I_6^{AAG}$ (\ref{IAleiner}) found in Ref. \onlinecite{AAG2}.

A much more important problem is, however, not in this numerical
discrepancy,
but rather in the fact that AAG evaluated the integrals
only in the limit $\omega\tau_H\gg 1$ and did not study the behavior of
the integrals (\ref{I16}) at lower frequencies $\omega\tau_H\lesssim 1$ at all.
Note that the low frequency behavior of these integrals
is crucially important because it determines the dependence
of the weak localization correction on $\tau_H$ at sufficiently low
temperatures $T\tau_H \lesssim 1$. [Let us remind the
reader that here we are discussing only the limit of strong magnetic fields
\cite{AAG2} $\tau_H \ll \tau_{\varphi}$ beyond which any perturbative (in the
interaction) calculation of the weak localization correction is meaningless.
Clearly, in this limit the condition $T \tau_H \lesssim 1$ is compatible with the condition
$T\tau_{\varphi} \gg 1$.]

It is easy to observe from (\ref{Iexac}) that
at low frequencies the integrals $I_1$, $I_3$, $I_4$, $I_5$ behave as
$\propto 1/\sqrt{i\omega\tau_H}$, the integral $I_2$ tends to a constant at $\omega\to 0$,
while the integral $I_6$ behaves as $I_6\propto 1/(i\omega\tau_H)^{3/2}$.
It implies that the contribution of the integrals $I_1$,..,$I_5$ to the
conductivity correction (\ref{CWL}) does not diverge as temperature goes to zero
but stays finite for any finite $\tau_H$ even at $T=0$. The contribution of the
integral $I_6$ diverges as $1/\sqrt{T}$ for low $T$. Note, that in the high
frequency limit (\ref{IAleiner}) all the integrals show the same
asymptotic behavior, and one could naively conclude that all these integrals are of
the same origin. In reality, however,
only (a part of) the integral $I_6$ can be
interpreted as the effect of interaction on the weak localization, while
all other terms actually contribute to dephasing, at least for not very high
temperatures $T\tau_H\lesssim 1$.

Let us split the function $I_6$ (\ref{Iexac}) into two terms which
give respectively  divergent and convergent contributions to the integral of $I_6(\omega)$ over $\omega$ at low frequencies
\begin{equation}
I_6(\omega)=\frac{\tau_H^2}{32D}\frac{1}{(i\omega\tau_H)^{3/2}}-
\frac{\tau_H^2}{32D}\frac{\sqrt{1+i\omega\tau_H}-1}
{(i\omega\tau_H)^{3/2}\sqrt{1+i\omega\tau_H}}
\label{I6}
\end{equation}
We will now treat only the first (divergent) term in this equation, while the
second (convergent) term will be added to the integrals $I_1$,..,$I_5$.
The divergent term yields the following contribution:
\begin{eqnarray}
\delta\sigma_{\rm CWL}&=&
\frac{\sigma}{2\pi\nu^2}{\rm Im}\int\frac{d\omega}{2\pi}
\left[\frac{d}{d\omega}\left(\omega\coth\frac{\omega}{2T}\right)\right]
8\frac{\tau_H^2}{32D}\frac{1}{(i\omega\tau_H)^{3/2}}
\nonumber\\
&=&
-\frac{e^2}{2\pi}\sqrt{\frac{D}{2\pi T}}\left(\frac{3\zeta(3/2)}{2}\right)
\frac{e^2}{\pi\sigma}\sqrt{D\tau_H}=
-\frac{1}{2}\delta\sigma_C(T)\frac{\delta\sigma_{WL}}{\sigma},
\label{divergent}
\end{eqnarray}
where
\begin{equation}
\delta\sigma_C(T)=
-\frac{e^2}{\pi}\sqrt{\frac{D}{2\pi T}}\left(\frac{3\zeta(3/2)}{2}\right)
\label{sigmaC}
\end{equation}
is the interaction correction to the conductivity, and
\begin{equation}
\delta\sigma_{WL}=-\frac{e^2}{\pi}\sqrt{D\tau_H}
\label{sigmaWL}
\end{equation}
is the weak localization correction. Within the validity range of the
perturbation theory in the interaction, the equation (\ref{divergent}) is valid
in the whole temperature interval from $T\tau_H\ll 1$ to $T\tau_H\gg 1$. The
physical origin of the correction (\ref{divergent}) is quite transparent. Indeed,
the interaction term (\ref{sigmaC}) contains the photon propagator
${\cal L}^R\propto 1/\sigma$. If one replaces the total conductivity
by the sum of the Drude conductivity and the weak localization correction, one
will immediately observe that in the first order in $\delta\sigma_{WL}$ the
interaction correction will be transformed as
$\delta\sigma_C\to \delta\sigma_C(1-\delta\sigma_{WL}/\sigma)$. Thus we arrive
at (\ref{divergent}).

This simple consideration clarifies the origin of the
correction (\ref{divergent}). It can be interpreted as the effect of weak
localization on the interaction correction. The same conclusion follows
if one considers the diagrams contributing to the integral $I_6$. They are
just the diagrams which yield the interaction correction multiplied by the weak
localization correction to the photon propagator. In other words, the
contribution (\ref{divergent}) originates from the second term in the
right-hand side of eq. (5.22) in Ref. \onlinecite{AAG2}, while all the
remaining contributions come from the first term, i.e. from the unrenormalized
photon propagator. Collecting all these contributions we express the final
result \cite{AAG2} for $\delta \sigma^{AAG}_{CWL}$ (\ref{CWL}) as a sum of
two terms
\begin{equation}
\delta \sigma_{CWL}^{AAG}=\delta \sigma_{CWL}+
\delta \sigma^{??},
\label{sumoftwo2}
\end{equation}
where $\delta \sigma_{CWL}$ is defined in (\ref{divergent}) and
\begin{equation}
\delta\sigma^{??}=\left\{
\begin{array}{ll}
\frac{e^2}{\pi}\frac{e^2}{\sigma_1}\left\{
\frac{3}{2\pi}D\tau_H+{\cal O}\left(D\tau_H\sqrt{T\tau_H}\right)\right\}, & T\tau_H\ll 1 \\
\frac{3}{2}\frac{e^2}{\pi}\sqrt{\frac{D}{2\pi T}}\left(\frac{3\zeta(3/2)}{2}
\right)
\frac{e^2}{\pi}\frac{\sqrt{D\tau_H}}{\sigma_1}, & T\tau_H\gg 1.
\end{array}\right.
\label{sigdephCWL}
\end{equation}
can be derived from the eq. (\ref{CWL}) combined with (\ref{Iexac}).

We observe that for $T\tau_H\lesssim 1$ the contribution
proportional to $\tau_H$ dominates and determines the behavior of $\delta
\sigma^{??}$ as a function of $\tau_H$. This contribution {\it cannot}
be interpreted as an
interaction correction since it grows {\it
faster} than $\sqrt{\tau_H}$ with increasing $\tau_H$.
For quasi-1d systems all contributions to the conductance growing faster
than $\sqrt{\tau_H}$ should be interpreted as {\it dephasing} terms
(see also our discussion in Section 2).
Thus -- although the correct and complete description of the
interaction-induced decoherence cannot
be obtained from the perturbation theory in the interaction -- the
conclusion about the existence of temperature-independent dephasing
in disordered conductors at low $T$ (determined by the
range of frequencies $\omega > T$) follows already
from the perturbative calculation \cite{AAG2}! This is in contrast with
the qualitative arguments about the absence of zero-temperature dephasing
presented in the same paper.

It also follows from the above consideration that the suggested by AAG 
splitting of the weak localization correction into two terms
(\ref{sumoftwo}) can hardly be justified in both limits $T\tau_H \gg 1$ and
$T\tau_H \ll 1$ even within the framework of their perturbation calculation
\cite{AAG2}.
Indeed, for $T\tau_H \gg 1$ the total expression for the weak localization
correction $\delta \sigma_{CWL}^{AAG}$ derived by AAG and defined
here in eqs. (\ref{sumoftwo}), (\ref{deph33}), (\ref{CWL}) and (\ref{I16})
can be expressed as a series expansion in $1/\sqrt{T\tau_H}$:
\begin{equation}
\delta\sigma_{WL}^{AAG}\simeq \frac{e^2}{\pi}\frac{e^2}{\sigma_1}
\frac{DT\tau_H^2}{4}\left(1+{\cal O}\left(\sqrt{\frac{1}{T\tau_H}}\right)\right).
\label{lTtH}
\end{equation}
In this limit $\delta \sigma_{CWL}^{AAG}$ is vanishingly small (it contributes
only to the next after the subleading term in (\ref{lTtH}) and can be safely
disregarded. Moreover, for $T\tau_H \gg 1$ {\it exactly the same} term
$\sim \delta \sigma_{CWL}^{AAG} \propto \sqrt{\tau_H/T}$
(with a slightly different numerical prefactor)
is contained in the expression for $\delta \sigma_{\rm deph}^{AAG}$
(\ref{deph}). Also due to this fact there are no reasons to distinguish the term
$\delta \sigma_{CWL}^{AAG}$ from the remaining contribution to
$\delta \sigma_{WL}^{AAG}$ in the above limit.

For $T\tau_H \sim 1$ both terms in (\ref{sumoftwo}) are of the same
order, and therefore their separation is not possible. Finally, in the
limit $T\tau_H \ll 1$ the result for $\delta \sigma_{WL}^{AAG}$ reads
\begin{equation}
\delta\sigma_{WL}^{AAG}= \delta \sigma_{CWL}+
\frac{e^2}{\pi}\frac{e^2}{\sigma_1}
\frac{3D\tau_H}{2\pi}(1+{\cal O} (\sqrt{T\tau_H})).
\label{sTtH}
\end{equation}
The origin of the term $\delta \sigma_{CWL} \propto \sqrt{\tau_H/T}$
(\ref{divergent}) was clarified above. It does not describe dephasing
and it is purely a matter of convention whether to include this term
into the perturbative weak localization or interaction corrections.
The second term in (\ref{sTtH}) is represented as a series expansion
in powers of $\sqrt{T\tau_H}$, it describes dephasing and remains
finite even at $T=0$. We emphasize again that in the limit $T\tau_H \ll 1$
the leading ``dephasing-type'' contribution to the perturbative weak localization
correction comes from $\delta \sigma_{CWL}^{AAG}$ and not from
the ``dephasing'' term \cite{AAG2} $\delta \sigma_{\rm deph}^{\rm AAG}$ which
only contributes to the higher order terms ($\sim (T\tau_H)^4$)
of the expansion of $\delta \sigma_{WL}^{AAG}$ in powers of $T\tau_H$ (cf. eq.
(\ref{deph})). Terms of the same order are also contained in the expression for
$\delta \sigma_{CWL}^{AAG}$. Thus the
splitting (\ref{sumoftwo}) is not justified also in the limit $T\tau_H \ll 1$.

\section{}

For reference purposes we present some rigorous expressions obtained within the
exactly solvable Caldeira-Leggett model \cite{CL} for a quantum particle with
coordinate $x$ interacting with an infinite bath of oscillators. The time evolution of
the density matrix of such a particle is defined by eq. (\ref{rho}) and
the kernel $J$ is given by the following path integral:
\begin{equation}
J(t,x_{1f},x_{2f},x_{1i},x_{2i})=
\int\limits^{x_{1f}}_{x_{1i}}{\cal D}x_1(t)
\int\limits^{x_{2f}}_{x_{2i}}{\cal D}x_2(t)
\exp\left(iS_0[x_1(t)]-iS_0[x_2(t)]-iS_R[x_1(t),x_2(t)]
-S_I[x_1(t),x_2(t)] \right),
\label{J15}
\end{equation}
where $S_0[x]=\int\limits_0^t dt'\;\; (m\dot x^2/2)$ is the action of a free
particle. The interaction part of the action has the following form
\begin{eqnarray}
S_R[x_1,x_2]&=&\int\limits_0^t dt_1\int\limits_0^{t_1} dt_2 \left\{
\alpha_R(t_1-t_2)x_1(t_1)x_1(t_2)-\alpha_R(t_1-t_2)x_2(t_1)x_2(t_2)+
\right.
\nonumber\\
&&
\left.
\alpha_R(t_1-t_2)x_1(t_1)x_2(t_2)-\alpha_R(t_1-t_2)x_2(t_1)x_1(t_2)
\right\}+
\frac{1}{2}\left(\int\frac{d\omega}{2\pi}\;\frac{2I(\omega)}{\omega}\right)
\int\limits_0^t dt' [x^2_1(t')-x^2_2(t')],
\label{SR15}
\end{eqnarray}
\begin{eqnarray}
S_I[x_1,x_2]&=&\int\limits_0^t dt_1\int\limits_0^{t_1} dt_2 \left\{
\alpha_I(t_1-t_2)x_1(t_1)x_1(t_2)+\alpha_I(t_1-t_2)x_2(t_1)x_2(t_2)-
\right.
\nonumber\\
&&
\left.
\alpha_I(t_1-t_2)x_1(t_1)x_2(t_2)-\alpha_I(t_1-t_2)x_2(t_1)x_1(t_2)
\right\}.
\label{SI15}
\end{eqnarray}
The kernels $\alpha_I$ and $\alpha_R$ are given by the following integrals:
\begin{eqnarray}
\alpha_R(t)&=&-i\int\frac{d\omega}{2\pi}\; I(\omega) e^{-i\omega t},
\label{alphaR}
\\
\alpha_I(t)&=&\int\frac{d\omega}{2\pi}\; I(\omega)
\coth\frac{\omega}{2T} e^{-i\omega t},
\label{alphaI}
\end{eqnarray}
where $I(\omega)$ is the spectral density of the oscillators. This function can
be arbitrary, but we will consider here only the case of Ohmic dissipation,
$I(\omega)=\eta\omega\theta(\omega_c-|\omega|)$, with $\omega_c$ being
the high cutoff frequency. This spectrum is the most relevant in view
of comparison to the disordered metal. The last term in the action $S_R$
(\ref{SR15}) compensates potential renormalization caused by the interaction
and maintains the translational invariance of the system.

The kernel $J$ (\ref{J15}) can be found exactly as the integrals over the
coordinates are Gaussian. One finds (see e.g. \cite{CL,Weiss})
\begin{eqnarray}
J(t,x_{1f},x_{2f},x_{1i},x_{2i})&=& \frac{\eta}{2\pi(1-e^{-\gamma t})}
\exp\left[ i\eta\frac{x^+_f x^-_f + e^{\gamma t}x^+_i x^-_i -e^{\gamma
t}x^+_f x^-_i- x^+_i x^-_f}{e^{\gamma t}-1} - \right.  \nonumber \\
&& \left. -\eta\big\{g_1(t){x^-_i}%
^2+g_2(t)(x^-_f-x^-_i)^2+g_3(t)x^-_i(x^-_f-x^-_i)\big\} \right],
\label{Jexp}
\end{eqnarray}
where
\begin{eqnarray}
g_1(t)&=&\frac{1}{2}\int\limits_0^t ds \int\limits_0^t ds^{\prime}\;\;
G(s-s^{\prime})\simeq Tt+\ln\frac{1-e^{-2\pi Tt}}{2\pi(T/\omega_c)}, \;\;\;
t\gg \omega_c^{-1}  \label{g1} \\
g_2(t)&=&\frac{1}{2}\int\limits_0^t ds \int\limits_0^t ds^{\prime}\;\; \frac{%
(e^{\gamma s}-1)G(s-s^{\prime})(e^{\gamma s^{\prime}}-1)}{(e^{\gamma t}-1)^2}
\label{g2} \\
g_3(t)&=&\int\limits_0^t ds \int\limits_0^t ds^{\prime}\;\; \frac{%
G(s-s^{\prime})(e^{\gamma s^{\prime}}-1)}{(e^{\gamma t}-1)},  \label{g3}
\end{eqnarray}
and
\begin{equation}
G(t)=\int\limits_{-\omega_c}^{\omega_c}\frac{d\omega}{2\pi}\;\; \omega \coth%
\frac{\omega}{2T} e^{-i\omega t}= -\frac{1}{\pi}\left(\frac{\pi T}{\sinh(\pi
Tt)}\right)^2.  \label{Gt}
\end{equation}

\section{}
As it was already pointed out, the authors of several recent papers
\cite{CI,VA,IFS} arrived at conclusions different from
ours \cite{GZ1,GZ2,GZ98} and, moreover, argued that our approach \cite{GZ2} is
not correct. On top of the standard arguments (which have been already discussed
in the bulk of this paper) the authors \cite{CI,VA,IFS} suggested 
various additional reasons which could invalidate our analysis.
In view of that we feel it will be appropriate to
address the arguments presented in the above papers. We believe
that this Appendix can be useful for the reader who would like to follow 
the details of the discussion around the problem of quantum decoherence
at low temperatures.

Cohen and Imry (CI) \cite{CI} proceeded within the Feynman-Vernon influence functional
formalism and found that the interference of any pair of time reversed
paths is suppressed due to interaction with an effective environment even at
$T=0$. This result (eq. (12) of Ref. \onlinecite{CI}) implies a nonzero decoherence rate
at $T=0$ and is consistent with our results \cite{GZ1,GZ2,GZ98}
as well with the results of other authors
\cite{LM,Wch9} obtained within the framework of the CL model. However, CI
argued that the saddle-point approximation they used
``cannot be trusted'' at low temperatures, and the $T$-independent contribution
to the dephasing rate ``should be omitted''. The CI's arguments in favor of this conclusion
follow the line of reasoning according to which a particle
with the energy $\sim T$ cannot excite environmental modes with energies exceeding
$T$.

The arguments \cite{CI} can not be accepted. Indeed,
if CI do not trust the saddle-point
approximation which gives nonzero dephasing down to $T=0$, this
could only imply that one should analyze the role of fluctuations around the saddle
points. The contribution
of non-saddle-point paths may only yield further suppression of quantum coherence
simply because the relevant saddle points provide the (local) minimum for
the action. The imaginary part of the effective action is {\it positive} $S_I >0$ for
{\it all} paths except for pairs of exactly equal paths in which case $S_I \equiv 0$.
[The latter paths do not contribute to the dephasing rate.] 
Thus the saddle point approximation may only
{\it underestimate} the dephasing rate.

We can add that in our problem the applicability of the saddle point
approximation cannot depend on temperature. This is particularly clear in
the weak interaction limit. In this case the saddle point paths are determined
only by the ``noninteracting'' parts of the action which do not depend
on temperature at all. The ``interacting'' contribution can then be treated
perturbatively in the exponent (cf. e.g. Sec. 5C), and this is a completely
legitimate procedure
controlled by a small parameter $1/p_Fl \ll 1$ in the case of disordered metals.
At sufficiently high $T$ this procedure yields the AAK's result \cite{AAK}. If one
accepts the saddle-point approximation at higher $T$, one should accept it also
at lower $T$: the saddle point paths are the same and the ``interacting''
contribution to the exponent may only decrease with decreasing $T$. The saddle 
point approximation which treats the interaction
term perturbatively in the exponent (and thus yields energy conservation on
the saddle point paths) is well known and was frequently used for the Feynman-Vernon
influence functionals, see e.g. Refs. \onlinecite{ChS,Legg}. 

CI \cite{CI} also
mentioned the simple perturbative results which -- according to them --
yield zero decoherence rate at $T=0$. As it was already explained above,
(i) the problem is essentially nonperturbative and no information
about the dephasing time $\tau_{\varphi}$ can be extracted from perturbation
theory in the interaction (Sec. 2) and (ii) at low temperatures the golden rule
approximation yields incorrect results even for perturbative terms 
(Sec. 4B). Eqs. (14), (15) of Ref. \onlinecite{CI} can be obtained
from the equation above eq. (14) of the same paper only within the golden
rule approximation (\ref{GRS}). Exact calculation leads to an additional term
(cf. eqs. (\ref{func}) and (\ref{funcCL})) which survives even at $T=0$ and diverges at
large times. Unfortunately this term is missing in eq. (14) of Ref. \onlinecite{CI}.

The same term is missing also in eq. (1) of the paper by Imry,
Fukuyama and Schwab (IFS) \cite{IFS}. Already this
fact as well as our analysis of Sec. 2 invalidates the ``proof that zero-point
motion does not dephase'' presented by IFS. Again, their eqs.
(1), (2) are obtained within the same golden rule approximation. Since IFS 
allow for a general form of the interaction matrix elements,
it should be legitimate to take ones e.g. for the CL model and to substitute them
into eq. (1) of Ref. \onlinecite{IFS}. Then one would immediately arrive at the
conclusion that no decoherence occurs in the CL model at $T=0$. Thus 
the proof of Ref. \onlinecite{IFS}
explicitly contradicts to the results obtained within the exactly solvable
Caldeira-Leggett model.

IFS also argued that
dephasing cannot occur at $T \to 0$ in equilibrium because ``in that limit
neither the electron nor the environment has any energy to exchange''.
It is well known, however, that the energy of a subsystem (an electron in our case)
interacting with other subsystem (other electrons) is not conserved {\it due to
interaction}. The energy exchange between different subsystems of a
closed system is, of course, possible even at $T=0$. In the presence of interaction
with any other quantum degrees of freedom an electron can be described only by
the density matrix \cite{LL} and is (obviously) not in its noninteracting ground state.
Therefore it always has energy to
exchange. The same is true for the environment: it is well known \cite{LP}
that an imaginary part of the dielectric susceptibility for a large system Im$\epsilon$
does not vanish even at $T=0$, thus implying the possibility of energy exchange.
The above argument \cite{IFS} 
disregards the interaction term in the Hamiltonian, and this is again nothing but the
golden rule approximation (see also Appendix B of Ref. \onlinecite{GZr2} for
further discussion of this point). Due to interaction the ground state
energy of the total system is different from (in our case larger than) the sum of
energies of its noninteracting parts, and the energy exchange is always possible.

IFS  also pointed out that our results \cite{GZ1,GZ2} are
``in disagreement with experiments'' of Ref. \onlinecite{Gersh}. 
The experimental results for the dephasing time $\tau_{\varphi}$ were
reported only in one (the second) out of three papers under Ref. \onlinecite{Gersh}.
In this experiment it was found that  at sufficiently low $T$ (below $\sim 1\div 3$ K)
``the dependence $L_{\varphi}(T)$ is flattened out''. 
The comparison between our results for $\tau_{\varphi}$ and the
experimental findings of the second Ref. \onlinecite{Gersh} was carried out in Ref.
\onlinecite{GZ98}. An {\it excellent} agreement was revealed 
for all three samples studied in the second paper
\cite{Gersh}, see Fig. 3 of Ref. \onlinecite{GZ98}. Therefore, it remains unclear
which of our results was meant by the authors \cite{IFS} to be in disagreement
with experiments \cite{Gersh}.

Since this point was not clarified by IFS we can try to conjecture
that they actually interpreted our conclusion \cite{GZ1} that
Coulomb interaction in weakly disordered quasi-1d metallic (many channel) conductors
(described within the standard Drude model) precludes the Thouless crossover into an
insulating state as contradicting to
a rapid growth of the wire resistance with decreasing temperature detected
in Ref. \onlinecite{Gersh} below $T \sim 1$ K. If so, we can only point out that
the above conclusion does not contradict to the experimental data 
but only to their interpretation in terms of the Thouless crossover adopted
in Ref. \onlinecite{Gersh}. The wire conductivity can be represented as
a sum of the Drude term $\sigma_1$, the interaction correction
$\delta \sigma_{int}(T)$ and the weak localization correction $\delta \sigma_{WL}(T,H)$:
$$
\sigma_1+\delta \sigma_{int}(T)+\delta \sigma_{WL}(T,H).
$$
Our analysis \cite{GZ1,GZ2} demonstrates that the last of these three terms
saturates at low $T$, and this is in agreement with the observations reported
in the second Ref. \onlinecite{Gersh}. However the total resistance may well keep
increasing at even lower $T$ because of the interaction term $\delta \sigma_{int}(T)$.
This scenario indeed precludes the ``noninteracting'' Thouless
crossover, but not the crossover into an insulating state due to interaction (e.g.
of the Coulomb blockade type). Since in Refs. \onlinecite{GZ1,GZ2} we did not
address the term $\delta \sigma_{int}(T)$ at all, one can hardly argue about any contradiction
between our results and the experimental data \cite{Gersh}.

Finally let us turn to the paper by Vavilov and Ambegaokar \cite{VA}. These authors
did not employ the ``orthodox'' golden rule approximation \cite{CI,IFS,RSC}
but rather attempted to analyze the problem by means of a high temperature expansion.
They also presented a critical analysis of our paper \cite{GZ2} (see Appendix C
of Ref. \onlinecite{VA}).

VA questioned the validity of our procedure  
which amounts to deriving a dephasing time $\tau_{\varphi}$
only from the terms in our expression for the effective action
\cite{GZ2} which dominate at sufficiently long times.
According to VA the dropped terms might be important
at times $t \sim \tau_{\varphi}$ (where $S_I(\tau_{\varphi}) \sim 1$) and due to that
the result for $\tau_{\varphi}$ could be different from ours \cite{GZ2}. Even
without making any calculation one can realize that the contribution of
these dropped terms,
if important, could only make the dephasing time {\it shorter} than that found
in our paper \cite{GZ2}. Indeed, if one assumes that taking all terms into
account one would obtain a longer dephasing time $\tau_{\varphi}' \gg \tau_{\varphi}$,
one would immediately arrive at a contradiction with the fact (acknowledged by VA)
that the dropped terms are unimportant at times $t \gg \tau_{\varphi}$: at least at times
$t \sim \tau_{\varphi}'$ (and, hence, at $t \gg \tau_{\varphi}$) these dropped terms
should still be significant. Since this is not the case, by neglecting the above
terms one actually gets an upper bound for $\tau_{\varphi}$. This is in contrast with
the VA's claim that $\tau_{\varphi}$ is parametrically longer than that found
in our paper \cite{GZ2} at low $T$.

The analysis presented in Sec. 3A of this paper fully confirms our previous results
\cite{GZ2}. {\it All} terms of our effective action \cite{GZ2} were explicitly taken
into account in eq. (\ref{fott}). Additional terms in the exponent (see eqs. (\ref{fquantum}),
(\ref{fthermal})-(\ref{deltaf3}))
indeed appear, but (i) they lead to further suppression of quantum coherence and
(ii) for all times $t \gtrsim \tau_{\varphi}$ they are small as compared
to the leading order terms which we already considered before \cite{GZ2}. 

The problem with ``unphysical divergences'' discussed in the beginning of Appendix C
of Ref. \onlinecite{VA} does not exist in our analysis either. This can be
observed already from the footnote 32 before eq. (76)
of Ref. \onlinecite{GZ2}: the actual low frequency cutoff in this equation is
at $\omega \sim 1/t$, see also eqs. (46), (47) of Ref. \onlinecite{GZ98}.
This cutoff is not imposed by hand, but rather follows from
the fact that the integral over $\omega$ in eq. (76) of Ref. \onlinecite{GZ2} and related
formulas is just the long time approximation of the sum over discrete
Fourier frequencies $\omega_n=2\pi n/t$. [The cutoff at $\sim 1/\tau_{\varphi}$ was
used e.g. by AAK \cite{AAK} and in many other papers, we have followed this
approximate and physically correct procedure in eq. (76) of Ref. \onlinecite{GZ2} 
just in order
to illustrate the similarity of the low frequency contribution to $\tau_{\varphi}$
found in Refs. \onlinecite{GZ2} and \onlinecite{AAK}.] The absence of any
``unphysical divergences'' in the exact expression (\ref{fott}) is also
completely transparent.  It is hard to understand why VA believe that
our results cannot be compared to those
obtained within the Caldeira-Leggett model. A close similarity between both problems
is obvious from our analysis presented in Sec. 5.

Another problem of VA with our analysis has to do
with the factor $(1-2n({\bbox p}, {\bbox r}))$ which appears in the real
part $S_R$ of our effective action  (see eq. (\ref{SR})).
VA stated that in Ref.\onlinecite{GZ2} we
``neglected the time dependence of the momentum''. This is not true.
The electron momentum changes its direction after each scattering event,
an this fact is explicitly taken into account in Ref. \onlinecite{GZ2}
where the classical electron trajectories in a disordered potential
were considered. No momentum conservation was imposed in our analysis,
rather the electron energy conservation on these trajectories was used.
This approximation is fully justified, since the saddle point paths
are determined by the ``noninteracting'' terms in the effective action
while ``interacting'' terms are treated as a perturbation in the exponent.
For these saddle point paths the factor $(1-2n({\bbox p}, {\bbox r}))$
(and not ${\bbox p}(t)$) does not depend on time. It is not quite clear why
VA suggest to consider odd functions of
time $n({\bbox p}(t))$. Such functions do not contribute to the exponent
at all. It is also not clear in which context the observation
\cite{VA} that for an odd function of time  $n({\bbox p}(t))$ the real part
of the action $S_R$ is of the same order as $S_I$ (taken at $T=0$)
could be important: there is no way how the term $iS_R$ can cancel $S_I$
for real and nonzero $S_R$ and $S_I$. The contribution
of the trajectories with nonzero $S_R$ to the path integral may only be
suppressed further due to the presence of the oscillating term $\exp (-iS_R)$.

Trying to justify their arguments VA presented several
equations (eqs. (C12-C15) of Ref. \onlinecite{VA}) which are somewhat reminiscent
to ones obtained e.g. in Sec. 4 of this paper. For instance, the combination
``coth+tanh'' appears in eq. (C15) of Ref. \onlinecite{VA} in a correct form (\ref{exact})
rather than in the form (\ref{GR}) used by some other authors. Unfortunately
VA did not evaluate their eq. (C15) but
just concluded ``we see that the contribution of high
frequencies is exponentially suppressed''. In Sec. 4B we have demonstrated
just the opposite: the high frequency contribution is not exponentially
suppressed even at $T \to 0$ and, moreover, it leads to the presence of
diverging terms already in the first order in the interaction (cf. eq. (\ref{func})).
Exactly the same terms are contained in eq. (C15) of Ref. \onlinecite{VA} and,
hence, the above statement \cite{VA} is explicitly incorrect. 
In Ref. \onlinecite{VA}
we also failed to find a proof of even stronger VA's statement that one can ``observe the
same type of cancellation of high frequency contributions to the conductivity
in any order of perturbation theory''. 

Having observed that none of our results \cite{GZ2} suffers from the VA's
``critique''  let us come to the analysis of the calculation \cite{VA}. In order
to understand the problems with this calculation we have to start from
another paper of the same authors \cite{VA1}. VA acknowledged that the paper 
\cite{VA1} ``is simply wrong, because it treats the phase
as a single particle and looses the physics associated with the exclusion
principle''. We cannot agree with the second part of this statement. In our
opinion, it was a good idea of Ref. \onlinecite{VA1} to apply the Caldeira-Leggett
model in order to qualitatively understand the experimental results \cite{Webb}.
It is not the absence of the exclusion principle in this model, it is the
calculation of Ref. \onlinecite{VA1} which is problematic. In particular, we
mention that according to Ref. \onlinecite{VA1} the real part of the
action $S_R$ is responsible for the low temperature saturation of $\tau_{\varphi}$
whereas the contribution of its imaginary part $S_I$ is not important at $T \to 0$.
A correct calculation demonstrates the opposite: both in the CL model and in
the case of a disordered metal the term $S_I$ is important, while the term $S_R$ does
not contribute to the exponent (and thus to the dephasing time) at all. The
term $S_R$ vanishes on the saddle-point trajectories both with and without the
exclusion principle, i.e. the latter does not play any crucial role in the
problem in question.

Revising their earlier paper, VA applied the same method
of calculating the path integral as one employed in Ref. \onlinecite{VA1}.
Should the term $S_R$ be kept in the exponent, the result \cite{VA1} would
follow again. In order to avoid that VA expanded the
exponent to the first order in $S_R$ and in the ``quantum'' part of $S_I$
and made use of the perturbation theory developed in Ref. \onlinecite{AAG2}.
Several comments are in order.

\begin{enumerate}

\item Even if the perturbative calculations \cite{VA} of the weak localization
correction in the high temperature limit were correct, the correct dephasing
time $\tau_{\varphi}$ could not be extracted from these calculations.
The reason for that is exactly the same as for the perturbative calculation
\cite{AAG2}: the $T$-independent linear in time perturbative contributions
coming from the Taylor expansion of the exponent and the pre-exponent exactly
cancel each other in the first order. As it was explained in Sec. 2 the
problem of finding the dephasing time $\tau_{\varphi}$ is essentially
nonperturbative and, hence, the method  \cite{VA} should fail already
at the point where the exponent was (partially) expanded in powers of the
interaction.

\item The perturbative analysis \cite{AAG2} cannot simply be repeated
for the situation considered by VA because the initial (non-perturbed)
propagators depend on the classical part of the fluctuating field. This implies
that the Fourier transformation in time cannot easily be performed, and
the whole calculation should be redone from the very beginning.

\item In order to obtain the ``classical'' part of the weak localization
correction VA used a minimization procedure, the
meaning of which remains unclear. Indeed, let us
rewrite eq. (92) of Ref. \onlinecite{VA} for the weak localization correction
in the presence of interaction (for the sake of transparency we keep
the notations adopted in the above paper)
\begin{equation}
\Delta\sigma_{wl}=\Delta \sigma_{wl}^{(0)}+\Delta \sigma_{deph}+\Delta \sigma_{deph}',
\label{blsh1}
\end{equation}
As compared to eq. (92) of Ref. \onlinecite{VA}, in (\ref{blsh1}) we omitted the last
(smallest) term $\Delta \sigma_{cwl}$ which is unimportant for our
discussion. According to VA the first term $\Delta \sigma_{wl}^{(0)}$
comes from the ``diagram without electron-electron interaction with finite
dephasing rate $\gamma$''. This should imply
that the rate $\gamma$ in Ref. \onlinecite{VA} is {\it not} due to the
electron-electron interaction
but rather due to some other dephasing mechanism. E.g. in the
presence of the magnetic field one has $\gamma \equiv 1/\tau_H$ and eq. (93)
of Ref. \onlinecite{VA} for $\Delta \sigma_{wl}^{(0)}$ is equivalent to our 
eq. (\ref{H}) or to the first
equation (4.2b) of Ref. \onlinecite{AAG2}. The second term $\Delta \sigma_{deph}$
is equivalent to the first term in eq. (4.13a) of Ref. \onlinecite{AAG2} (or to the first term
in eq. (\ref{AAG155}) of the present paper)) provided one assumes $\gamma \equiv 1/\tau_H$.
This term already originates from the electron-electron interaction (classical
Nyquist noise), but again the decoherence rate $\gamma$ is {\it not} related
to this interaction but rather is due to some other mechanism. We also note
that the term $\Delta \sigma_{deph}$ written in the form of eq. (96) of
Ref. \onlinecite{VA} makes sense only if it is small 
$\Delta \sigma_{deph} \ll \Delta \sigma_{wl}^{(0)}$.

The next step of the authors \cite{VA} is not easy to understand. They {\it minimize} the
sum of the first two terms in (\ref{blsh1}) with respect to the dephasing rate
$\gamma$ and in this way they ``self-consistently'' derive
$\gamma \sim 1/\tau_{\varphi}^{AAK} \propto T^{2/3}$. Unfortunately VA did not
clarify (i) why the result (\ref{blsh1}) should be minimized with
respect to $\gamma$ and (ii) why the rate $\gamma$ depends now on the
electron-electron interaction though initially it was assumed that $\gamma$ is
{\it not} related to the electron-electron interaction at all. [Actually VA minimized
only the two first terms in (\ref{blsh1}). This implies that these terms can be
of the same order in contradiction to the inequality
$\Delta \sigma_{deph} \ll \Delta \sigma_{wl}^{(0)}$].
No explanation of the above procedure was given in Ref. \onlinecite{VA}.

\item VA found a negative ``quantum'' correction $\Delta \sigma_{deph}'$, see e.g. eqs. 
(88-89) of Ref. \onlinecite{VA}. From this result VA concluded
that the dephasing rate \cite{AAK} ``is overestimated, and the quantum correction
suppresses it''. In other words, according to the authors \cite{VA} quantum noise --
at least at not very low $T$ -- acts ``against'' the classical noise.
No physical reasons supporting this strange result was presented in Ref.
\onlinecite{VA}. Fortunately there is no need to look for such reasons. It is
obvious e.g. from our eq. (\ref{fthermal}) that the correction to the classical
dephasing rate \cite{AAK} is {\it positive} at any relevant temperature, i.e.
quantum noise may only enhance the dephasing effect of the classical noise.
Below we will demonstrate that no conclusion about both the sign and the value 
of the quantum correction to the classical dephasing rate can be drawn from
the expression for the leading order high temperature correction 
$\Delta \sigma_{deph}'$.

\item VA's procedure can easily be tested with the aid of the results
derived in the present paper. Since the authors \cite{VA} assume ``that the electron-electron
interaction is the only mechanism of decoherence'' and consider only the
high temperature limit, in our formulas of Sec. 2-4 we can put
$1/\tau_H=0$ and keep only the term
\begin{equation}
f_{1cl}(t)= \frac43\frac{e^2}{\sigma_1}\sqrt{\frac{D}{\pi}}Tt^{3/2}
\label{fcl}
\end{equation}
in the exponent of (\ref{dsigWL}) expanding this exponent to
the first order in $f_1(t)-f_{1cl}(t)$.
We also expand the pre-exponent $A_1(t)$ to the first order in the interaction
$A_1(t)\simeq A_1^{(0)}(t)+A_1^{(1)}(t)$, where $A_1^{(0)}(t)=1/(2\sqrt{\pi Dt})$.
This procedure should exactly correspond to the high
temperature expansion of Ref. \onlinecite{VA}. Then for the weak localization correction
(\ref{blsh1}) (which we now denote as $\delta \sigma_{WL}$) we will obtain
\begin{equation}
\delta\sigma_{WL} = -\frac{e^2\sqrt{D}}{\pi^{3/2}}
\int\limits_0^{+\infty} \frac{dt}{\sqrt{t}} e^{-f_{1cl}(t)}+
\frac{e^2\sqrt{D}}{\pi^{3/2}}
\int\limits_0^{+\infty} \frac{dt}{\sqrt{t}}\left(f_1(t)-f_{1cl}(t)-
\frac{A_1^{(1)}(t)}{A_1^{(0)}(t)}\right) e^{-f_{1cl}(t)}.
\label{chVA1}
\end{equation}
Here the first integral determines the ``classical'' part of the weak localization
correction while the second integral gives the ``quantum'' correction. Both integrals
can easily be evaluated. Combining eqs. (\ref{fthermal}), (\ref{dephthermal1}) and
(\ref{tanhthermal02}) (the last equation allows to evaluate the term
$A_1^{(1)}(t)/A_1^{(0)}(t)$) and keeping only the leading corrections
from the high temperature expansion we obtain
\begin{equation} f_1(t)-f_{1cl}(t)-
\frac{A_1^{(1)}(t)}{A_1^{(0)}(t)} \simeq
\frac{\zeta (1/2)}{\sqrt{2}}\frac{e^2}{\sigma_1}\sqrt{\frac{D}{\pi}}t\sqrt{T}.
\label{chVA2}
\end{equation}
As before, the $T$-independent linear in time terms contained in $f_1(t)$ (\ref{fthermal})
and in $A_1^{(1)}(t)/A_1^{(0)}(t)$ exactly cancel each other. This cancellation
illustrates again why
the correct dephasing time at low $T$ as well as the leading order high temperature
correction to the classical dephasing rate cannot be recovered by means
of a high temperature expansion for the conductance. From (\ref{fcl})-(\ref{chVA2}) we find
\begin{equation}
\delta\sigma_{WL} \simeq  -\left(\frac{2e^4\sigma_1D}{9\pi^4T}\right)^{1/3}\Gamma (1/3)
+\frac{\zeta (1/2)}{(2\pi )^{3/2}}
\frac{e^2\sqrt{D}}{\sqrt{T}}.
\label{chVA3}
\end{equation}
 
\end{enumerate}

We conclude that
the procedure developed by VA in Ref. \onlinecite{VA} does not provide correct
information about the interaction-induced dephasing in disordered conductors.


\end{document}